%% file: main.tex
\documentclass[lettersize,journal]{IEEEtran}
\usepackage{amsmath,amsfonts}
\usepackage{algorithmic}
\usepackage{algorithm}
\usepackage{array}
\usepackage[caption=false,font=normalsize,labelfont=sf,textfont=sf]{subfig}
\usepackage{textcomp}
\usepackage{stfloats}
\usepackage{url}
\usepackage{verbatim}
\usepackage{graphicx}
\usepackage{cite}
\usepackage[acronym]{glossaries}
\usepackage{amsmath}
\usepackage{booktabs}
\usepackage{multirow}
\usepackage{makecell}
\usepackage{textgreek}
\usepackage[inline]{enumitem}
\usepackage{tabularx}
\usepackage{threeparttable}
\usepackage[colorlinks=faluse]{hyperref}
\usepackage[capitalise,noabbrev,nameinlink]{cleveref}
\crefname{appendix}{Appendix}{Appendices}

\makeatletter
\newcommand\Autoref[1]{\@first@ref#1,@}
\def\@throw@dot#1.#2@{#1}% discard everything after the dot
\def\@set@refname#1{%
    \edef\@tmp{\getrefbykeydefault{#1}{anchor}{}}%
    \xdef\@tmp{\expandafter\@throw@dot\@tmp.@}%
    \ltx@IfUndefined{\@tmp autorefnameplural}%
         {\def\@refname{\@nameuse{\@tmp autorefname}s}}%
         {\def\@refname{\@nameuse{\@tmp autorefnameplural}}}%
}
\def\@first@ref#1,#2{%
  \ifx#2@\autoref{#1}\let\@nextref\@gobble% only one ref, revert to normal \autoref
  \else%
    \@set@refname{#1}%  set \@refname to autoref name
    \@refname~\ref{#1}% add autoefname and first reference
    \let\@nextref\@next@ref% push processing to \@next@ref
  \fi%
  \@nextref#2%
}
\def\@next@ref#1,#2{%
   \ifx#2@ and~\ref{#1}\let\@nextref\@gobble% at end: print and+\ref and stop
   \else, \ref{#1}% print  ,+\ref and continue
   \fi%
   \@nextref#2%
}
\makeatother

\crefformat{equation}{(#2#1#3)}
\Crefformat{equation}{(#2#1#3)}
\crefrangeformat{equation}{(#3#1#4) to (#5#2#6)}
\crefmultiformat{equation}{(#2#1#3)}{~and (#2#1#3)}{, (#2#1#3)}{~and (#2#1#3)}
\Crefmultiformat{equation}{(#2#1#3)}{~and (#2#1#3)}{, (#2#1#3)}{~and (#2#1#3)}

\graphicspath{ {./figures} }

\newacronym{noma}{NOMA}{non-orthogonal multiple access}
\newacronym{oma}{OMA}{orthogonal multiple access}
\newacronym{pd-noma}{PD-NOMA}{Power-Domain NOMA}
\newacronym{cd-noma}{CD-NOMA}{Code-Domain NOMA}
\newacronym{bs}{BS}{base station}
\newacronym{ue}{UE}{user equipment}
\newacronym{awgn}{AWGN}{additive white gaussian noise}
\newacronym{sic}{SIC}{successive interference cancellation}
\newacronym{mld}{MLD}{maximum likelihood detector}
\newacronym{ber}{BER}{bit error rate}
\newacronym{pep}{PEP}{pairwise error probability}
\newacronym{bpsk}{BPSK}{binary phase-shift keying}
\newacronym{qam}{QAM}{quadrature amplitude modulation}
\newacronym{pdf}{PDF}{probability density function}
\newacronym{cdf}{CDF}{cumulative distribution function}
\newacronym{ccdf}{CCDF}{complementary cumulative distribution function}
\newacronym{ser}{SER}{symbol error rate}
\newacronym{iot}{IoT}{internet of things}
\newacronym{qos}{QoS}{quality of service}
\newacronym{mimo}{MIMO}{multiple-input multiple-output}
\newacronym{snr}{SNR}{signal-to-noise ratio}
\newacronym{sinr}{SINR}{signal-to-interference-plus-noise ratio}
\newacronym{qpsk}{QPSK}{Quadrature Phase Shift Keying}
\newacronym{uwoc}{UWOC}{underwater wireless optical communication}
\newacronym{auv}{AUV}{autonomous underwater vehicles}
\newacronym{ind}{i.n.d.}{independent and non-identically distributed}
\newacronym{sc}{SC}{superposition coding}
\newacronym{iid}{i.i.d}{independent and identically distributed}

\DeclareMathOperator*{\argmin}{arg\,min}

\begin{document}

\title{Closed-Form BER Analysis for Uplink NOMA with Dynamic SIC Decoding}

\author{Hequn Zhang,~Qu Luo ~\IEEEmembership{Member,~IEEE},~Pei Xiao,~\IEEEmembership{Senior Member,~IEEE,}~Yue Zhang,~\IEEEmembership{Senior Member,~IEEE,}~and~Huiyu Zhou

\thanks{Hequn Zhang is with the School of Engineering, University of Leicester, LE1 7RH
Leicester, UK (e-mail: hz148@leicester.ac.uk).}
\thanks{Qu Luo and Pei Xiao are with 5GIC $\&$ 6GIC, Institute for Communication Systems (ICS) of University of Surrey,
Guildford, GU2 7XH, UK (e-mail:  q.u.luo@surrey.ac.uk, p.xiao@surrey.ac.uk).}
\thanks{Yue Zhang is with the Institute for Communication Systems and Measurement of China, Chengdu 610095, China (e-mail: zhangyue@icsmcn.cn).}
\thanks{Huiyu Zhou is with the School of Computing and Mathematical Sciences, University of Leicester, LE1 7RH
Leicester, UK (e-mail: hz143@leicester.ac.uk).}
}
\maketitle

\begin{abstract}
This paper, for the first time, presents a closed-form error performance analysis of uplink power-domain non-orthogonal multiple access (PD-NOMA) with dynamic successive interference cancellation (SIC) decoding, where the decoding order is adapted to the instantaneous channel conditions. We first develop an analytical framework that characterizes how dynamic ordering affects error probabilities in uplink PD-NOMA systems. For a two-user system over independent and non-identically distributed Rayleigh fading channels, we derive closed-form probability density functions (PDFs) of ordered channel gains and the corresponding unconditional pairwise error probabilities (PEPs). To address the mathematical complexity of characterizing ordered channel distributions, we employ a Gaussian fitting to approximate truncated distributions while maintaining analytical tractability. Finally, we extend the bit error rate analysis for various $M$-quadrature amplitude modulation schemes (QAM) in both homogeneous and heterogeneous scenarios. Numerical results validate the theoretical analysis and demonstrate that dynamic SIC eliminates the error floor issue observed in fixed-order SIC, achieving significantly improved performance in high signal-to-noise ratio regions. Our findings also highlight that larger power differences are essential for higher-order modulations, offering concrete guidance for practical uplink PD-NOMA deployment.
\end{abstract}


\section{Introduction}
\IEEEPARstart{N}{OMA} is an advanced multiple access technique designed to enhance spectral efficiency and provide massive connectivity for future wireless communication systems\cite{9693417,10024901,ett4289}. Unlike traditional \gls{oma} schemes, which allocate time/frequency resources in an orthogonal manner, \gls{noma} enables multiple users to share the same resources non-orthogonally, facilitating overloaded multi-user communications\cite{8370069}. The existing \gls{noma} schemes can be broadly classified into two categories:  power domain NOMA (PD-NOMA) and code-domain \gls{noma}. The former multiplexes users by allocating different power levels\cite{6692652}, while the latter employs specific codes or sequences\cite{9369968}. Among these two approaches, this paper focuses on \gls{pd-noma}. Owing to its distinct advantages, such as low detection complexity and high spectral efficiency, \gls{pd-noma} has been widely studied in both industry and academia. 

In \gls{pd-noma}, user signals are superimposed at the transmitter using superposition coding and decoded at the receiver via \gls{sic}, which sequentially removes stronger signals based on their decoding order. In downlink scenarios, the decoding order is explicitly controlled by the \gls{bs} through power allocation, and all superimposed signals experience the same propagation channel, making the system analytically tractable. In contrast, uplink \gls{noma} presents fundamentally different challenges: user signals undergo independent and often \gls{ind} fading channels, and the decoding order depends dynamically on instantaneous channel realizations. This randomness disrupts conventional analytical tools based on fixed ordering or \gls{iid} assumptions, making it difficult to derive closed-form \gls{pep} and \gls{ber} expressions. \textit{This paper derives closed-form \gls{ber} expressions for uplink \gls{pd-noma} systems with dynamic \gls{sic} decoding.}

\subsection{Related works}
Accurate \gls{ber} analysis plays a critical role in the design and optimization of \gls{noma} systems, as it provides essential insights into system reliability and facilitates practical design. The unconditional \gls{pep} and union bound computation provide tractable and tight performance estimates for \gls{ber} analysis~\cite{8501953}. However, due to the mathematical complexity involved, most existing studies have focused on downlink \gls{noma} systems, where analytical simplifications are feasible. In downlink \gls{pd-noma}, all user signals propagate through the same fading channel from \gls{bs} to \gls{ue}, and the decoding order is explicitly controlled through power allocation. These features facilitate the derivation of closed-form \gls{pep} expressions. The work in~\cite{iet-com.2018.5278} evaluated downlink \gls{ber} performance under the simplifying assumption that users follow a fixed decoding order based on average channel gains. However, this approach neglects the inherent randomness of the fading channels, which can cause deviations from the average-based order. More rigorous studies such as~\cite{8501953,8843850} derived closed-form \gls{pep} expressions for downlink \gls{noma} systems over Nakagami-$m$ fading by integrating the $Q$-function over the ordered channel gain distributions. Similarly,~\cite{8732367} extended this approach to cognitive radio networks using \gls{noma}. 

In contrast, several studies have analyzed the \gls{ber} performance of uplink \gls{pd-noma} systems. The work in~\cite{10873839} investigated a general uplink scenario with \gls{sic}-based decoding, assuming \gls{iid} fading channels and a fixed decoding order. It should be noted that unlike the downlink, the channels between each \gls{ue} and the \gls{bs} are independently and often \gls{ind}, resulting in complex joint distributions of user signal powers. Therefore, the fixed-order assumption in~\cite{10873839} may not be practical for uplink environments. 
To partially address the dynamic nature of decoding,~\cite{10873839} investigated the general \gls{ber} performance of uplink \gls{noma} with \gls{sic}, where the authors assume that the channel distribution of each user is \gls{iid} and the decoding order is fixed. However, since the propagation paths of different \glspl{ue} typically differ significantly in uplink scenarios, the \gls{iid} assumption is not valid in most cases. The research work~\cite{7959539} considered dynamic \gls{sic} for uplink \gls{noma} systems with \gls{ind} fading channels. Rather than deriving ordered \glspl{pdf} from the underlying channel distributions, it directly assumes exponential distributions for the instantaneous received power without establishing their connection to the original channel statistics, the method sacrifices generality and may not accurately represent practical channel conditions. Several studies investigated the \gls{ber} performance of uplink \gls{noma} systems, but the complexity of uplink channels has prevented the derivation of closed-form \gls{pep} expressions~\cite{10531681,8771371}. For instance,~\cite{10531681} analyzed autonomous underwater vehicles-based underwater wireless optical communication with uplink \gls{noma} and derived the ordered \gls{pdf} and \gls{cdf}, but the inherent complexity of the underwater wireless optical communication channel precludes obtaining closed-form unconditional \gls{pep} expressions. Similarly,~\cite{8771371} examined the upper bound \gls{ber} performance under the assumption that \glspl{ue} are arranged in decreasing order of their estimated channel gains, yet it also fails to provide closed-form \gls{pep} expressions.

\subsection{Motivations and Contributions}
The decoding order in uplink \gls{pd-noma} is inherently stochastic, as it depends on the instantaneous channel realizations of independently fading users. The \gls{ind} nature of uplink channels, where signals from different \glspl{ue} experience independent fading with distinct statistical parameters, leads to complex joint distributions that hinder accurate modeling of ordered channel gains. Existing uplink \gls{noma} studies either rely on oversimplified assumptions such as \gls{iid} channels and fixed decoding orders~\cite{10873839}, or fail to derive closed-form expressions necessary for practical system design~\cite{10531681,8771371}. Even attempts to incorporate dynamic ordering~\cite{7959539} resort to ad-hoc distributional assumptions without establishing their connection to the underlying channel statistics. Despite the critical importance of dynamic ordering in uplink \gls{noma}, existing literature has not provided closed-form \gls{pep} expressions that properly account for the stochastic variations in decoding order. 

To address this research gap, this paper presents a comprehensive analytical framework for uplink \gls{noma} systems employing dynamic \gls{sic} decoding. We rigorously analyze how dynamic ordering transforms the statistical properties of channel distributions and quantify its impact on system error performance. Our main contributions are as follows:
\begin{enumerate}
  \item We propose a unified analytical framework for BER analysis in uplink \gls{noma} systems with dynamic \gls{sic}. The framework rigorously accounts for all possible decoding orders and their associated probabilities, enabling accurate performance evaluation under practical \gls{ind} fading conditions.
  
  \item We derive the exact closed-form \glspl{pdf} of ordered channel gains for a two-\gls{ue} uplink \gls{noma} system over \gls{ind} Rayleigh fading channels. By addressing the mathematical complexity arising from the non-identical channel distributions, we employ Gaussian fitting to approximate truncated distributions and subsequently obtain the first closed-form \gls{pep} expressions for uplink \gls{noma} with dynamic \gls{sic} decoding, revealing that dynamic ordering eliminates the error floor phenomenon inherent in fixed-order systems.
  
  \item We extend our analytical framework to derive closed-form \gls{ber} expressions for practical $\mathcal{M}$-\gls{qam} modulation schemes (\gls{bpsk}, 4\gls{qam}, 16\gls{qam}, 64\gls{qam}) in both homogeneous and heterogeneous configurations. The analysis quantifies the minimum power separation requirements for different modulation orders and provides design guidelines for reliable \gls{sic} operation in uplink \gls{noma} systems.

  \item We provide extensive numerical results that validate our theoretical analysis across diverse system configurations. The simulations confirm the accuracy of our closed-form expressions and demonstrate that dynamic \gls{sic} achieves significant performance improvements over fixed-order decoding, particularly in eliminating error floors at high \gls{snr} regimes while maintaining excellent agreement between theoretical predictions and Monte Carlo simulations.
\end{enumerate}

\subsection{Organization}
The rest of this paper is organized as follows: \autoref{sec:system_model} presents the uplink \gls{noma} system with dynamic \gls{sic} decoding; \autoref{sec:2_ue_ul_tx_rayleigh} presents the closed-form \gls{pdf} and \gls{pep} of 2 \glspl{ue}; \autoref{sec:ber_m_qam} discusses the \gls{ber} with higher modulation orders; \autoref{sec:numerical_results} presents the numerical results; and conclusions are given in \autoref{sec:conclusion}. The appendices at the end of this paper provide detailed mathematical derivations and proofs.

\section{System model}\label{sec:system_model}
In an uplink \gls{noma} system, a \gls{bs} receives signals from $N$ \glspl{ue}, with both the \gls{bs} and \glspl{ue} equipped with a single antenna. The received signal can be expressed as\cite{8876877}:
{\small
\begin{align}
  y=\sum_{n=1}^{N}\sqrt{p_n}h_nx_n+w,
  \label{eq:ul_noma_rx_signal}
\end{align}
}%
where $x_n$ is the modulated symbol of \gls{ue} $n$, $p_n$ denotes the normalized transmission power coefficient of \gls{ue} $n$, satisfying $\sum_{n=1}^{N}p_n = 1$, $w\sim\mathcal{CN}\left(0, N_0\right)$ denotes the \gls{awgn}, and $h_n$ represents the channel coefficient between \gls{ue} $n$ and the \gls{bs}. We assume that the channels of \glspl{ue} are \gls{ind} and, to focus solely on the \gls{sic} decoding performance without effects from channel estimation imperfections, there is no channel estimation error at the receiver side (i.e., all channel coefficients are perfectly estimated and recovered).

\gls{sic} is applied to decode the signal of each \gls{ue} at the \gls{bs} side, with the decoding order optimized based on the instantaneous channel gains. In dynamic \gls{sic}, the \gls{bs} sorts the received signals according to their instantaneous channel gains $|h_n|$ in descending order, rather than following a predetermined sequence based on \gls{ue} indices or average channel conditions. This adaptive ordering ensures that the strongest signal is decoded first, minimizing interference for subsequent decoding stages.

To formalize this process, we use the notation $(k)$ to denote the $k$-th decoding order. For example, (1) represents the first decoded \gls{ue} with the strongest instantaneous channel gain, (2) corresponds to the second decoded \gls{ue} with the second strongest channel gain, and so on. Notably, the mapping between decoding order and \gls{ue} index $n$ varies with each channel realization. For instance, if the instantaneous channel gains satisfy $|h_2|>|h_3|>|h_1|$, then the decoding proceeds as: order (1) corresponds to \gls{ue} 2, order (2) to \gls{ue} 3, and order (3) to \gls{ue} 1. This dynamic adaptation to channel conditions is the key distinction from fixed-order \gls{sic}, where the decoding sequence remains constant regardless of instantaneous channel variations.

Given that \gls{sic} imperfectly cancels interference, residual interference may persist after decoding the previous \gls{ue}. Additionally, with interference from undecoded signals, the received signal decoded in the $(k)$-th order can be expressed as~\cite{9115309}:
{\small
\begin{align}
y_{(k)}&=\underbrace{\sqrt{p_{(k)}}h_{(k)}{x}_{(k)}}_{\text{desired signal}}\nonumber\\
  &+\underbrace{\sum_{i=1}^{k-1}\sqrt{p_{(i)}}h_{(i)}\Delta_{(i)}}_{\text{residual interference}}+\underbrace{\sum_{j=k+1}^{N}\sqrt{p_{(j)}}h_{(j)}{x}_{(j)}}_{\text{interference}}+w\nonumber\\
  &=\sqrt{p_{(k)}}h_{(k)}{x}_{(k)}+\zeta_{(k)}+w,
  \label{eq:ul_noma_rx_signal_k}
\end{align}
}%
where $\sqrt{p_{(k)}}$ and $h_{(k)}$ are the power coefficient and channel of the desired signal. $\sum_{i=1}^{k-1}\sqrt{p_{(i)}}h_{(i)}\Delta_{(i)}$ is the residual interference from previously decoded signals, where $\Delta_{(i)}=|x_{(i)}|-|\check{x}_{(i)}|$ is the residual estimation error and $x_{(i)}, \check{x}_{(i)}$ are the transmitted signal and the decoded signal. $\sum_{j=k+1}^{N}\sqrt{p_{(j)}}h_{(j)}{x}_{(j)}$ is the interference from signals yet to be decoded. The total interference can be denoted as $\zeta_{(k)}=\sum_{i=1}^{k-1}\sqrt{p_{(i)}}h_{(i)}\Delta_{(i)}+\sum_{j=k+1}^{N}\sqrt{p_{(j)}}h_{(j)}{x}_{(j)}$. The decoding order is determined by the channel gains, such that $|h_{(i)}|\geq|h_{(k)}|\geq|h_{(j)}|$, for any $i$ and $j$.
Similar to ~\cite{7676258}, we employ a power allocation strategy where users with stronger instantaneous channel gains receive higher transmission power, specifically $p_{(1)} > p_{(2)} > \ldots > p_{(N)}$. Since the decoding order adapts to instantaneous channel conditions, when $|h_n|$ is the $k$-th largest channel gain, \gls{ue} $n$ transmits with power $p_n = p_{(k)}$. This approach    ensures sufficient power separation for reliable SIC operation.

At the receiver side, after the \gls{bs} determines the decoding order based on the instantaneous channel gains, the modulated symbol of $(k)$-th signal can be decoded by deploying \gls{mld}:
{\small
\begin{align}
\hat{x}_{(k)}&=\argmin_{\tilde{x}_{(k)}\in\mathbb{X}_{(k)}}|y_{(k)}-\sqrt{p_{(k)}}h_{(k)}\tilde{x}_{(k)}|^2,
  \label{eq:mld}
\end{align}
}%
where $\mathbb{X}_{(k)}$ denotes the set of modulation symbols of the $(k)$-th signal. After decoding the $(k)$-th signal, the signal of the $(k+1)$-th \gls{ue} can be decoded by cancelling the decoded $(k)$-th signal by $y_{(k+1)}=y_{(k)}-\sqrt{p_{(k)}}h_{(k)}\hat{x}_{(k)}$. This process continues until all \glspl{ue} are decoded.

\section{Error Rate Analysis of \gls{ue} $n$}
In this section, we analyze the error performance of an arbitrary \gls{ue} $n$ in the uplink \gls{noma} system with dynamic \gls{sic} decoding.

Since error probabilities depend on the decoding order, the average error probability of \gls{ue} $n$ must consider all possible decoding sequences. Let $A_{n}$ denote the event that \gls{ue} $n$ is decoded incorrectly, and let $B_{n, (k)}$ denote the event that \gls{ue} $n$ is decoded in the $k$-th position. Using Bayes' theorem, the error probability of \gls{ue} $n$ can be expressed as:
{\small
\begin{align}
P\left(A_{n}\right)&=\sum_{k=1}^{N}P\left(A_{n}\mid B_{n, (k)}\right)P\left(B_{n, (k)}\right),
  \label{eq:tot_p}
\end{align}
}%
where $P(A_{n} \mid B_{n, (k)})$ is the conditional probability of \gls{ue} $n$ being decoded incorrectly given that it is decoded in the $k$-th position, and $P(B_{n, (k)})$ is the probability that \gls{ue} $n$ has the $k$-th largest instantaneous channel gain among all \glspl{ue}.

This paper incorporates error propagation effects in the \gls{sic} detection process. When an error occurs during the decoding of the $(k)$-th order signal, it propagates and impacts the subsequent decoding of the $(k+1)$-th order signal~\cite{9115309}. To accurately model this phenomenon, the error probability of the $(k+1)$-th order decoding must account for the cumulative effects of all previous decoding errors through
{\small
\begin{align}
  P\left(A_{n}\mid B_{n, (k+1)}\right) &= P\left(A_{n}\mid \bar{A}_{(k)}, B_{n, (k+1)}\right) P\left(\bar{A}_{(k)}\mid B_{(k)}\right)\nonumber\\
  &\quad + P\left(A_{n}\mid A_{(k)}, B_{n, (k+1)}\right) P\left(A_{(k)}\mid B_{(k)}\right),
  \label{eq:err_prop_k_plus_1}
\end{align}
}%
where $\bar{A}_{(k)}$ and $A_{(k)}$ denote the events that the $k$-th decoded signal (from the previous user) is decoded correctly and incorrectly, respectively, and $P\left(\bar{A}_{(k)}\mid B_{(k)}\right)=1-P\left(A_{(k)}\mid B_{(k)}\right)$. This decomposition captures the error propagation effect: the error probability of user $n$ decoded at position $(k+1)$ depends on whether the previous user was correctly decoded. For the special case of $k=1$ (first decoding position), no error propagation occurs, yielding $P\left(A_{n}\mid B_{n, (1)}\right)$ directly.

To evaluate the error probability in \cref{eq:tot_p}, we decompose the computation into two fundamental components: the conditional error probability $P(A_{n} \mid B_{n, (k)})$, which quantifies the error performance given a specific decoding position, and the decoding order probability $P(B_{n, (k)})$, which determines the likelihood of \gls{ue} $n$ being decoded at position $k$.

\subsection{Calculation of $P\left(A_{n}\mid B_{n, (k)}\right)$}
To analyze the error performance under dynamic decoding order, we employ the conditional \gls{pep} $P\left(x_{(k)}\rightarrow \check{x}_{(k)}\mid\mathbf{h}\right)$ to characterize the error probability $P\left(A_{n}\mid B_{n, (k)}\right)$~\cite{485720}. This conditional \gls{pep} represents the probability that the transmitted symbol $x_{(k)}$ is erroneously decoded as $\check{x}_{(k)}$ when \gls{ue} $n$ occupies the $k$-th position in the decoding order.

Let $\mathbf{h}=\{h_{(i)}, i=1, \dots, N\}$ denote the vector of channel gains sorted in decreasing order of magnitude, where $|h_{(k)}|$ corresponds to the channel gain of the \gls{ue} decoded at position $k$. Following the \gls{mld} criterion in \eqref{eq:mld}, the conditional \gls{pep} is derived as:
{\small
\begin{align}  
  P\left(A_{n}\mid B_{n, (k)}\right)&=P\left(x_{(k)}\rightarrow \check{x}_{(k)}\mid\mathbf{h}\right)\nonumber\\  
  &=P\left(2\mathfrak{R}\left\{\xi_{(k)}w\right\}\geq|\xi_{(k)}|^2+2\mathfrak{R}\left\{\xi_{(k)}\zeta_{(k)}\right\}\right),
  \label{eq:cond_pep}
\end{align}
}%
where $\xi_{(k)}=\sqrt{p_{(k)}}h_{(k)}\Delta_{(k)}$ denotes the residual of the desired signal with $\Delta_{(k)}=x_{(k)}-\check{x}_{(k)}$ being the symbol error. Since the noise $w\sim\mathcal{CN}\left(0, N_0\right)$, the term $2\mathfrak{R}\{\xi_{(k)}w\}$ follows a Gaussian distribution $\mathcal{N}\left(0, 2|\xi_{(k)}|^2N_0\right)$.
Consequently, the conditional \gls{pep} can be expressed using the Q-function. Define $z_{(k)}$ as:
{\small
\begin{align}
  z_{(k)} = \frac{|\xi_{(k)}|+2\mathfrak{R}\left\{\zeta_{(k)}\right\}}{\sqrt{2N_0}}.
  \label{eq:z_k_def}
\end{align}
}%
Then, the conditional \gls{pep} becomes
{\small
\begin{align}
  P\left(x_{(k)}\rightarrow \check{x}_{(k)}\mid\mathbf{h}\right) = Q\left(z_{(k)}\right) = Q\left(\frac{|\xi_{(k)}|+2\mathfrak{R}\left\{\zeta_{(k)}\right\}}{\sqrt{2N_0}}\right).
  \label{eq:cpep_q}
\end{align}
}%
To obtain the unconditional error probability $P\left(A_{n}\mid B_{n, (k)}\right)$, we integrate the Q-function over the distribution of $z_{(k)}$:
{\small
\begin{align}
  P\left(A_{n}\mid B_{n, (k)}\right)=\int_{0}^{\infty}Q\left(z\right)f_{\mathcal{Z}_{(k)}}\left(z\right)dz,
  \label{eq:upep_integ}
\end{align}
}%
where $f_{\mathcal{Z}_{(k)}}\left(z\right)$ denotes the \gls{pdf} of the random variable $z_{(k)}$. Deriving this \gls{pdf} requires analyzing the ordered channel statistics and the interference structure. We address this challenge for the two-\gls{ue} case in \autoref{sec:probability_an_bnk}, where we derive the closed-form expressions for these \glspl{pdf}.

\subsection{Calculation of $P\left(B_{n, (k)}\right)$}\label{sec:P_Bnk}
The event $B_{n, (k)}$ depends on the instantaneous channel gains of $N$ \glspl{ue} arranged in decreasing order of magnitude, where index $(k)$ denotes the $k$-th largest channel gain. Specifically, $B_{n, (k)}$ represents the event that \gls{ue} $n$ has the $k$-th largest channel gain among all \glspl{ue}, which occurs when $|h_n|$ lies between the $(k-1)$-th and $(k+1)$-th ordered statistics. The probability of this event can be expressed as:
{\small
\begin{align}
  P\left(B_{n, (k)}\right)&=P\left(|h_{(k+1)}|\leq|h_{n}|<|h_{(k-1)}|\right)\nonumber\\
  &=P\left(|h_{(k+1)}|\leq|h_{n}|\right)-P\left(|h_{(k-1)}|\leq|h_{n}|\right),
  \label{eq:P_Bnk}
\end{align}
}%
where $P\left(|h_{(r)}|\leq|h_n|\right)$ denotes the probability that the $r$-th order statistic of the remaining $N-1$ \glspl{ue} (excluding \gls{ue} $n$) is less than or equal to $|h_n|$. Special attention must be given to the boundary cases: when $k=1$ (strongest channel), we have $P\left(B_{n, (1)}\right)=P\left(|h_{(2)}|\leq|h_n|\right)$, and when $k=N$ (weakest channel), we have $P\left(B_{n, (N)}\right)=1-P\left(|h_{(N-1)}|\leq|h_n|\right)$.

To obtain $P\left(|h_{(r)}|\leq|h_{n}|\right)$, we integrate over all possible values of $|h_n|$, weighted by its probability density. Specifically, for each realization $x$ of $|h_n|$, we compute the probability that the $(r)$-th order statistic does not exceed $x$. This yields:
{\small
\begin{align}
  P\left(|h_{(r)}|\leq|h_{n}|\right) = \int_{0}^{\infty}P\left(|h_{(r)}|\leq x\right)f_{|h_{n}|}\left(x\right)dx.
  \label{eq:P_hr_leq_hn}
\end{align}
}%
Here, $(r)$ denotes the decoding position in the range $1\leq r\leq N-1$, and $F_{|h_{(r)}|}\left(x\right) = P\left(|h_{(r)}|\leq x\right)$ represents the \gls{cdf} of the $(r)$-th order statistic. For \gls{ind} variables, this \gls{cdf} can be computed using order statistics theory~\cite[Theorem 4.1]{f2273ed0-b48a-3ace-9640-7dc852794849}. The general expression and two special cases are derived in~\cref{app:proof_F_Hk}.

\subsection{Calculation of \gls{ser} of \gls{ue} $n$}\label{sec:ser_ue_n}
The probability $P\left(A_{n}\right)$ derived thus far represents the error probability for only one specific combination of desired, residual, and interference signals. To compute the \gls{ser}, we must account for all possible symbol combinations in the system. We define $\mathbf{S}_{i}=\{\Delta_{(1)}, \ldots, \Delta_{(k)}, x_{(k+1)}, \ldots, x_{(N)}\}$ as the $i$-th combination of residual errors and interference symbols, where $\Delta_{(j)}=x_{(j)}-\check{x}_{(j)}$ denotes the residual error for the $j$-th decoded symbol, with both $x_{(j)}$ (transmitted) and $\check{x}_{(j)}$ (detected) belonging to the constellation set $\mathbb{X}_{(j)}$. Let $\mathbb{S}=\{\mathbf{S}_{1}, \mathbf{S}_{2}, \ldots\}$ represent the set of all possible combinations, whose cardinality $|\mathbb{S}|$ depends on both the number of \glspl{ue} $N$ and their respective modulation schemes.

For $\mathcal{M}$-\gls{qam} schemes, the constellation exhibits symmetry between in-phase and quadrature components, allowing us to analyze only the in-phase component without loss of generality. We define $\mathbb{X}_{(j)}$ as the set of amplitudes for in-phase components of all possible symbols:
{\small
\begin{align}
  \mathbb{X}_{(j)}=
  \begin{cases}
  \{\pm(2\ell-1)d : 1\leq \ell\leq \sqrt{\mathcal{M}_{(j)}}\} & \text{for } \mathcal{M}_{(j)}\geq 4 \\
  \{\pm d\} & \text{for } \mathcal{M}_{(j)}=2
  \end{cases},
\end{align}
}%
where the scaling factor $d$ ensures unit average energy per symbol and is given by~\cite{752121}:
{\small
\begin{align}
  d=\sqrt{\frac{3E_b\log_2\mathcal{M}_{(j)}}{2(\mathcal{M}_{(j)}-1)}}.
  \label{eq:d}
\end{align}
}%
Here, $E_b$ denotes the energy per bit, and the factor $\log_2\mathcal{M}_{(j)}$ accounts for the number of bits per symbol.

Having defined all possible symbol combinations, the \gls{ser} of \gls{ue} $n$ is obtained by summing the error probabilities across all combinations~\cite{9351759}:
{\small
\begin{align}
  P_{\text{SER}}(n)=\sum_{\mathbf{S}_{i}\in\mathbb{S}}P\left(A_{n}\right).
  \label{eq:ser}
\end{align}
}%

\section{Two-\glspl{ue} Uplink Transmission over Rayleigh Channels}\label{sec:2_ue_ul_tx_rayleigh}
This section derives the closed-form \gls{pep} for a two-\gls{ue} uplink \gls{noma} system over \gls{ind} Rayleigh fading channels. The channel coefficients $h_1$ and $h_2$ follow Rayleigh distributions with parameters $\sigma_1$ and $\sigma_2$, where $\sigma_i^2$ represents the average channel power of \gls{ue} $i$ (i.e., $\mathbb{E}[|h_i|^2] = \sigma_i^2$). Since $\sigma_1 \neq \sigma_2$, the channels are \gls{ind}. Without loss of generality, we assume $\sigma_1^2 > \sigma_2^2$, indicating that \gls{ue} 1 has a stronger average channel condition.

Due to the dynamic \gls{sic} decoding, each \gls{ue} can be decoded in either first or second position depending on instantaneous channel realizations. Using the law of total probability from \cref{eq:tot_p}, the \gls{pep} for \gls{ue} 1 and \gls{ue} 2 can be expressed as:
{\small
\begin{equation}
\begin{aligned}
  P(A_{1}) &= P(A_{1}\mid B_{1, (1)})P(B_{1, (1)}) + P(A_{1}\mid B_{1, (2)})P(B_{1, (2)}), \\
  P(A_{2}) &= P(A_{2}\mid B_{2, (1)})P(B_{2, (1)}) + P(A_{2}\mid B_{2, (2)})P(B_{2, (2)}), \label{eq:P_A1_A2}
\end{aligned}
\end{equation}
}%
where $A_n$ denotes the error event for \gls{ue} $n$, and $B_{n, (k)}$ represents the event that \gls{ue} $n$ is decoded in the $k$-th order.

When a \gls{ue} is decoded second, its error probability depends on whether the first \gls{ue} was correctly decoded. This error propagation effect, formalized in \cref{eq:err_prop_k_plus_1}, yields:
\begin{equation}
\small
\begin{aligned}
  P(A_{1}\mid B_{1, (2)}) &= P(A_{1}\mid \bar{A}_{2}, B_{1, (2)}) P(\bar{A}_{2}\mid B_{2, (1)}) \\
  &\quad + P(A_{1}\mid A_{2}, B_{1, (2)}) P(A_{2}\mid B_{2, (1)}), \\
  P(A_{2}\mid B_{2, (2)}) &= P(A_{2}\mid \bar{A}_{1}, B_{2, (2)}) P(\bar{A}_{1}\mid B_{1, (1)}) \\
  &\quad + P(A_{2}\mid A_{1}, B_{2, (2)}) P(A_{1}\mid B_{1, (1)}), \label{eq:P_A1_B12_P_A2_B22}
\end{aligned}
\end{equation}%
where $\bar{A}_n$ denotes the complement of $A_n$ (i.e., correct decoding of \gls{ue} $n$), and we use the fact that $P(\bar{A}_n\mid B_{n, (k)}) = 1 - P(A_n\mid B_{n, (k)})$.

To evaluate the error probabilities in \cref{eq:P_A1_A2}, we need to determine two key components: the conditional error probabilities $P(A_n\mid B_{n, (k)})$ and the decoding order probabilities $P(B_{n, (k)})$. The following subsections derive closed-form expressions for these probabilities by leveraging the statistical properties of the ordered channel gains in the Rayleigh fading channels.

\subsection{Calculation of  $P\left(A_{n}\mid B_{n, (k)}\right)$}\label{sec:probability_an_bnk}
From \cref{eq:upep_integ}, the closed-form expression for $P\left(A_{n}\mid B_{n, (k)}\right)$ is obtained by evaluating the integral of the Q-function $Q\left(z_{(k)}\right)$ and the \gls{pdf} $f_{\mathcal{Z}_{(k)}}\left(z_{(k)}\right)$. In the two-\gls{ue} scenario, each \gls{ue} can be decoded in one of two orders: $(1)$ first (stronger channel) or $(2)$ second (weaker channel), yielding error probabilities $P\left(A_{n}\mid B_{n, (1)}\right)$ and $P\left(A_{n}\mid B_{n, (2)}\right)$.

For \glspl{ue} $n$ and $m$ where $n, m\in\{1, 2\}$, $n\neq m$, the normalized detection statistics are:
{\small
\begin{equation}
\begin{aligned}
z_{n, (1)} &= \frac{|\xi_{n, (1)}|+2\mathfrak{R}\left\{\zeta_{n, (1)}\right\}}{\sqrt{2N_0}}, \\
z_{n, (2)} &= \frac{|\xi_{n, (2)}|+2\mathfrak{R}\left\{\zeta_{n, (2)}\right\}}{\sqrt{2N_0}},
\end{aligned}
\end{equation}
}%
where:
{\small
\begin{equation}
\begin{aligned}
    \xi_{n, (k)} &= \sqrt{p_{(k)}}h_{n, (k)}\Delta_{n, (k)} \quad \text{desired signal residual}, \\
    \zeta_{n, (1)} &= \sqrt{p_{(2)}}h_{m, (2)}{x}_{m, (2)} \quad \text{interference from \gls{ue} $m$}, \\
    \zeta_{n, (2)} & = \begin{cases}
    \sqrt{p_{(1)}}h_{m, (1)}\Delta_{m, (1)} & \text{for incorrect decoding (${A}_{m}$)}, \\
    0 & \text{for correct decoding ($\bar{A}_{m}$)}.
\end{cases}
\end{aligned}
\end{equation}
}%

The closed-form \glspl{pdf} of $z_{n, (1)}$ and $z_{n, (2)}$ are essential for computing the probability $P\left(A_{n}\mid B_{n, (k)}\right)$. Through the detailed derivations presented in~\cref{app:pdf_zn1_zn2}, we obtain the following closed-form expressions. For the first decoding order, the \gls{pdf} of $z_{n, (1)}$ is given in \cref{eq:cf_pdf_z_1}. For the second decoding order, the \gls{pdf} of $z_{n, (2)}$ depends on whether the previous UE was correctly decoded: when $\zeta_{n, (2)}\neq0$ (incorrect decoding), the \gls{pdf} is given in \cref{eq:cf_pdf_z_2_incor}, while when $\zeta_{n, (2)}=0$ (correct decoding), it simplifies to the expression in \cref{eq:cf_pdf_z_2_cor}.

\begin{figure*}[!t]
  {\small
  \begin{align}\label{eq:cf_pdf_z_1}
    f_{\mathcal{Z}_{n, (1)}}\left(z\right) &= \frac{2a_1\sqrt{2N_0p_{(2)}\pi}\mathfrak{R}\left\{x_{m, (2)}\right\}}{\sqrt{p_{(1)}\Delta_{n, (1)}}} \exp\left(-\frac{b_1^2}{c_1^2}\right) \left( \sqrt{p_{(1)}}|\Delta_{n, (1)}|c_1^2\sqrt{2N_0}z \left( \frac{\mathcal{E}\left(\sigma_{n}\right)D_2\left(\sigma_{n}\right)\sigma_{n}^2}{D_1\left(\sigma_{n}\right)} - \frac{\mathcal{E}\left(\sigma_{m}\right)D_2\left(\sigma_{m}\right)\sigma_{m}^2}{D_1\left(\sigma_{m}\right)} \right) \right. \nonumber \\
    & \left. + \left( p_{(1)}|\Delta_{n, (1)}|^2 \left( -\frac{2b_1\mathcal{E}\left(\sigma_{n}\right)\sigma_{n}^3}{D_1\left(\sigma_{n}\right)^{3/2}} + \frac{2b_1\mathcal{E}\left(\sigma_{m}\right)\sigma_{m}^3}{D_1\left(\sigma_{m}\right)^{3/2}} \right) \right.\right.\nonumber\\
    & \left.\left. + 4p_{(2)}\left(\mathfrak{R}\left\{x_{m, (2)}\right\}\right)^2c_1^2\sqrt{2N_0}z \left( -\frac{\mathcal{E}\left(\sigma_{n}\right)\sigma_{n}}{D_1\left(\sigma_{n}\right)^{3/2}} + \frac{\mathcal{E}\left(\sigma_{m}\right)\sigma_{m}}{D_1\left(\sigma_{m}\right)^{3/2}} \right) \right) |c_1| \right),
  \end{align}
  }%
  where
  {\small
  \begin{equation}
  \begin{aligned}
    &\mathcal{E}\left(\sigma\right) = \exp\left(\frac{p_{(1)}|\Delta_{n, (1)}|^2b_1^2\sigma^2 + c_1^2\left(4\sqrt{2N_0p_{(2)}}\mathfrak{R}\left\{x_{m, (2)}\right\}b_1 - 2N_0z\right)z}{D_1}\right), D_1\left(\sigma\right) = 4p_{(2)}\left(\mathfrak{R}\left\{x_{m, (2)}\right\}\right)^2c_1^4 + p_{(1)}|\Delta_{n, (1)}|^2c_1^2\sigma^2 \\
    &\text{and }D_2\left(\sigma\right) = \sqrt{\frac{1}{p_{(2)}\left(\mathfrak{R}\left\{x_{m, (2)}\right\}\right)^2c_1^2}+\frac{4}{p_{(1)}|\Delta_{n, (1)}|^2\sigma_{n}^2}}\sigma^2.\nonumber
  \end{aligned}
  \end{equation}
  }

  % IEEE uses as a separator
  \hrulefill
  % The spacer can be tweaked to stop underfull vboxes.
  \vspace*{4pt}
\end{figure*}

\begin{figure*}[!t]
  {\small
  \begin{align}\label{eq:cf_pdf_z_2_incor}
    f_{\mathcal{Z}_{n, (2)}}\left(z\right)&=\sum_{i=1}^{N_\mathcal{G}}\frac{a_i\sqrt{2N_0p_{(2)}}|\Delta_{n, (2)}c_i|\sigma_{n}^2\mathcal{E}_{1, i}}{\left(p_{(2)}|\Delta_{n, (2)}|^2\sigma_{n}^2+4p_{(1)}\left(\mathfrak{R}\{\Delta_{m, (1)}\}\right)^2c_i^2\right)^{3/2}}\left(-2\sqrt{p_{(1)}\pi}\mathfrak{R}\left\{\Delta_{m, (1)}\right\}b_i\mathcal{E}_{2, i}+\sqrt{\pi}\mathcal{E}_{2, i}\sqrt{2N_0}z+\right.\nonumber\\
    &\left.4p_{(1)}\left(\mathfrak{R}\{\Delta_{m, (1)}\}\right)^2c_i^2\sqrt{\frac{1}{p_{(2)}|\Delta_{n, (2)}|^2\sigma_{n}^2}+\frac{1}{4p_{(1)}\left(\mathfrak{R}\{\Delta_{m, (1)}\}\right)^2c_i^2}}+\sqrt{\pi}\mathcal{E}_{2, i}\left(2\sqrt{p_{(1)}}\mathfrak{R}\left\{\Delta_{m, (1)}\right\}b_i-\sqrt{2N_0}z\right)\mathcal{E}_{3, i}\right),
  \end{align}
  }%
  where
  {\small
  \begin{equation}
  \begin{aligned}
    &\mathcal{E}_{1, i}=\exp\left(-\frac{\left(\sqrt{2N_0}z-2\sqrt{p_{(1)}}\mathfrak{R}\left\{\Delta_{m, (1)}\right\}b_i\right)^2}{4p_{(1)}\left(\mathfrak{R}\{\Delta_{m, (1)}\}\right)^2c_i^2}\right), \mathcal{E}_{2, i}=\exp\left(\frac{p_{(2)}|\Delta_{n, (2)}|^2\sigma_{n}^2\left(\sqrt{2N_0}z-2\sqrt{p_{(1)}}\mathfrak{R}\left\{\Delta_{m, (1)}\right\}b_i\right)^2}{4p_{(1)}\left(\mathfrak{R}\{\Delta_{m, (1)}\}\right)^2c_ip_{(2)}^2|\Delta_{n, (2)}|^2\sigma_{n}^2+16p_{(1)}^2\left(\mathfrak{R}\{\Delta_{m, (1)}\}\right)^4c_i^4}\right)\\
    &\text{and }\mathcal{E}_{3, i}=\text{erf}\left(\frac{2\sqrt{p_{(1)}}\mathfrak{R}\left\{\Delta_{m, (1)}\right\}b_i-\sqrt{2N_0}z}{4p_{(1)}\left(\mathfrak{R}\{\Delta_{m, (1)}\}\right)^2c_i^2\sqrt{\frac{1}{p_{(2)}|\Delta_{n, (2)}|^2\sigma_{n}^2}+\frac{1}{4p_{(1)}\left(\mathfrak{R}\{\Delta_{m, (1)}\}\right)^2c_i^2}}}\right).\nonumber
  \end{aligned}
  \end{equation}
  }
  % IEEE uses as a separator
  \hrulefill
  % The spacer can be tweaked to stop underfull vboxes.
  \vspace*{4pt}
\end{figure*}

{\small
\begin{align}
  f_{\mathcal{Z}_{n, (2)}}\left(z\right)&=\sqrt{2N_0}f_{|\xi_{n, (2)}|}\left(\sqrt{2N_0}z\right)\nonumber\\
  &=\frac{4N_0z}{p_{(2)}|{\Delta}_{n, (2)}|^2\sigma_{n}^2}\exp\left(-\frac{2N_0z^2}{p_{(2)}|{\Delta}_{n, (2)}|^2\sigma_{n}^2}\right).
  \label{eq:cf_pdf_z_2_cor}
\end{align}
}%

The derivation of \cref{eq:cf_pdf_z_1,eq:cf_pdf_z_2_incor,eq:cf_pdf_z_2_cor} involves analyzing the component random variables $|h_{n, (k)}|$ and $\mathfrak{R}\{h_{m, (k)}\}$, fitting truncated Gaussian distributions, and solving the resulting convolution integrals. The complete derivation details, including the intermediate \glspl{pdf} and Gaussian fitting coefficients, are provided in~\cref{app:pdf_zn1_zn2}.

The final step in deriving the closed-form unconditional \gls{pep} involves solving the integral in \cref{eq:upep_integ}. We substitute $f_{\mathcal{Z}_{n, (k)}}\left(z\right)$ and employ the approximate Q-function $Q\left(x\right)\approx\frac{1}{12}\exp\left(-\frac{x^2}{2}\right)+\frac{1}{4}\exp\left(-\frac{2x^2}{3}\right)$~\cite{1210748}. These expressions were derived using symbolic computation tools~\cite{Mathematica}.

\subsubsection{Case 1: \gls{ue}~$n$ with Decoding Order $(1)$}
For the first decoding order, we obtain the closed-form of $P\left(A_{n}\mid B_{n, (1)}\right)$ by substituting \cref{eq:cf_pdf_z_1} into \cref{eq:upep_integ} and evaluating the integral. The resulting expression is given by
{\small
\begin{align}\label{eq:cf_pep_z_1}
  P\left(A_{n}\mid B_{n, (1)}\right) &= \int_{0}^{\infty}Q\left(z\right)f_{\mathcal{Z}_{n, (1)}}\left(z\right)dz \nonumber\\
  &= \frac{\sqrt{N_0\pi}\mathfrak{R}\left\{x_{m, (2)}\right\}a_1\exp\left(-\frac{b_1^2}{c_1^2}\right)}{3\sqrt{2}\Delta_{n, (1)}\left(\sigma_{n}^2-\sigma_{m}^2\right)} \nonumber \\
  &\quad \times \left[G_1\left(\sigma_{n}\right)+G_2\left(\sigma_{n}\right)-G_1\left(\sigma_{m}\right)-G_2\left(\sigma_{m}\right)\right],
\end{align}
}%
where $G_1\left(\sigma\right)$ is given by:
{\small
\begin{multline}\label{eq:G1}
  G_1\left(\sigma\right) = \gamma_1(\sigma) D_1\left(\sigma\right) \\
  \times \left(\frac{\sqrt{3}}{2\sqrt{2}}\sqrt{p_{(1)}}|\Delta_{n, (1)}|\sigma\alpha_1(\sigma) - \beta_1(\sigma)\right) \\
  - \delta_1(\sigma) F_1\left(\sigma\right),
\end{multline}
}%
and $G_2\left(\sigma\right)$ is:
{\small
\begin{multline}\label{eq:G2}
  G_2\left(\sigma\right) = \gamma_2(\sigma) D_2\left(\sigma\right) \\
  \times \left(p_{(1)}|\Delta_{n, (1)}|^2\sigma\alpha_1(\sigma) - \beta_2(\sigma)\right) \\
  - \delta_2(\sigma) F_1\left(\sigma\right).
\end{multline}
}%
The intermediate terms of $G_i(\sigma)$, $F_i$, $D_i$, $\gamma_i$, $\delta_i$, $\alpha_i$, and $\beta_i$ for $i \in \{1, 2\}$ are defined in~\cref{app:cf_pep_2_ues}.

\subsubsection{Case 2: \gls{ue}~$n$ with Decoding Order $(2)$}
For the second decoding order, $P\left(A_{n}\mid B_{n, (2)}\right)$ must account for error propagation as specified in \cref{eq:err_prop_k_plus_1}. This yields two distinct cases:
\begin{itemize}
    \item When the previous \gls{ue} is correctly decoded ($\zeta_{n, (2)}=0$): $P\left(A_{n}\mid \bar{A}_{(1)}, B_{n, (2)}\right)$
    \item When the previous \gls{ue} is incorrectly decoded ($\zeta_{n, (2)}\neq0$): $P\left(A_{n}\mid A_{(1)}, B_{n, (2)}\right)$
\end{itemize}

\paragraph{Sub-case 2a: \gls{ue}~$m$ Decoded Incorrectly}
If the first-decoded \gls{ue}~$m$ experiences a decoding error, the \gls{pep} for \gls{ue}~$n$ is
{\small
\begin{align}\label{eq:cf_pep_z_2_inc}
  P\left(A_{n}\mid B_{n, (2)}\right) &= \int_{0}^{\infty}Q\left(z\right)f_{\mathcal{Z}_{n, (2)}}\left(z\right)dz \nonumber\\
  &= \sum_{i=1}^{3}\left[\mathcal{S}_{i, 1} + \mathcal{S}_{i, 2} + \mathcal{S}_{i, 3}\right],
\end{align}
}%
where $\mathcal{S}_{i, 1}$ is given by:
{\small
\begin{align}\label{eq:S_i1}
    \mathcal{S}_{i, 1} &= \frac{a_i D_2}{72\sqrt{2}\Omega_i^{3/2}} \left[T_{i, 1}^{(1)} + T_{i, 2}^{(1)} - T_{i, 3}^{(1)}\right] \nonumber \\
    &\quad + \frac{a_i D_2}{16\sqrt{3}\Omega_i^{3/2}} \left[T_{i, 1}^{(2)} + T_{i, 2}^{(2)} - T_{i, 3}^{(2)}\right],
\end{align}
}%
$\mathcal{S}_{i, 2}$ is defined as:
{\small
\begin{align}\label{eq:S_i2}
    \mathcal{S}_{i, 2} &= \frac{a_i D_2 e_{i, 5}}{12} \left[\Theta_{i, 1} + \Theta_{i, 2}\right] \nonumber \\
    &\quad + \frac{a_i D_2 e_{i, 5}}{8} \left[\Theta_{i, 3} + \Theta_{i, 4}\right],
\end{align}
}%
and $\mathcal{S}_{i, 3}$ is:
{\small
\begin{align}\label{eq:S_i3}
    \mathcal{S}_{i, 3} &= \frac{a_i D_2 e_{i, 8}}{8\Omega_i^{3/2}} \left[\Lambda_{i, 1} + \Lambda_{i, 2}\right] \nonumber \\
    &\quad + \frac{a_i D_2 e_{i, 8}}{24\Omega_i^{3/2}} \left[\Lambda_{i, 3} + \Lambda_{i, 4}\right].
\end{align}
}%
The intermediate terms $T_{i, j}^{(k)}$, $\Theta_{i, j}$, $\Lambda_{i, j}$, $a_i$, $D_2$, $\Omega_i$, and $e_{i, j}$ appearing in $\mathcal{S}_{i, 1}$, $\mathcal{S}_{i, 2}$, and $\mathcal{S}_{i, 3}$ are defined in~\cite{app:cf_pep_2_ues}.

\paragraph{Sub-case 2b: \gls{ue}~$m$ Decoded Correctly}
If the first-decoded \gls{ue}~$m$ is decoded correctly, the \gls{pep} for \gls{ue}~$n$ is
{\small
\begin{align}
  P\left(A_{n} \mid B_{n, (2)}\right) &= \int_{0}^{\infty} Q(z) f_{\mathcal{Z}_{n, (2)}}(z) \, dz \nonumber\\
  &= \frac{1}{12 + 6\Delta_{n, (2)}^2\sigma_{n}^2} + \frac{3}{12 + 8\Delta_{n, (2)}^2\sigma_{n}^2}.
  \label{eq:cf_pep_z_2_cor}
\end{align}
}%

\subsection{Calculation of  $P\left(B_{n, (k)}\right)$}
In this subsection, we derive the probability $P\left(B_{n, (k)}\right)$ that each \gls{ue} is decoded in a specific order for the two-\gls{ue} case. These probabilities depend on the instantaneous channel realizations and determine the likelihood of each decoding scenario.

For the two-\gls{ue} system, the decoding order is determined by comparing the instantaneous channel gains $|h_1|$ and $|h_2|$. Specifically:
\begin{itemize}
    \item $B_{1, (1)}$: \gls{ue} 1 is decoded first when $|h_1|\geq|h_2|$
    \item $B_{1, (2)}$: \gls{ue} 1 is decoded second when $|h_2|>|h_1|$
\end{itemize}

To compute these probabilities, we substitute the appropriate \glspl{cdf} and \glspl{pdf} into \cref{eq:P_Bnk}. Using~\cref{eq:proof_F_Hk} for the complementary \glspl{cdf} $\bar{F}_{1, (1)}$ and $\bar{F}_{2, (1)}$, and integrating with respect to the \glspl{pdf} of $|h_2|$ and $|h_1|$ respectively, we obtain:
{\small
\begin{equation}
\begin{aligned}
  P\left(B_{1, (1)}\right) &= \int_{0}^{\infty} \left[1-F_{|h_1|}\left(x\right)\right]f_{|h_{2}|}\left(x\right)dx=\frac{\sigma_{1}^2}{\sigma_1^2+\sigma_2^2}, \\
  P\left(B_{1, (2)}\right) &= \int_{0}^{\infty} \left[1-F_{|h_2|}\left(x\right)\right]f_{|h_{1}|}\left(x\right)dx=\frac{\sigma_{2}^2}{\sigma_1^2+\sigma_2^2}. \label{eq:P_B11_P_B12}
\end{aligned}
\end{equation}
}%
For \gls{ue} 2, the probabilities follow from symmetry. Since $B_{2, (1)}$ represents the event $|h_2|\geq|h_1|$, we have $P\left(B_{2, (1)}\right)=P\left(B_{1, (2)}\right)=\frac{\sigma_{2}^2}{\sigma_1^2+\sigma_2^2}$. Similarly, $P\left(B_{2, (2)}\right)=P\left(B_{1, (1)}\right)=\frac{\sigma_{1}^2}{\sigma_1^2+\sigma_2^2}$. The detailed derivations are provided in~\cref{app:cf_P_Bnk}.

These closed-form expressions enable us to compute the average \gls{pep} for both \glspl{ue} by substituting into \cref{eq:P_A1_A2}, accounting for all possible decoding orders weighted by their respective probabilities.

\section{\gls{ber} Analysis for $\mathcal{M}$-\gls{qam} Schemes}\label{sec:ber_m_qam}
In \Autoref{sec:system_model,sec:2_ue_ul_tx_rayleigh}, we derived closed-form \gls{pep} expressions for uplink \gls{noma} systems with dynamic \gls{sic} decoding, providing comprehensive symbol-level error analysis for the two-\gls{ue} scenario. While \gls{pep} captures symbol error characteristics, practical systems require bit-level performance metrics. This section extends our framework to derive average \gls{ber} expressions for Gray-coded $\mathcal{M}$-\gls{qam} constellations ($\mathcal{M} \in \{4, 16, 64\}$). By leveraging the signal-space properties of Gray mapping and the ordered channel statistics developed earlier, we establish the relationship between symbol and bit error probabilities under dynamic decoding conditions, enabling accurate performance prediction for practical modulation schemes.

To evaluate the average \gls{ber} for systems employing $\mathcal{M}$-\gls{qam} modulation, we must analyze the error probability of each bit individually due to the Gray-coded constellation structure. Following the approach in~\cite{752121}, we leverage the signal-space representation to derive the coherent detection statistics for each bit position. 

For a desired symbol $x_{(k)}$ decoded at the $k$-th position, the bit-level error analysis builds upon the symbol error probability framework established in \cref{eq:cond_pep}. Specifically, the error probability for bit $b_m$ within symbol $x_{(k)}$ can be expressed as~\cite{1611074}:
{\small
\begin{align}
  P_{(k), \mathbf{S}_{i}, b_m}&=P\left(2\mathfrak{R}\left\{w\right\}\geq 2\Delta_{b_m}\rho_{(k)}+\mathcal{I}_{(k)}\right)\nonumber\\
  &=Q\left(\frac{2\Delta_{b_m}\rho_{(k)}+\mathcal{I}_{(k)}}{\sqrt{2N_0}}\right),
  \label{eq:cpep_per_bit}
\end{align}
}%
where $\rho_{(k)}=\sqrt{p_{(k)}}|h_{(k)}|$ represents the effective channel gain, and $\mathbf{S}_{i}$ denotes a specific combination of residual errors and interference signals from other users. The parameter $\Delta_{b_m}$ represents the minimum Euclidean distance from the transmitted symbol to the decision boundary for bit $b_m$, which varies depending on the constellation point and will be detailed for each \gls{qam} scheme. The aggregate interference term is given by:
{\small
\begin{align}
  \mathcal{I}_{(k)}=2\mathfrak{R}\left\{\sum_{i=1}^{k-1}\Delta_{(i)}\rho_{(i)}+\sum_{j=k+1}^{N}x_{(j)}\rho_{(j)}\right\},
  \label{eq:interference_term}
\end{align}
}%
where $\Delta_{(i)}$ represents residual errors from previously decoded users and $x_{(j)}$ denotes undecoded symbols that contribute to interference.

Having derived the error probability for individual bits $b_m$, we now extend our analysis to the average \gls{ber} of the system. The average \gls{ber} requires evaluating the error probability across all possible symbol combinations in the constellation space.

For the $k$-th decoded user, the set of all possible symbol combinations is denoted as $\mathbb{S}$, which includes residual errors from previously decoded users ($j=1, \ldots, k-1$) and interference from undecoded users ($j=k+1, \ldots, N$). The cardinality of this set is given by:
{\small
\begin{align}
  |\mathbb{S}|=\prod_{j=1}^{k-1}\left(\sqrt{\mathcal{M}_{(j)}}-1\right)\prod_{j=k+1}^{N}\frac{1}{2}\sqrt{\mathcal{M}_{(j)}},
  \label{eq:S_cardinality}
\end{align}
}%
where $\sqrt{\mathcal{M}_{(j)}}-1$ represents the number of possible error patterns for the $j$-th decoded symbol (corresponding to decision boundaries in the constellation), and $\frac{1}{2}\sqrt{\mathcal{M}_{(j)}}$ accounts for the in-phase component symbols of undecoded users.

The average \gls{ber} for the $k$-th decoded user is then obtained by averaging over all symbol combinations and bit positions:
{\small
\begin{align}
  P_{(k)}=\frac{1}{|\mathbb{S}|}\sum_{\mathbf{S}_{i}\in\mathbb{S}}\sum_{m=1}^{\frac{1}{2}\log_2\mathcal{M}_{(k)}}P_{(k), \mathbf{S}_{i}, b_m},
  \label{eq:cpep_avg_ber}
\end{align}
}%
where $P_{(k), \mathbf{S}_{i}, b_m}$ denotes the error probability of bit $b_m$ under symbol combination $\mathbf{S}_{i}$. The constellation symmetry between in-phase and quadrature components in $\mathcal{M}$-\gls{qam} allows us to consider only $\frac{1}{2}\log_2\mathcal{M}_{(k)}$ bits (corresponding to the in-phase component) rather than the full $\log_2\mathcal{M}_{(k)}$ bits per symbol.

To obtain the unconditional \gls{ber}, we define $z_{(k)}$ as the argument of the Q-function in \cref{eq:cpep_per_bit} and integrate over its \gls{pdf}, following the approach established in \cref{eq:upep_integ}. This integration yields the closed-form expression for $P_{(k), \mathbf{S}_{i}, b_m}$, which represents the bit error probability for a specific symbol combination $\mathbf{S}_{i}$ and bit position $b_m$. Subsequently, the average \gls{ber} for any \gls{ue} decoded at the $(k)$-th position is computed by substituting these closed-form expressions into \cref{eq:cpep_avg_ber}.

The subsequent analysis extends this framework to practical Gray-coded $\mathcal{M}$-\gls{qam} constellations, specifically examining 4\gls{qam}, 16\gls{qam}, and 64\gls{qam} schemes. We select 16\gls{qam} as our primary example for detailed exposition, as it provides an optimal balance between analytical tractability and practical relevance, while 4\gls{qam} offers limited insight due to its simplicity and 64\gls{qam} introduces excessive complexity that may obscure the fundamental principles. The error probability derivations for 4\gls{qam} and 64\gls{qam} follow analogous procedures, and thus we present their final expressions and key insights without redundant mathematical details.

The following subsection demonstrates the application of our general framework to the 16\gls{qam} constellation, illustrating the systematic approach for deriving bit-level error probabilities in Gray-coded systems.

\subsection{Gray-coded 16QAM Analysis}\label{sec:gray_16qam}
The 16\gls{qam} constellation employs Gray coding to minimize bit errors, ensuring that adjacent constellation points differ by only a single bit. This property is crucial for minimizing the \gls{ber} when symbol errors occur due to noise or interference.

\begin{figure}[htb]
  \centering
  \includegraphics[width=2.5in]{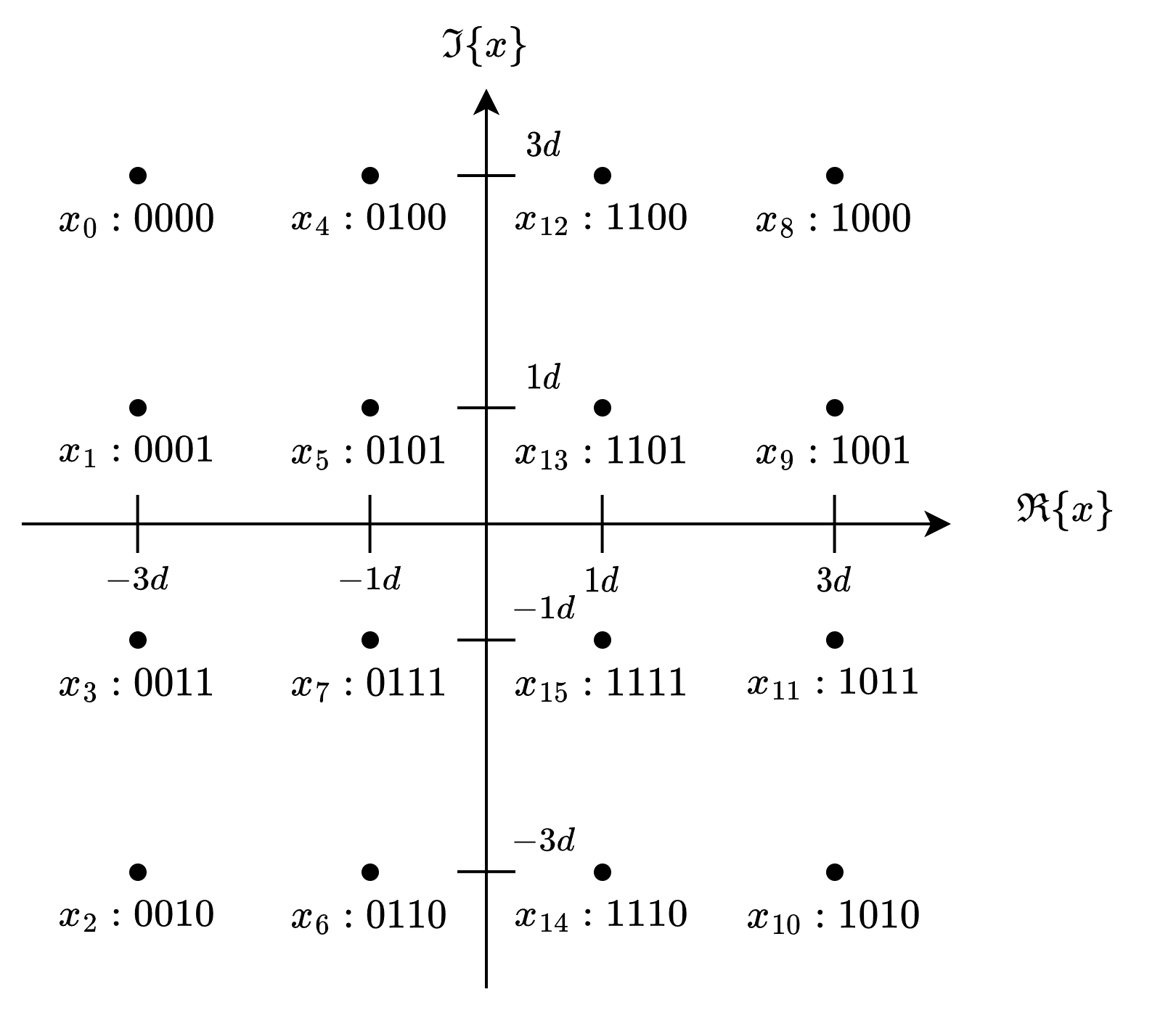}
  \caption{16QAM constellation with Gray coding. Adjacent constellation points differ by only one bit, minimizing the bit error probability when symbol errors occur.}
  \label{fig:16QAM}
\end{figure}

\autoref{fig:16QAM} illustrates the 16\gls{qam} constellation with Gray-coded bit mapping. Each constellation point is uniquely identified by a 4-bit sequence $b_{1}b_{2}b_{3}b_{4}$, where:
\begin{itemize}
    \item Bits $b_{1}b_{2}$ determine the in-phase (I) component position
    \item Bits $b_{3}b_{4}$ determine the quadrature (Q) component position
\end{itemize}
The constellation points are uniformly spaced with distance $2d$ between adjacent points, where $d$ is the normalization factor ensuring unit average symbol energy as defined in \cref{eq:d}.

We first analyze the error probabilities for the in-phase bits $b_1b_2$. 

\subsubsection{Analysis of bit $b_1$}
For bit $b_1=0$, an error occurs when the transmitted symbol belongs to the set $x\in\{4,5,7,6\}$ but is detected as $\check{x}\in\{12,13,15,14\}$ or $\{8,9,11,10\}$. Two distinct error scenarios arise:
\begin{itemize}
    \item \textbf{Near-boundary error}: When $x\in\{4,5,7,6\}$, the minimum error distance to the decision boundary $\mathfrak{R}\{x\}=0$ is $\Delta_{b_1=0}^{(1)}=d$, where $d$ is defined in \cref{eq:d}.
    \item \textbf{Far-boundary error}: When $x\in\{0,1,3,2\}$, the error distance increases to $\Delta_{b_1=0}^{(2)}=3d$.
\end{itemize}

These error distances are illustrated in \autoref{subfig:16QAM_b1-0}. According to \cref{eq:cpep_per_bit}, the corresponding error probabilities are:
{\small
\begin{equation}
\begin{aligned}
  P_{(k),b_1=0}' &= P\left(2\mathfrak{R}\{w\} \geq 2d\rho_{(k)} + \mathcal{I}_{(k)}\right), \\
  P_{(k),b_1=0}'' &= P\left(2\mathfrak{R}\{w\} \geq 6d\rho_{(k)} + \mathcal{I}_{(k)}\right).
\end{aligned}
\end{equation}
}%

The quadrature symmetry of the constellation ensures that bit $b_3$ exhibits identical error characteristics to $b_1$, with the decision boundary shifted to $\mathfrak{I}\{x\}=0$. Therefore, $P_{(k),b_3} = P_{(k),b_1}$.

The total conditional \gls{ber} for bits $b_1$ and $b_3$ is obtained by averaging over all possible transmitted symbols:
{\small
\begin{align}
  P_{(k),b_1} = P_{(k),b_3} &= \frac{1}{2}\bigg[Q\left(\frac{2d\rho_{(k)} + \mathcal{I}_{(k)}}{\sqrt{2N_0}}\right) \nonumber\\
  &\quad + Q\left(\frac{6d\rho_{(k)} + \mathcal{I}_{(k)}}{\sqrt{2N_0}}\right)\bigg]. \label{eq:p_b1_b3_total}
\end{align}
}%

\subsubsection{Analysis of bit $b_2$}
\begin{figure*}[!t]
  \centering
  \captionsetup[subfloat]{labelfont=scriptsize,textfont=scriptsize}
  \subfloat[Error distances of symbols with $b_1=0$ to their decision boundary\label{subfig:16QAM_b1-0}]{%
    \includegraphics[width=0.32\textwidth]{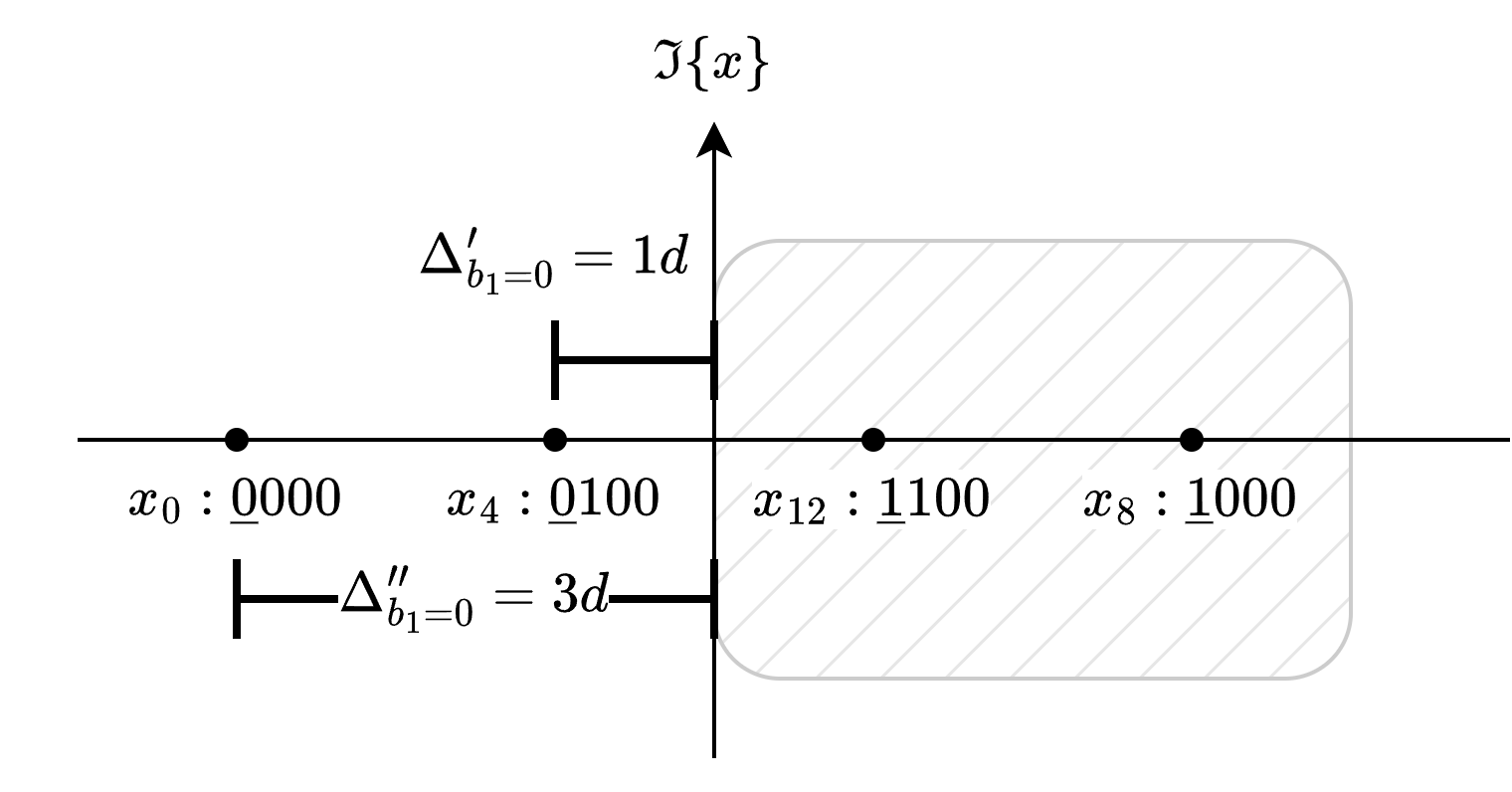}%
  }%
  \hfill
  \subfloat[Error distances of symbols with $b_2=1$ to their decision boundaries\label{subfig:16QAM_b2-1}]{%
    \includegraphics[width=0.32\textwidth]{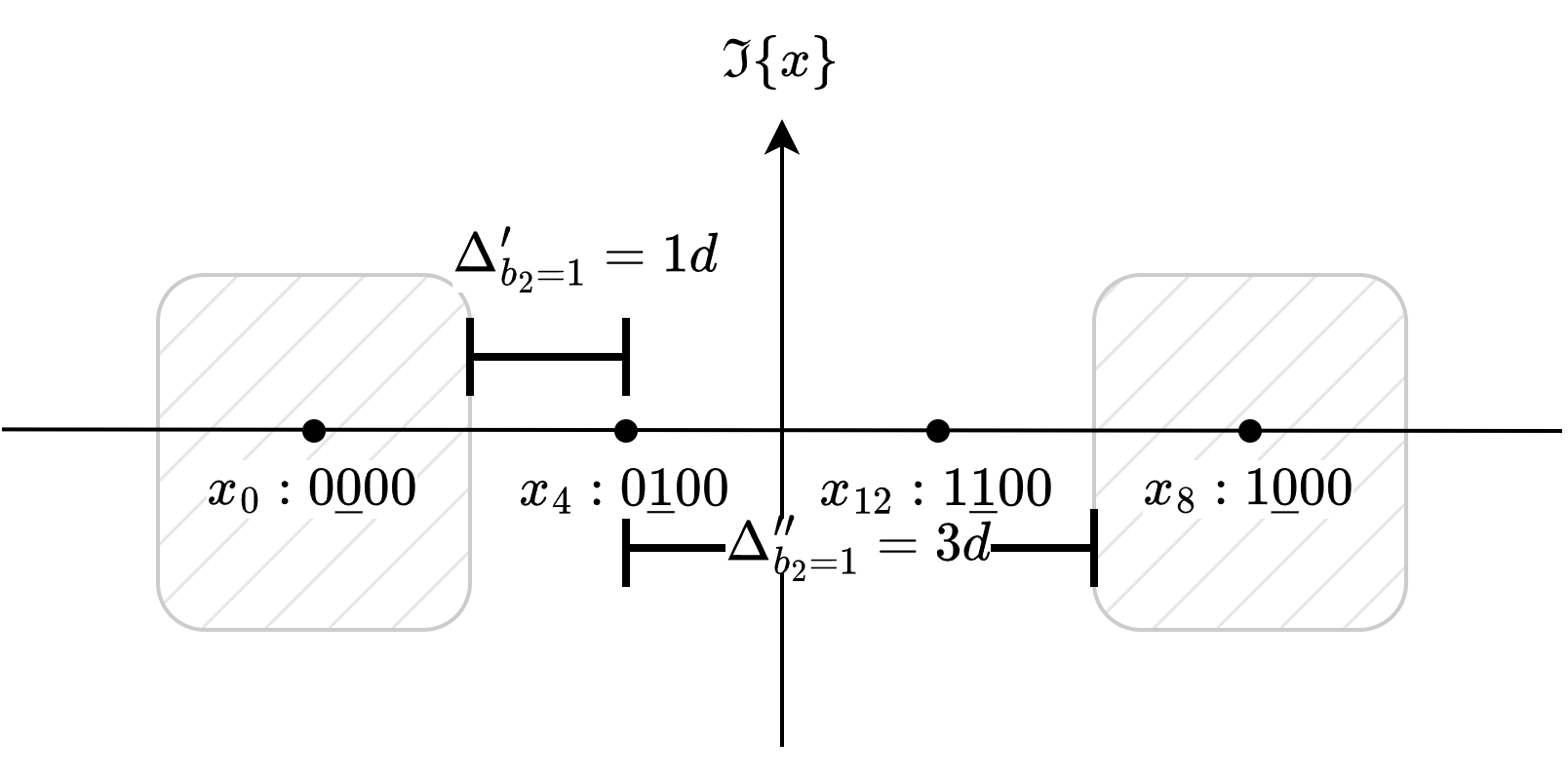}%
  }%
  \hfill
  \subfloat[Error distances of symbols with $b_2=0$ to their decision boundaries\label{subfig:16QAM_b2-0}]{%
    \includegraphics[width=0.32\textwidth]{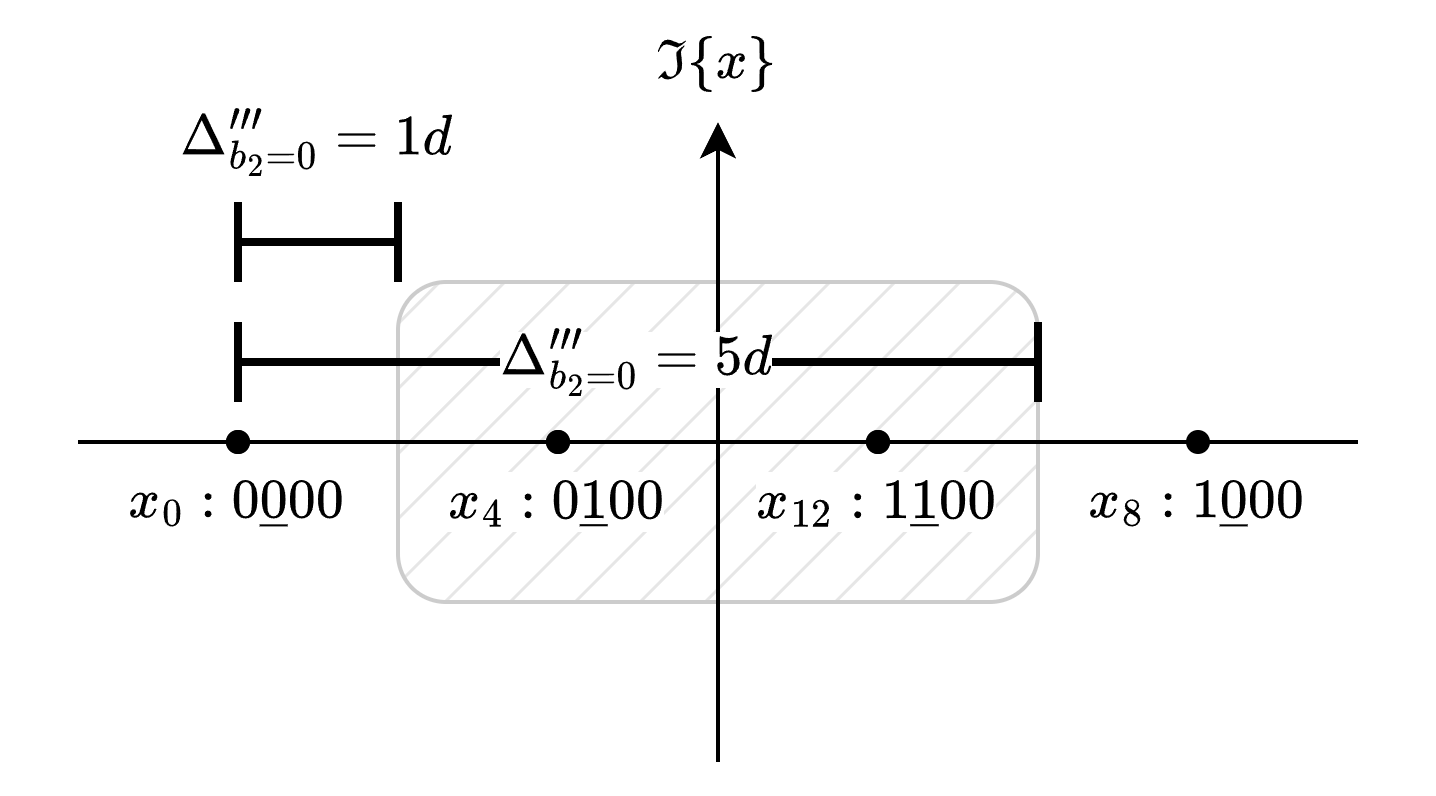}%
  }%
  \caption{Error distances between transmitted 16-QAM symbols and decision boundaries for different bit values in Gray-coded constellation mapping}
  \label{fig:16qam_err_dist}
\end{figure*}

The error analysis for bit $b_2$ is more complex due to multiple decision boundaries. We consider two cases:

\textbf{Case 1: $b_2=1$ decoded as 0}
Two error scenarios exist, as shown in \autoref{subfig:16QAM_b2-1}:
\begin{itemize}
    \item When $x\in\{4,5,7,6\} \cup \{12,13,15,14\}$ is detected as $\check{x}\in\{0,1,3,2\}$: $\Delta_{b_2=1}^{(1)}=d$.
    \item When detected as $\check{x}\in\{8,9,11,10\}$: $\Delta_{b_2=1}^{(2)}=3d$.
\end{itemize}

\textbf{Case 2: $b_2=0$ decoded as 1}
When $x\in\{0,1,3,2\} \cup \{8,9,11,10\}$ is detected as $\check{x}\in\{4,5,7,6\} \cup \{12,13,15,14\}$, the signal must cross one of two decision boundaries:
\begin{itemize}
    \item Inner boundary at $\mathfrak{R}\{x\}=-2d$: $\Delta_{b_2=0}^{(1)}=d$.
    \item Outer boundary at $\mathfrak{R}\{x\}=2d$: $\Delta_{b_2=0}^{(2)}=5d$.
\end{itemize}

These scenarios are illustrated in \autoref{subfig:16QAM_b2-0}. The corresponding error probabilities of two cases are:
{\small
\begin{equation}
\begin{aligned}
  P_{(k),b_2=0}' &= P\left(2d\rho_{(k)} + \mathcal{I}_{(k)} \leq 2\mathfrak{R}\{w\} \leq 10d\rho_{(k)} + \mathcal{I}_{(k)}\right) \\
  &= Q\left(\frac{2d\rho_{(k)} + \mathcal{I}_{(k)}}{\sqrt{2N_0}}\right) - Q\left(\frac{10d\rho_{(k)} + \mathcal{I}_{(k)}}{\sqrt{2N_0}}\right), \\
  P_{(k),b_2=1}' &= Q\left(\frac{2d\rho_{(k)} + \mathcal{I}_{(k)}}{\sqrt{2N_0}}\right), \\
  P_{(k),b_2=1}'' &= Q\left(\frac{6d\rho_{(k)} + \mathcal{I}_{(k)}}{\sqrt{2N_0}}\right).
\end{aligned}
\end{equation}
}%

Again, quadrature symmetry ensures that bit $b_4$ shares identical error characteristics with $b_2$, yielding $P_{(k),b_4} = P_{(k),b_2}$.

The total conditional \gls{ber} for bits $b_2$ and $b_4$ is:
{\small
\begin{align}
  P_{(k),b_2} &= P_{(k),b_4} = \frac{1}{2}\bigg[2Q\left(\frac{2d\rho_{(k)} + \mathcal{I}_{(k)}}{\sqrt{2N_0}}\right) \nonumber\\
  &\quad + Q\left(\frac{6d\rho_{(k)} + \mathcal{I}_{(k)}}{\sqrt{2N_0}}\right) - Q\left(\frac{10d\rho_{(k)} + \mathcal{I}_{(k)}}{\sqrt{2N_0}}\right)\bigg]. \label{eq:p_b2_b4_total}
\end{align}
}%

The Gray coding symmetry ensures that the error patterns of bit $b_2$ inherently capture all possible error scenarios for bit $b_1$, eliminating redundant calculations. Consequently, the average \gls{ber} can be computed by considering only the error distances associated with bit $b_2$. 

Substituting the error distances from \autoref{tab:mqam_err_dist} into \cref{eq:cpep_avg_ber}, the conditional average \gls{ber} for a user with modulation order $\mathcal{M}_{(k)}=16$ is expressed as:
{\small
\begin{align}
  P_{(k)}&=\frac{1}{|\mathbb{S}|}\sum_{\mathbf{S}_{i}\in\mathbb{S}}P_{(k),b_2}\nonumber\\
  &=\frac{1}{|\mathbb{S}|}\sum_{\mathbf{S}_{i}\in\mathbb{S}}\frac{1}{2}\bigg[2Q\left(\frac{2d\rho_{(k)}+\mathcal{I}_{(k)}}{\sqrt{2N_0}}\right)\nonumber\\
  &\quad+Q\left(\frac{6d\rho_{(k)}+\mathcal{I}_{(k)}}{\sqrt{2N_0}}\right)-Q\left(\frac{10d\rho_{(k)}+\mathcal{I}_{(k)}}{\sqrt{2N_0}}\right)\bigg].
  \label{eq:avg_ber_16qam_app}
\end{align}
}%
The derivation methodology for obtaining closed-form unconditional \gls{pep} expressions follows the same approach as detailed in the main text, where we integrate the Q-function over the \gls{pdf} of the random variable $z_{(k)}$.

\subsection{BER Expressions for $\mathcal{M}$-QAM Schemes}
Having illustrated the detailed analysis methodology through the 16\gls{qam} example, we now present the consolidated \gls{ber} expressions for all practical $\mathcal{M}$-\gls{qam} constellations. The error distance analysis for Gray-coded $\mathcal{M}$-\gls{qam} constellations leverages the symmetry properties and structured bit mapping to derive \gls{ber} expressions. For each modulation scheme, we analyze the minimum Euclidean distances from transmitted symbols to decision boundaries, which directly determine the Q-function arguments in the \gls{ber} expressions.

\autoref{tab:mqam_err_dist} summarizes the minimum error distances for different bit positions across 4\gls{qam}, 16\gls{qam}, and 64\gls{qam} schemes. The Gray coding symmetry and quadrature independence allow us to simplify the error analysis by considering only the in-phase component bits. The 16\gls{qam} analysis is presented in detail in \autoref{sec:gray_16qam}, while the complete derivation methodology for 4\gls{qam} and 64\gls{qam} is provided in~\cref{app:ber_of_m_qam}.

Substituting these error distances into the general \gls{ber} framework established in \cref{eq:cpep_avg_ber}, we obtain the following closed-form expressions for the average \gls{ber}:

\textbf{4\gls{qam}:} The single error distance per bit yields:
{\small
\begin{align}
  P_{(k)}&=\frac{1}{|\mathbb{S}|}\sum_{\mathbf{S}_{i}\in\mathbb{S}}Q\left(\frac{2d\rho_{(k)}+\mathcal{I}_{(k)}}{\sqrt{2N_0}}\right).
  \label{eq:avg_ber_4qam}
\end{align}
}%

\textbf{16\gls{qam}:} Accounting for multiple decision boundaries:
{\small
\begin{align}
  P_{(k)}&=\frac{1}{|\mathbb{S}|}\sum_{\mathbf{S}_{i}\in\mathbb{S}}\frac{1}{2}\bigg[2Q\left(\frac{2d\rho_{(k)}+\mathcal{I}_{(k)}}{\sqrt{2N_0}}\right)\nonumber\\
  &\quad+Q\left(\frac{6d\rho_{(k)}+\mathcal{I}_{(k)}}{\sqrt{2N_0}}\right)-Q\left(\frac{10d\rho_{(k)}+\mathcal{I}_{(k)}}{\sqrt{2N_0}}\right)\bigg].
  \label{eq:avg_ber_16qam}
\end{align}
}%

\textbf{64\gls{qam}:} With increased constellation density:
{\small
\begin{align}
  P_{(k)}&=\frac{1}{|\mathbb{S}|}\sum_{\mathbf{S}_{i}\in\mathbb{S}}\frac{1}{6}\bigg[4Q\left(\frac{2d\rho_{(k)}+\mathcal{I}_{(k)}}{\sqrt{2N_0}}\right)+4Q\left(\frac{6d\rho_{(k)}+\mathcal{I}_{(k)}}{\sqrt{2N_0}}\right)\nonumber\\
  &\quad+Q\left(\frac{18d\rho_{(k)}+\mathcal{I}_{(k)}}{\sqrt{2N_0}}\right)-Q\left(\frac{26d\rho_{(k)}+\mathcal{I}_{(k)}}{\sqrt{2N_0}}\right)\bigg],
  \label{eq:avg_ber_64qam}
\end{align}
}%
where $d$ is the normalization factor defined in \cref{eq:d}, and the coefficients in each expression correspond to the frequency of each error type weighted by the inclusion-exclusion principle for overlapping decision regions.

% Table generated by Excel2LaTeX from sheet 'Horizontal'
\begin{table}[htb]
  \centering
  \caption{Minimum error distances for different bit positions in Gray-coded $\mathcal{M}$-QAM constellations}
  \begin{tabular}{@{}lccc@{}}
    \toprule
    \textbf{Modulation} & \textbf{Bit} $\boldsymbol{b_1}$ & \textbf{Bit} $\boldsymbol{b_2}$ & \textbf{Bit} $\boldsymbol{b_3}$ \\
    \midrule
    4\gls{qam} & $\Delta' \geq d$ & --- & --- \\
    \midrule
    16\gls{qam} & \begin{tabular}[c]{@{}l@{}}
              $\Delta' \geq d$ \\
              $\Delta'' \geq 3d$
              \end{tabular} & 
              \begin{tabular}[c]{@{}l@{}}
              $\Delta' \geq d$ \\
              $\Delta'' \geq 3d$ \\
              $d \leq \Delta''' \leq 5d$
              \end{tabular} & 
              --- \\
    \midrule
    64\gls{qam} & \begin{tabular}[c]{@{}l@{}}
              $\Delta' \geq d$ \\
              $\Delta'' \geq 3d$ \\
              $\Delta''' \geq 5d$ \\
              $\Delta^{(4)} \geq 7d$
              \end{tabular} & 
              \begin{tabular}[c]{@{}l@{}}
              $\Delta' \geq d$ \\
              $\Delta'' \geq 3d$ \\
              $\Delta''' \geq 5d$ \\
              $\Delta^{(4)} \geq 7d$ \\
              $d \leq \Delta^{(5)} \leq 9d$ \\
              $3d \leq \Delta^{(6)} \leq 11d$
              \end{tabular} & 
              \begin{tabular}[c]{@{}l@{}}
              $\Delta' \geq d$ \\
              $\Delta'' \geq 3d$ \\
              $d \leq \Delta''' \leq 5d$ \\
              $3d \leq \Delta^{(4)} \leq 7d$ \\
              $\Delta^{(5)} \geq 9d$ \\
              $\Delta^{(6)} \geq 11d$ \\
              $9d \leq \Delta^{(7)} \leq 13d$
              \end{tabular} \\
    \bottomrule
  \end{tabular}%
  \label{tab:mqam_err_dist}%
\end{table}%

\section{Numerical Results}\label{sec:numerical_results}
This section presents theoretical and simulated \gls{ber} results for uplink \gls{noma} systems employing dynamic \gls{sic} decoding orders. We consider a two-\gls{ue} scenario where users transmit over independent Rayleigh fading channels with distinct average channel gains: $h_1\sim\mathcal{R}\left(\sigma_1\right)$ and $h_2\sim\mathcal{R}\left(\sigma_2\right)$. To facilitate \gls{sic} detection, we apply power coefficients $p_{(1)}$ and $p_{(2)}$ to the first and second decoded signals, respectively, where the first decoded signal receives higher power allocation to minimize interference during successive decoding.

Both \glspl{ue} can employ identical or different modulation schemes from the set $\mathcal{M}_n\in\{2, 4, 16, 64\}$, corresponding to \gls{bpsk}, 4\gls{qam}, 16\gls{qam}, and 64\gls{qam}. The simulated and theoretical BER results are denoted as ``Sim.'' and ``Theo.'', respectively. We first examine the \gls{ber} performance when instantaneous channel conditions satisfy $|h_1|>|h_2|$ and $|h_2|>|h_1|$ separately in order to reveal the impact of channel ordering on individual user performance. Second, we compare fixed and dynamic \gls{sic} decoding strategies to validate the accuracy of our theoretical framework while confirming the performance advantages of adaptive decoding order. Finally, we investigate heterogeneous systems where \glspl{ue} employ different modulation schemes.

\subsection{BER Performance of Two UEs Under Different Channel Ordering Scenarios}\label{sec:ber_performance_of_2_ues_when_h1_g_h2_and_h2_g_h1}
\begin{figure*}[htb]
  \centering
  \captionsetup[subfloat]{labelfont=scriptsize, textfont=scriptsize}
  \subfloat[\gls{ber} of UE considering both Case 1 and 2]{%
    \includegraphics[width=0.3\textwidth, trim=50 170 50 180, clip]{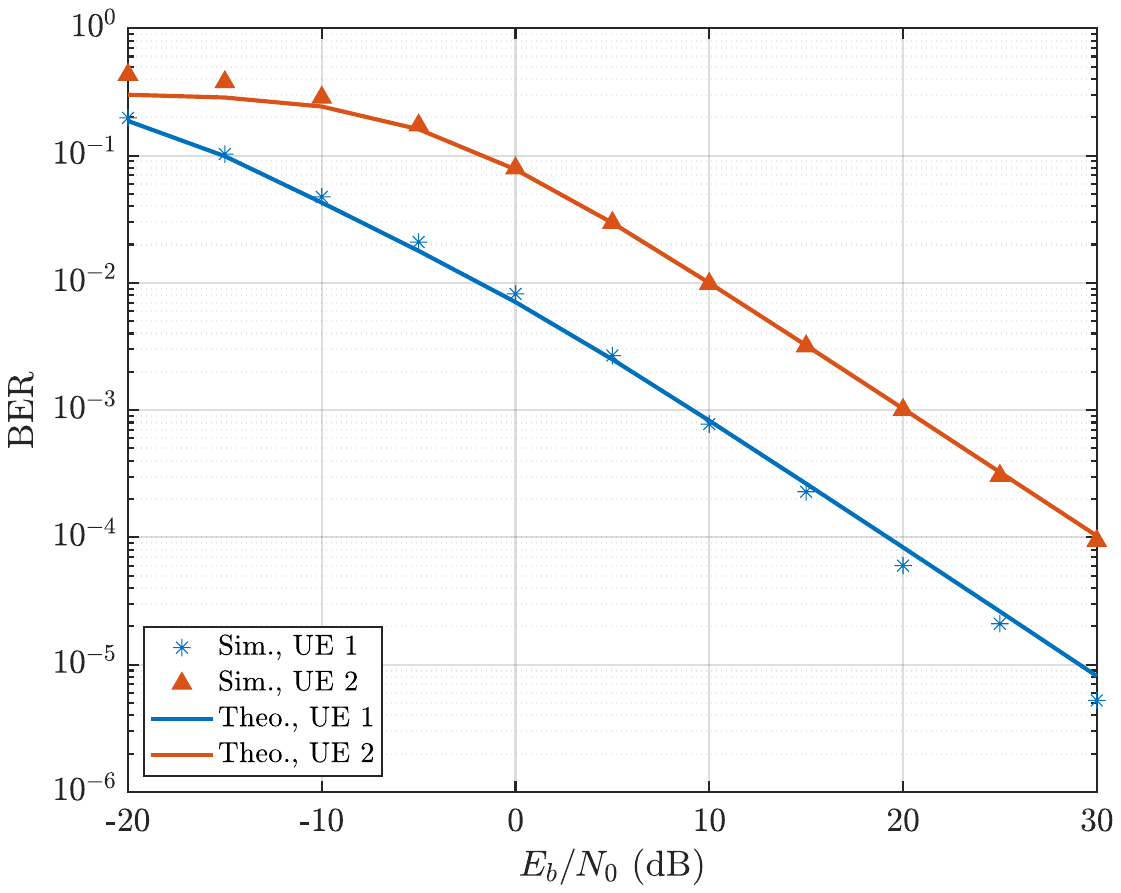}
    \label{fig:BER_of_2_UEs_both}
  }
  \hfill
  \subfloat[Case 1: $|h_1| > |h_2|$ only]{%
    \includegraphics[width=0.3\textwidth, trim=50 170 50 180, clip]{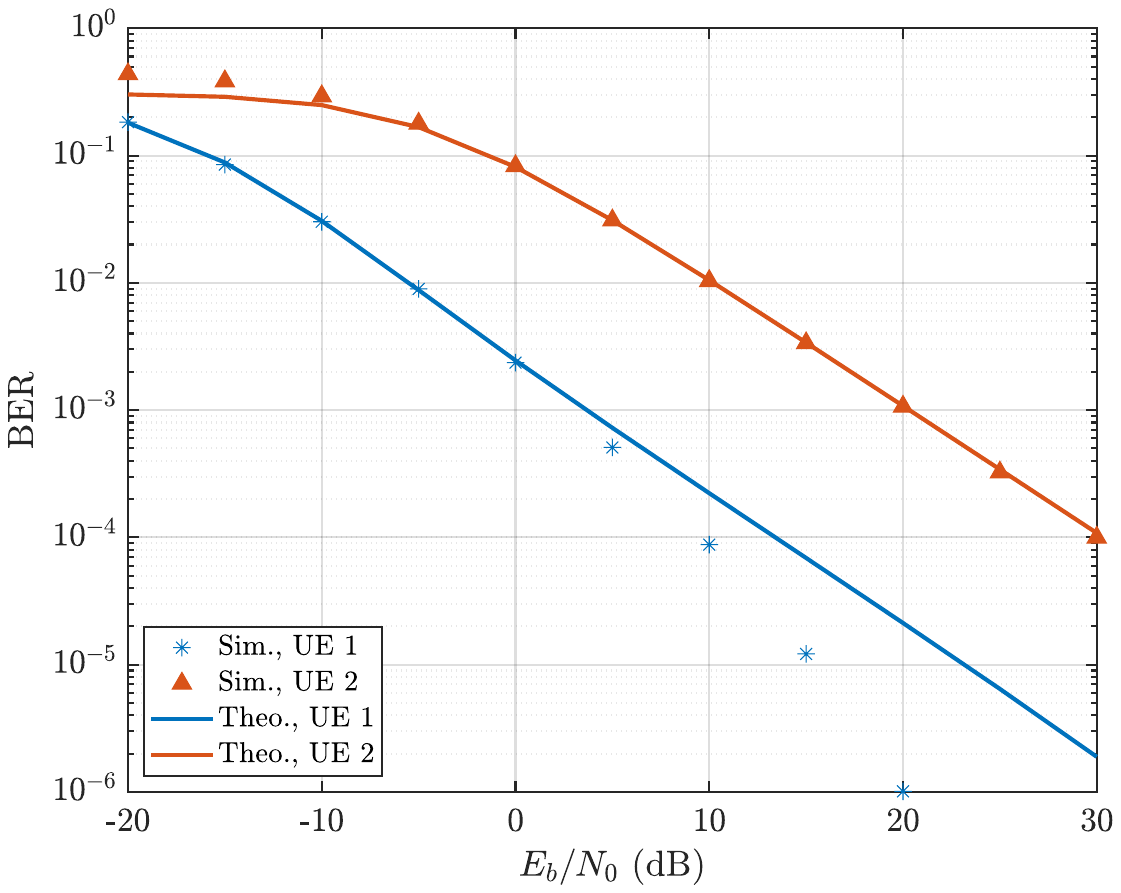}
    \label{fig:BER_of_2_UEs_h1_g_h2}
  }
  \hfill
  \subfloat[Case 2: $|h_2| > |h_1|$ only]{%
    \includegraphics[width=0.3\textwidth, trim=50 170 50 180, clip]{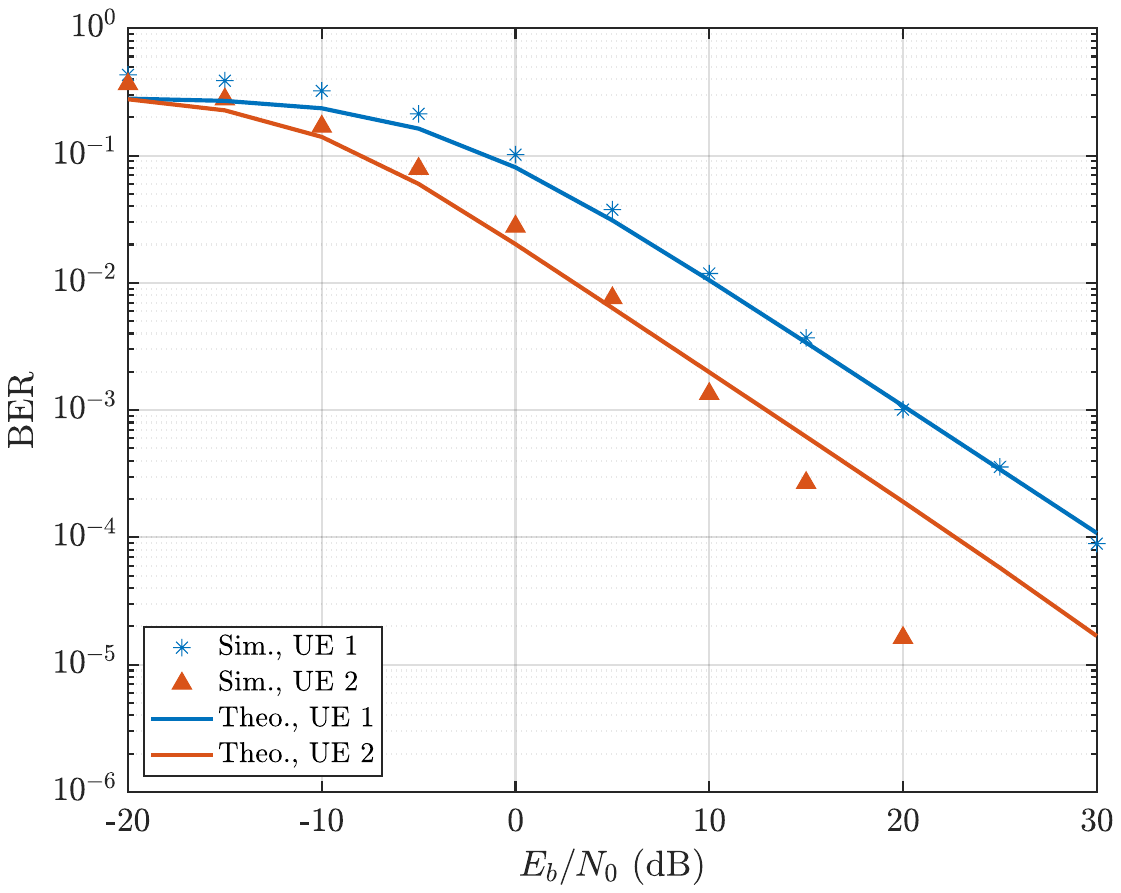}
    \label{fig:BER_of_2_UEs_h2_g_h1}
  }
  \caption{\gls{ber} performance of \gls{ue} 1 and \gls{ue} 2 with dynamic \gls{sic} decoding, where their power allocation are $p_{(1)} = -2.22$ dB, $p_{(2)} = -3.98$ dB and channel parameters $\sigma_1^2 = 20$ dB, $\sigma_2^2 = 7.96$ dB, respectively.}
  \label{fig:BER_of_2_UEs}
\end{figure*}

\autoref{fig:BER_of_2_UEs} presents the \gls{ber} performance analysis for a two-\gls{ue} system employing \gls{bpsk} modulation, where each \gls{ue} operates with distinct average channel gains and power allocation coefficients. As established in \autoref{sec:2_ue_ul_tx_rayleigh}, the dynamic \gls{sic} decoding order depends on instantaneous channel gains, resulting in two mutually exclusive scenarios:
\begin{itemize}
  \item $|h_1| > |h_2|$: \gls{ue} 1 is decoded first, followed by \gls{ue} 2
  \item $|h_2| > |h_1|$: \gls{ue} 2 is decoded first, followed by \gls{ue} 1
\end{itemize}
This dynamic ordering implies that each \gls{ue} experiences two distinct detection conditions: being decoded either first (with only interference from undecoded signals) or second (with potential error propagation from the first decoded signal). The average \gls{ber} for each \gls{ue} is obtained by combining the performance across both scenarios, weighted by their respective probabilities of occurrence.

\autoref{fig:BER_of_2_UEs_both} demonstrates the aggregate \gls{ber} performance for both \glspl{ue}, incorporating all possible decoding orders. The close agreement between theoretical predictions and simulation results across the entire \gls{snr} range validates the accuracy of our analytical framework. To provide deeper insights, \autoref{fig:BER_of_2_UEs_h1_g_h2} and \autoref{fig:BER_of_2_UEs_h2_g_h1} isolate the performance under scenarios 1 and 2, respectively. In \autoref{fig:BER_of_2_UEs_h1_g_h2}, considering only the instances where $|h_1|>|h_2|$, \gls{ue} 1 consistently benefits from being decoded first, resulting in lower \gls{ber} compared to its aggregate performance in \autoref{fig:BER_of_2_UEs_both}. Conversely, \gls{ue} 2 suffers from error propagation as it is always decoded second, leading to significantly degraded performance. The opposite behavior is observed in \autoref{fig:BER_of_2_UEs_h2_g_h1}, where \gls{ue} 2 enjoys the advantage of first decoding position.

The above results yield  the following key insights:
\begin{itemize}
  \item \textbf{Impact of Channel Fading}: Instantaneous fading can reverse the channel ordering despite \gls{ue} 1's superior average conditions, significantly impacting error performance when average channel gains are comparable.
  
  \item \textbf{High-\gls{snr} Behavior}: Theoretical and simulated \gls{ber} diverge slightly at high \gls{snr} due to inter-user interference dominance. The model slightly overestimates interference impact, suggesting potential refinements in modeling $\mathfrak{R}\{h_{m, (2)}\}$.
  
  \item \textbf{Performance Inversion}: \gls{ue} 2 achieves lower \gls{ber} than \gls{ue} 1 in \autoref{fig:BER_of_2_UEs_h2_g_h1} when $|h_2| > |h_1|$, benefiting from first decoding position. Both ordering scenarios must be considered since $\sigma_1^2 > \sigma_2^2$ does not guarantee $|h_1| > |h_2|$ for all realizations.
\end{itemize}

\subsection{Impact of Varying Power Coefficients on BER}\label{sec:varying_power_coeff}
This section investigates the alignment between theoretical predictions and simulation results under varying power allocation strategies. We examine the \gls{ber} performance of a two-\gls{ue} system where the power ratio $10\log_{10}(p_1/p_2)$ varies from 0.01 to 0.5 for $p_2$, with $p_1 = 1-p_2$. The system operates under heterogeneous channel conditions with average channel gains $\sigma_1^2=10$ dB and $\sigma_2^2=0$ dB for \gls{ue} 1 and \gls{ue} 2, respectively. Performance evaluation is conducted at $E_b/N_0 = 20$ dB.

\begin{figure}[htbp]
    \centering
    \includegraphics[width=0.8\columnwidth, trim=50 180 50 180, clip]{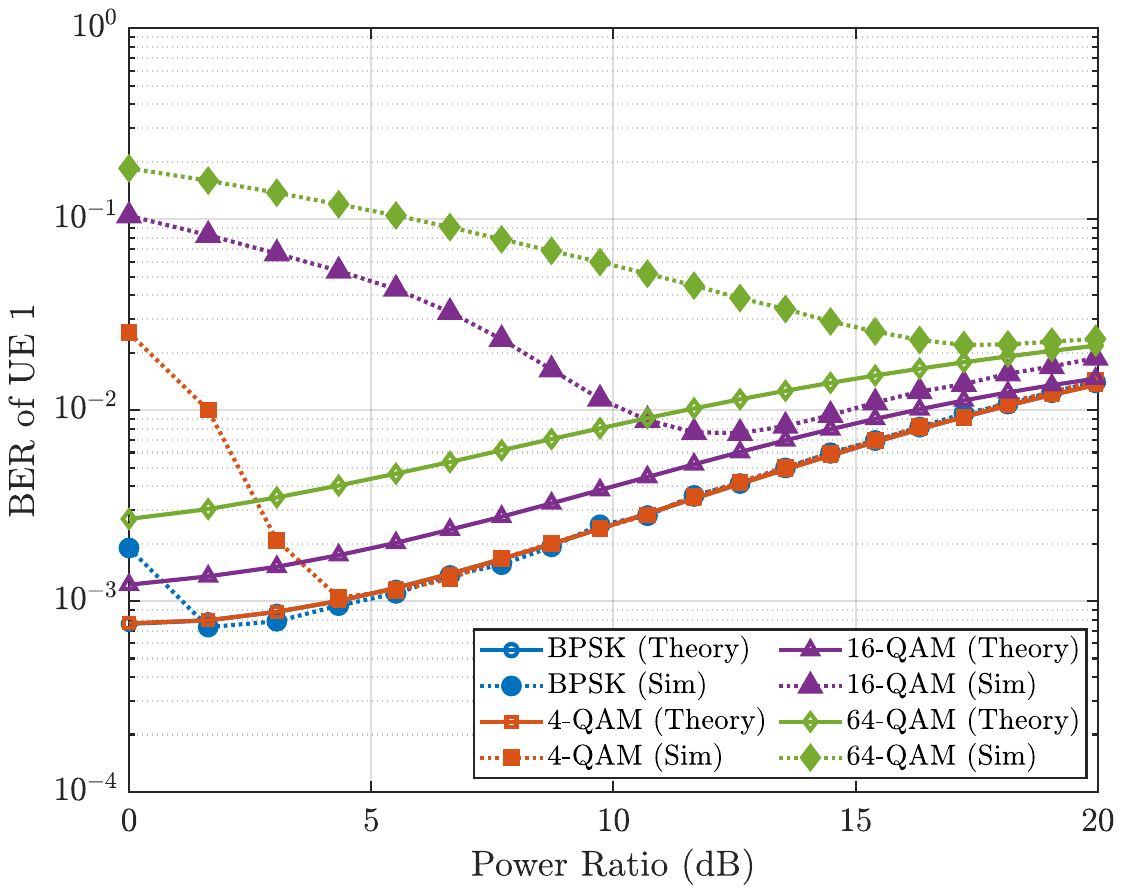}
    \caption{\gls{ber} of UE 1 versus power ratio $10\log_{10}(p_1/p_2)$ for BPSK, 4-QAM, 16-QAM, and 64-QAM at $E_b/N_0 = 20$ dB. Solid lines: theoretical \gls{pep} analysis; dotted lines with markers: Monte Carlo simulations. The two-user \gls{noma} system has noise variance parameters $\sigma_1^2 = 10$ dB and $\sigma_2^2 = 0$ dB.}
    \label{fig:noma_ber_vs_power_ratio_dB_all_modulations}
\end{figure}

\autoref{fig:noma_ber_vs_power_ratio_dB_all_modulations} presents the \gls{ber} performance across different modulation schemes under varying power allocation strategies. The results reveal important characteristics of the power-performance trade-off in \gls{pd-noma} systems:

\begin{itemize}
  \item \textbf{Theoretical-Simulation Alignment}: Strong agreement between theoretical and simulation results exists when the power ratio exceeds a specific threshold. Below this threshold, simulated \gls{ber} deteriorates sharply while theoretical predictions remain optimistic. This divergence highlights a fact that our theoretical calculation assumes the power differentiation between \glspl{ue} is sufficient.
  \item \textbf{Power Ratio Requirements}: Each modulation scheme demands a minimum power ratio for reliable \gls{sic} operation: \gls{bpsk} (1.63 dB), 4\gls{qam} (4.33 dB), 16\gls{qam} (12.61 dB), and 64\gls{qam} (19.05 dB). Below these thresholds, inter-user interference dominates as the receiver cannot distinguish between overlapping symbols, causing \gls{sic} failure regardless of \gls{snr}%\cite{7676258}
  . The increasing power requirements for higher-order modulations directly correlate with their reduced minimum constellation distances.
\end{itemize}
These findings demonstrate that power allocation in uplink \gls{noma} must balance spectral efficiency with decodability. While higher-order modulations offer greater throughput, their stringent power separation requirements may limit practical deployment in power-constrained uplink scenarios.

\subsection{Impact of Varying Average Channel Gains on BER}\label{sec:varying_sigma2}
Next, we  examine  how channel gain differences between \glspl{ue} influence error performance in uplink \gls{noma} systems. We consider a two-\gls{ue} system where \gls{ue} 2 serves as the reference with a fixed average channel gain of $\sigma_2^2 = 10$ dB. The average channel gain of \gls{ue} 1 ($\sigma_1^2$) increases from 10 dB to 50 dB. To minimize inter-user interference and focus on channel effects, we employ asymmetric power allocation with $p_1 = -0.04$ dB and $p_2 = -20$ dB, ensuring a sufficiently large power gap between users.

\begin{figure}[!t]
    \centering
    \includegraphics[width=0.8\columnwidth, trim=50 170 50 180, clip]{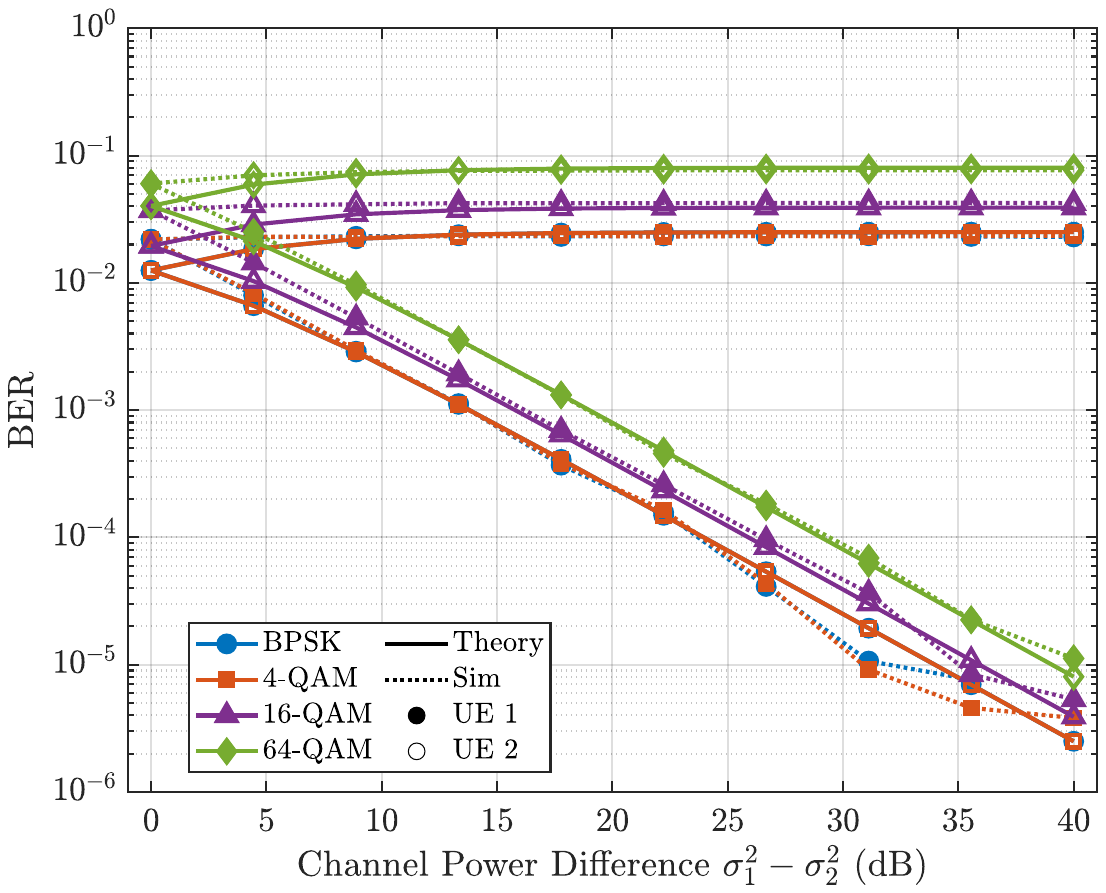}
    \caption{\gls{ber} performance versus channel power difference ($\sigma_1^2 - \sigma_2^2$) for \gls{noma} system with various modulation schemes. System parameters: $p_1 = -0.04$ dB, $p_2 = -20$ dB, with \gls{ue} 2 channel power fixed at $\sigma_2^2 = 10$ dB, $E_b/N_0 = 20$ dB.}
    \label{fig:noma_ber_channel_diff}
\end{figure}

In  \autoref{fig:noma_ber_channel_diff}, the main observations are as follows:
\begin{itemize}
  \item \gls{ue} 1's \gls{ber} decreases monotonically with channel gain difference, while \gls{ue} 2's \gls{ber} initially increases (0-10 dB) then stabilizes. This occurs because dynamic \gls{sic} allows \gls{ue} 2 to decode first when instantaneous fading favors it—more likely with similar average gains, rare with large disparities.
  \item Our framework accurately predicts performance for channel differences larger than 10 dB across all modulations. Below 10 dB, frequent channel crossovers between similarly-matched \glspl{ue} reduce accuracy.
\end{itemize}

These findings reveal that our theoretical framework accurately predicts performance when channel gain differences exceed 10 dB, but degrades for closely matched \gls{ue} conditions. Maintaining adequate channel diversity through user pairing or scheduling enhances both system performance and theoretical prediction reliability.

\subsection{BER Performance Comparison of SIC with Fixed and Dynamic Decoding Orders}
\begin{figure*}[htb]
  \centering
  \captionsetup[subfloat]{labelfont=scriptsize, textfont=scriptsize}
  \begin{tabular}{@{}c@{\hspace{5mm}}c@{}}
    \subfloat[BPSK: $p_{(1)}=-2.22\text{ dB}$, $p_{(2)}=-3.98\text{ dB}$, $\sigma_1^2=20\text{ dB}$, $\sigma_2^2=7.96\text{ dB}$]{%
      \includegraphics[width=0.35\textwidth, trim=50 170 50 180, clip]{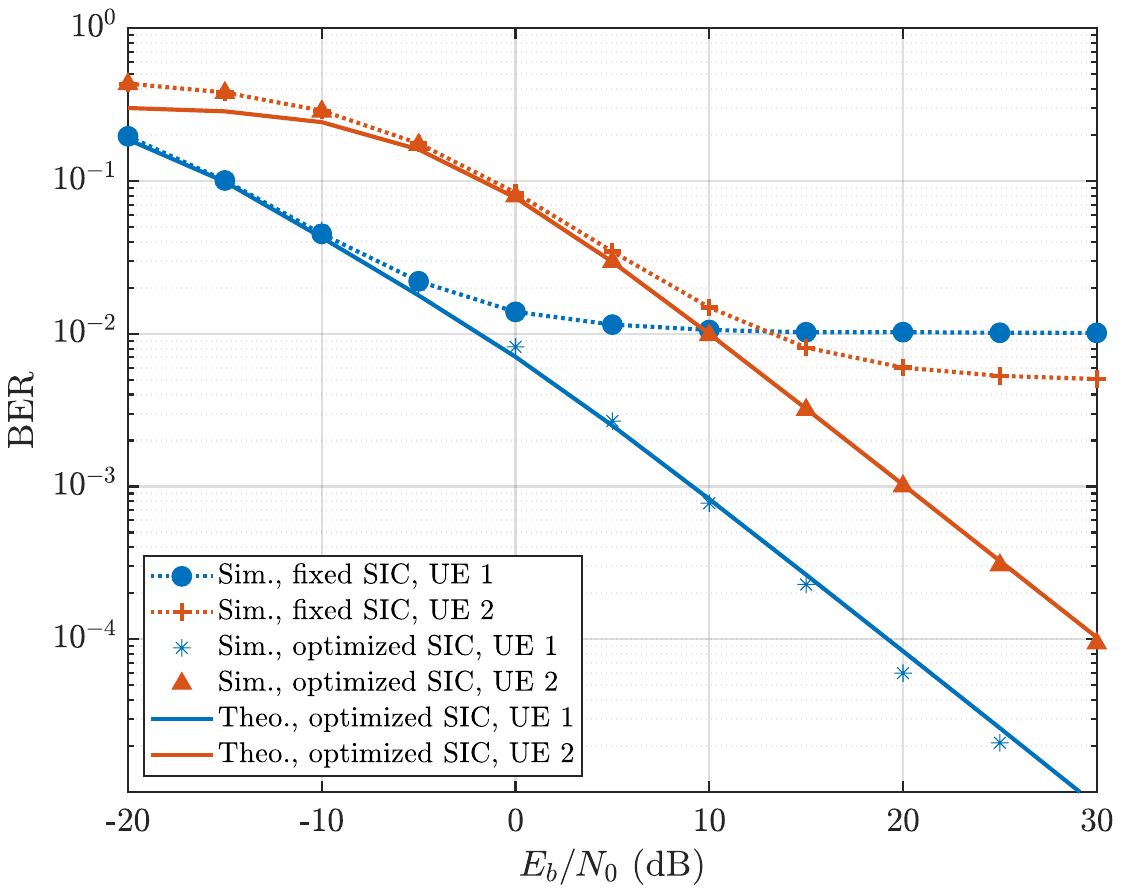}
      \label{fig:BER_fixed_SIC_vs_dynamic_SIC_bpsk}
    } &
    \subfloat[4\gls{qam}: $p_{(1)}=-0.46\text{ dB}$, $p_{(2)}=-10\text{ dB}$, $\sigma_1^2=20\text{ dB}$, $\sigma_2^2=7.96\text{ dB}$]{%
      \includegraphics[width=0.35\textwidth, trim=50 170 50 180, clip]{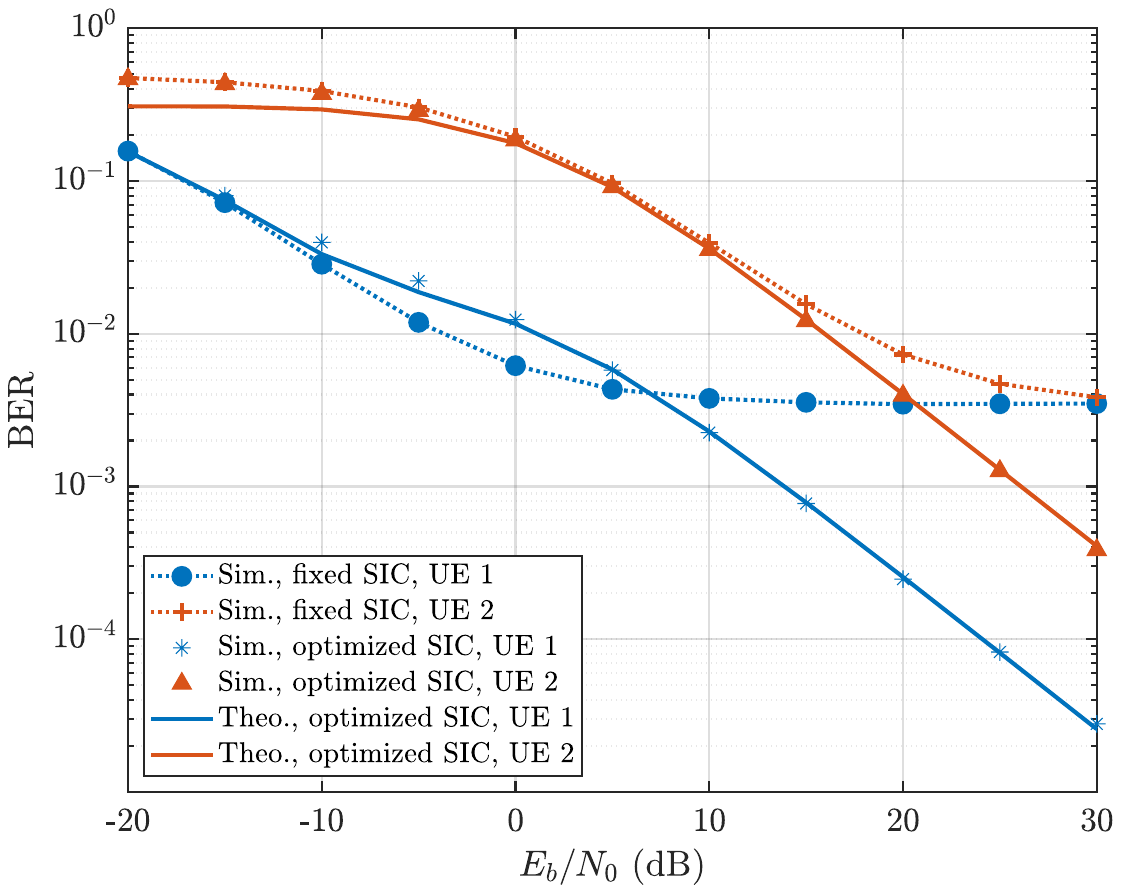}
      \label{fig:BER_fixed_SIC_vs_dynamic_SIC_4qam}
    } \\
    \subfloat[16\gls{qam}: $p_{(1)}=-0.04\text{ dB}$, $p_{(2)}=-20\text{ dB}$, $\sigma_1^2=38.06\text{ dB}$, $\sigma_2^2=26.02\text{ dB}$]{%
      \includegraphics[width=0.35\textwidth, trim=50 170 50 180, clip]{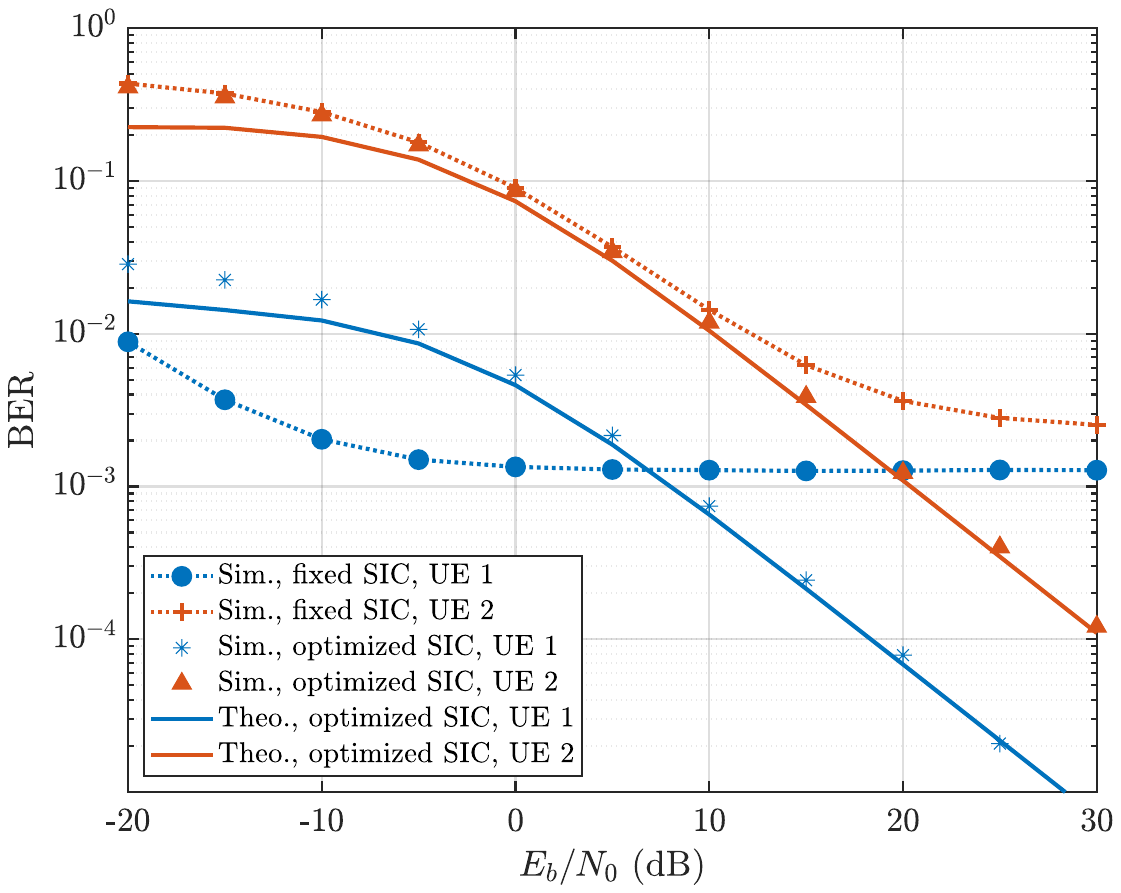}
      \label{fig:BER_fixed_SIC_vs_dynamic_SIC_16qam}
    } &
    \subfloat[64\gls{qam}: $p_{(1)}=-0.04\text{ dB}$, $p_{(2)}=-20\text{ dB}$, $\sigma_1^2=38.06\text{ dB}$, $\sigma_2^2=26.02\text{ dB}$]{%
      \includegraphics[width=0.35\textwidth, trim=50 170 50 180, clip]{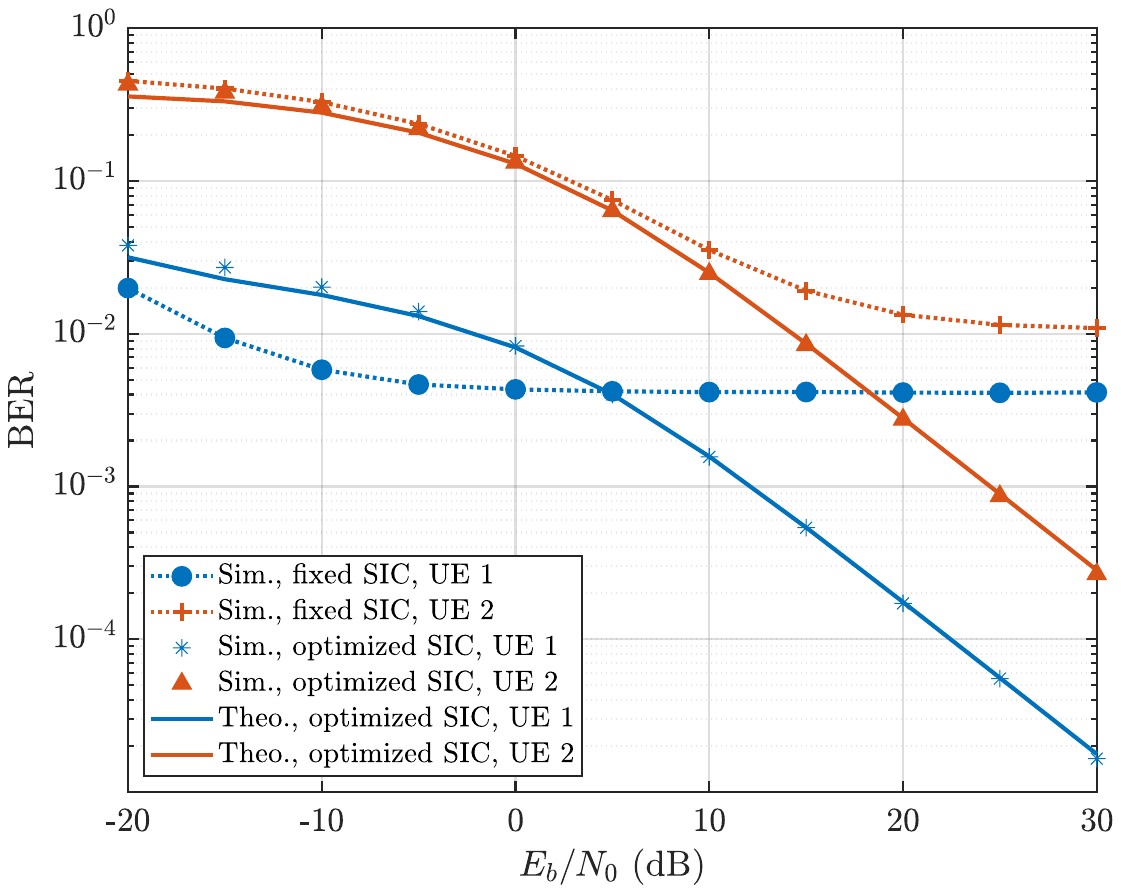}
      \label{fig:BER_fixed_SIC_vs_dynamic_SIC_64qam}
    }
  \end{tabular}
  \caption{BER of fixed \gls{sic} vs dynamic \gls{sic} with \gls{bpsk}, 4\gls{qam}, 16\gls{qam} and 64\gls{qam}.}
  \label{fig:BER_fixed_SIC_vs_dynamic_SIC}
\end{figure*}

\autoref{fig:BER_fixed_SIC_vs_dynamic_SIC} compares the \gls{ber} performance between fixed and dynamic \gls{sic} decoding for \gls{bpsk}, 4\gls{qam}, 16\gls{qam}, and 64\gls{qam}. Fixed \gls{sic}, where \gls{ue} 1 maintains decoding priority based on average channel gain, exhibits error floors at high \gls{snr} when instantaneous channel fluctuations violate the assumed order. Dynamic \gls{sic} eliminates these error floors by adapting to instantaneous channel realizations.

Notably, fixed \gls{sic} creates an asymmetric error mechanism: \gls{ue} 2 paradoxically outperforms \gls{ue} 1 at high \gls{snr} because \gls{ue} 1 remains interference-limited while \gls{ue} 2 becomes noise-limited after successful \gls{ue} 1 decoding. Higher-order modulations (16/64\gls{qam}) require increased power separation ($p_{(1)}=-0.04$ dB, $p_{(2)}=-20$ dB) due to reduced constellation distances, yet error floors persist. Dynamic \gls{sic} consistently achieves error-free performance with \gls{ber} decreasing continuously with \gls{snr}, validating our theoretical framework across all modulation schemes.

\subsection{BER Performance Analysis for Heterogeneous Modulation Schemes}
To demonstrate the generality and practical applicability of our analytical framework, we extend the performance evaluation to heterogeneous scenarios where \glspl{ue} employ different modulation orders. This analysis addresses real-world deployments where users may have varying quality of service requirements or channel conditions that necessitate different modulation schemes. Note that modulation orders are pre-assigned based on average channel gains (\gls{ue} 1 with $\sigma_1^2 > \sigma_2^2$ uses higher-order modulation), while dynamic modulation adaptation is left for future work.

\begin{figure*}[htb]
  \captionsetup[subfloat]{labelfont=scriptsize, textfont=scriptsize}
  \subfloat[Configuration 1: 4-QAM/BPSK heterogeneous system]{%
    \includegraphics[width=0.3\textwidth, trim=50 170 50 180, clip]{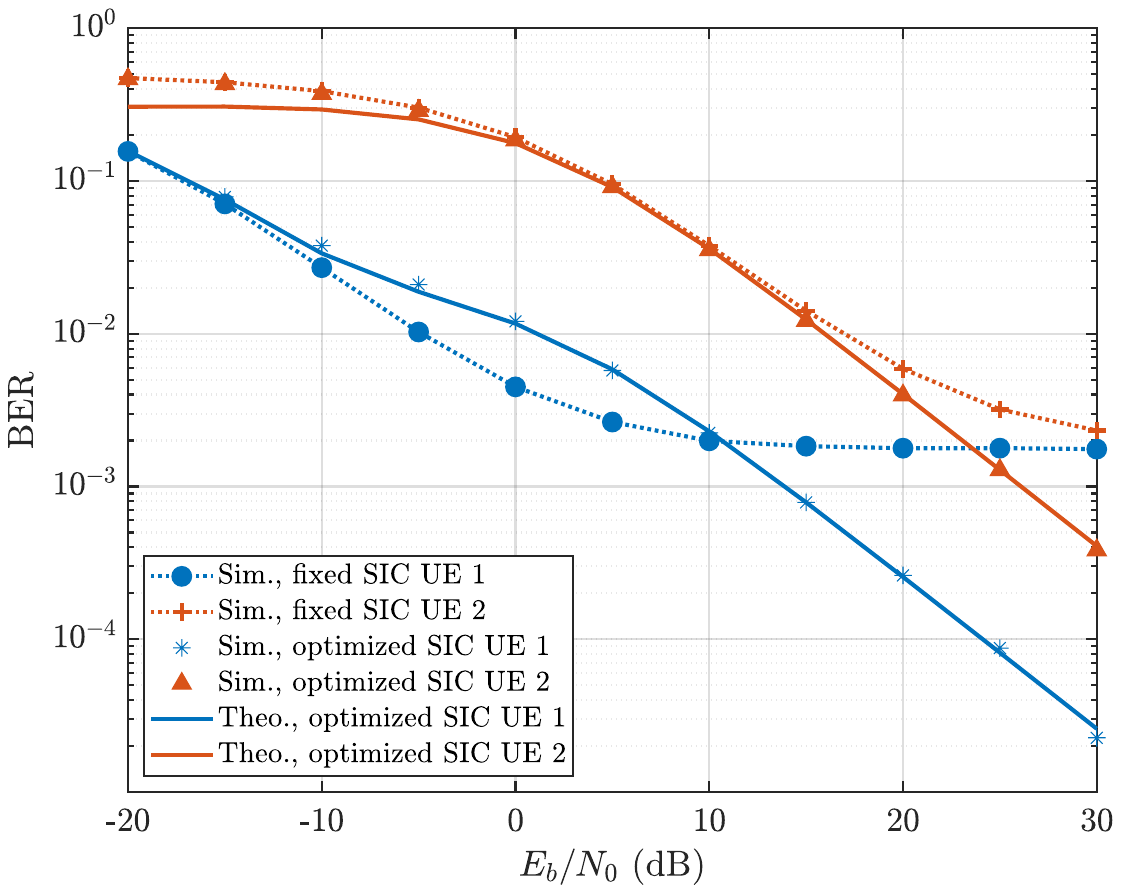}
    \label{fig:ber_het_config1}
  }
  \hfill
  \subfloat[Configuration 2: 16-QAM/4-QAM heterogeneous system]{%
    \includegraphics[width=0.3\textwidth, trim=50 170 50 180, clip]{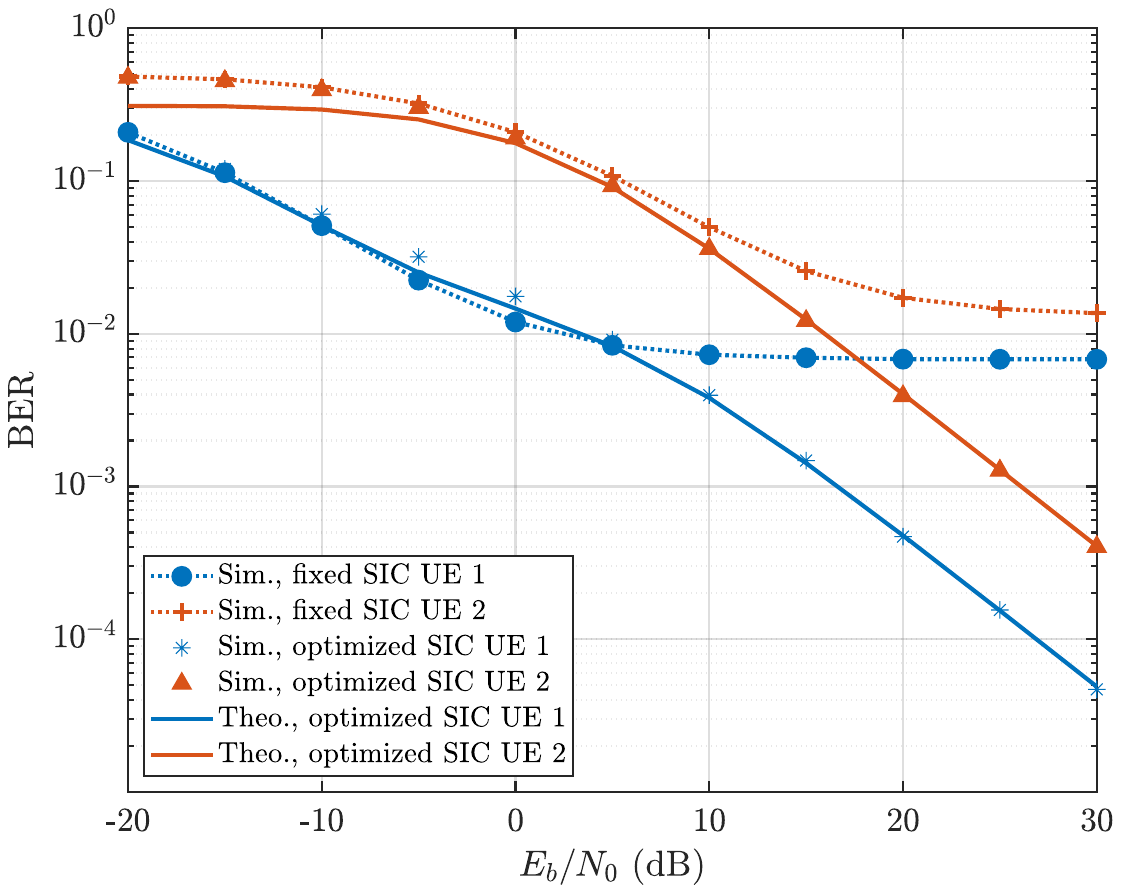}
    \label{fig:ber_het_config2}
  }
  \hfill
  \subfloat[Configuration 3: 64-QAM/BPSK heterogeneous system]{%
    \includegraphics[width=0.3\textwidth, trim=50 170 50 180, clip]{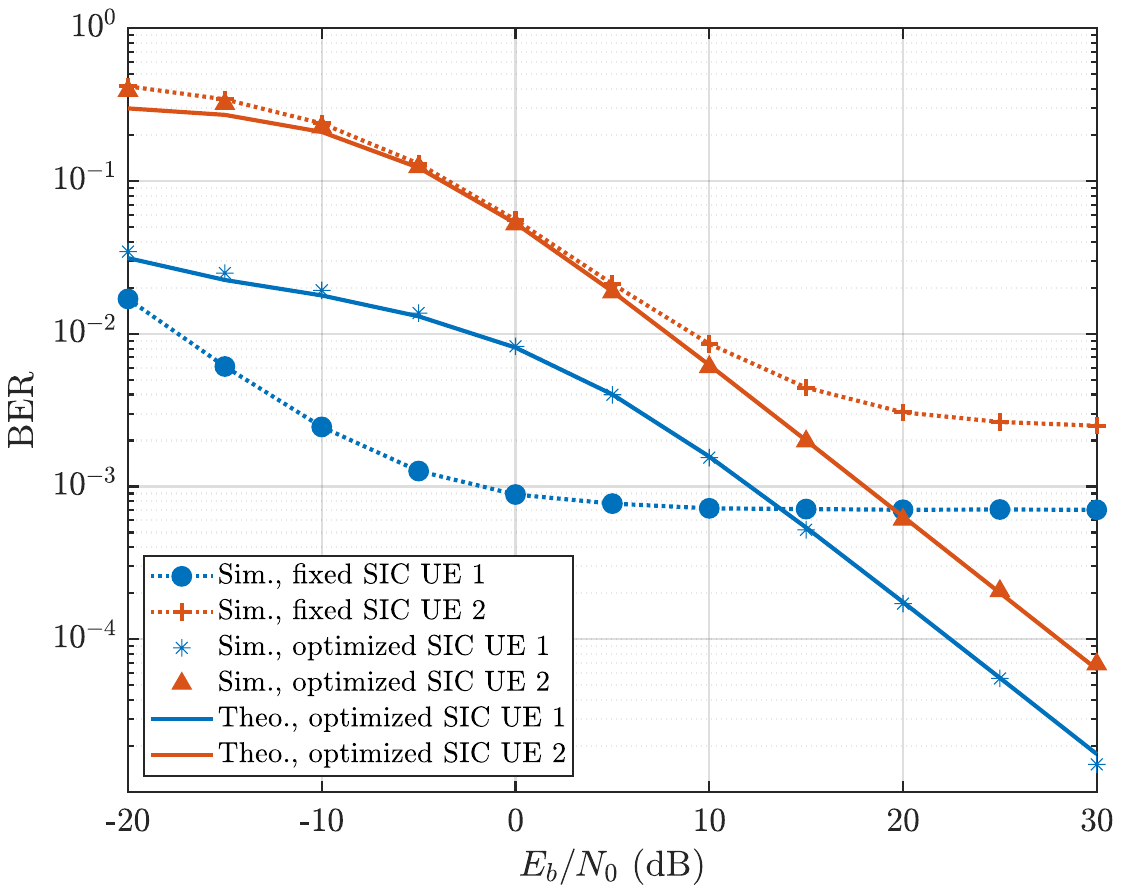}
    \label{fig:ber_het_config3}
  }
  \caption{BER performance comparison between fixed and dynamic \gls{sic} decoding for heterogeneous modulation schemes. Solid lines represent theoretical predictions while markers denote simulation results. Dynamic \gls{sic} consistently outperforms fixed \gls{sic} across all configurations, with the performance gap widening as modulation order disparity increases.}
  \label{fig:BER_fixed_SIC_vs_dynamic_SIC_Diff_M_QAM}
   % This paper \cite{1611074} use similar way to get the range of error distance
\end{figure*}

\autoref{fig:BER_fixed_SIC_vs_dynamic_SIC_Diff_M_QAM} presents the \gls{ber} performance comparison between fixed and dynamic \gls{sic} decoding for three representative heterogeneous configurations. The system parameters, detailed in \autoref{tab:modulation_configs}, were selected to explore diverse operational scenarios while maintaining practical relevance. From \autoref{fig:BER_fixed_SIC_vs_dynamic_SIC_Diff_M_QAM}, it demonstrates that:
\begin{itemize}
  \item \textbf{Universal validity}: Our theoretical framework accurately predicts \gls{ber} performance across all heterogeneous configurations with excellent theoretical-simulation agreement.
  
  \item \textbf{Dynamic \gls{sic} superiority}: Dynamic \gls{sic} eliminates error floors observed in fixed \gls{sic}, with benefits increasing for larger modulation order disparities.
  
  \item \textbf{Power-modulation trade-off}: Extreme modulation heterogeneity requires careful power allocation and channel separation to balance user fairness and performance.
\end{itemize}

These results validate our framework's applicability to practical heterogeneous deployments with diverse user requirements.

\begin{table}[htb]
  \centering
  \caption{Heterogeneous modulation configurations and system parameters for comparative performance evaluation}
  \begin{tabular}{c|cc|cc|cc}
    \toprule
    \textbf{Config.} & \multicolumn{2}{c|}{\textbf{Modulation}} & \multicolumn{2}{c|}{\textbf{Power (dB)}} & \multicolumn{2}{c}{\textbf{Channel (dB)}} \\
    \cmidrule(lr){2-3} \cmidrule(lr){4-5} \cmidrule(lr){6-7}
    & $\mathcal{M}_1$ & $\mathcal{M}_2$ & $p_{(1)}$ & $p_{(2)}$ & $\sigma_1^2$ & $\sigma_2^2$ \\
    \midrule
    1 & 4\gls{qam} & \gls{bpsk} & $-0.46$ & $-10$ & $20$ & $7.96$ \\
    2 & 16\gls{qam} & 4\gls{qam} & $-0.46$ & $-10$ & $20$ & $7.96$ \\
    3 & 64\gls{qam} & \gls{bpsk} & $-0.04$ & $-20$ & $38.06$ & $26.02$ \\
    \bottomrule
  \end{tabular}
  \label{tab:modulation_configs}
\end{table}

\section{Conclusion}\label{sec:conclusion}
This paper presents a comprehensive error performance analysis of uplink \gls{noma} systems employing dynamic \gls{sic} decoding, where the decoding order adapts to instantaneous channel conditions. We derive closed-form \gls{pep} expressions that capture the fundamental impact of dynamic ordering on system performance, addressing a critical gap in the theoretical understanding of uplink \gls{noma}. We first derive the exact closed-form \glspl{pdf} for ordered channel gains in \gls{ind} Rayleigh fading environments. To maintain analytical tractability while achieving high accuracy, a Gaussian approximation framework is developed for the truncated channel gain distributions. Finally, closed-form BER expressions are derived for various modulation schemes (i.e., \gls{bpsk}, 4\gls{qam}, 16\gls{qam}, and 64\gls{qam}) in both homogeneous and heterogeneous scenarios.

Our evaluation demonstrates that: 
\begin{enumerate*}[label=(\roman*)]
  \item Fixed \gls{sic} decoding suffers from an inherent error floor at high \gls{snr} that persists regardless of transmission power, while dynamic \gls{sic} eliminates this fundamental limitation to achieve error floor-free performance;
  \item Dynamic \gls{sic} delivers substantial \gls{ber} improvements across all modulation schemes, enabling reliable communication in medium to high \gls{snr} regimes where fixed ordering fails;
  \item Our analytical framework accurately predicts dynamic \gls{sic} performance in heterogeneous scenarios where \glspl{ue} employ different modulation orders, validating its applicability to practical deployments with diverse user requirements.
\end{enumerate*}
These findings provide theoretical confirmation that adapting the decoding order to instantaneous channel conditions is essential for reliable uplink \gls{noma} communications. 
%The strong agreement between theoretical predictions and simulations across diverse scenarios validates both our mathematical framework and the practical benefits of dynamic \gls{sic}.

%Future work will extend this framework to multi-user scenarios ($N > 2$), incorporate imperfect channel state information effects, and develop adaptive power allocation strategies that jointly optimize transmission parameters based on modulation schemes and instantaneous channel conditions. Machine learning approaches could also be explored to enhance decoding order prediction and power allocation decisions in real-time systems.

\begin{appendices}
\crefalias{section}{appendix}
\input{app_proof_F_Hk.tex}
\input{app_pdf_zn1_zn2.tex} % merged app_pdf_xi_zeta to here
\input{app_cf_pep_2_ues.tex} %F->C
\input{app_cf_P_Bnk.tex} %G->D
\input{app_ber_of_m_qam.tex} %H->E
\input{app_pdf_hnk.tex}% C->F
\input{app_truncated_Rhm.tex}%D->G

\end{appendices}

\bibliographystyle{IEEEtran}
\bibliography{IEEEabrv,main}

% \newpage

\vfill

\end{document}

%% file: app_proof_F_Hk.tex
\section{General Distribution Function of $F_{|h_{(k)}|}\left(x\right)$}\label{app:proof_F_Hk}
As mentioned in \autoref{sec:P_Bnk}%\cite[III-B]{zhang2025error_noma}
, the probability of $B_{n,(k)}$ can be calculated by integrating $x$ over the \gls{cdf} of the $r$-th order statistic $F_{|h_{(r)}|}\left(x\right)$ for $k+1\leq r\leq N$. Recall that the instantaneous channel gains of the $N$ \gls{ue}s, $|h_1|,\dots,|h_N|$, are \gls{ind} random variables with \gls{cdf}s $F_{|h_1|},\dots,F_{|h_N|}$, respectively. According to \cite[Section 5.4]{david2003order}, the special case where $r=1$ for order statistics of \gls{ind} random variables can be expressed as:
{\small
\begin{align}
  F_{|h_{(1)}|}\left(x\right) &= P\left(|h_{(1)}|\leq x\right)= 1 - P\left(|h_{(1)}|> x\right)\nonumber\\
  &=1 - P\left(|h_{1}|> x,\dots,|h_{N}|> x\right)\nonumber\\
  &= 1 - \prod_{i=1}^{N}P\left(|h_{i}| > x\right)\nonumber\\
  &=1 - \prod_{i=1}^{N}\left[1-F_{|h_i|}\left(x\right)\right].
  \label{eq:F_H1}
\end{align}
}
Similarly, the special case where $r=N$ can be expressed as:
{\small
\begin{align}
  F_{|h_{(N)}|}\left(x\right) &= P\left(|h_{(N)}|\leq x\right)= P\left(|h_{1}|\leq x,\dots,|h_{N}|\leq x\right)\nonumber\\
  &=\prod_{i=1}^{N}P\left(|h_{i}|\leq x\right) = \prod_{i=1}^{N}F_{|h_i|}\left(x\right).
  \label{eq:F_HN}
\end{align}
}
For the general case where $1<r<N$, the \gls{cdf} of the $r$-th order statistic $|h_{(r)}|$ can be expressed as follows \cite[Theorem 4.1]{f2273ed0-b48a-3ace-9640-7dc852794849}:
{\small
\begin{align}
  F_{|h_{(r)}|}\left(x\right)&=P\left(|h_{(r)}|\leq x\right)\nonumber\\
  &=\sum_{i=r}^{N}P\left(\text{exactly } i\text{ of the }|h_1|,\dots,|h_N|\text{ are }\leq x\right)\nonumber\\
  &=\sum_{i=r}^{N}\sum_{S_i}\prod_{j\in S_i}F_{|h_{j}|}\left(x\right) \prod_{j\notin S_i}\left[1-F_{|h_{j}|}\left(x\right)\right]\nonumber\\
  &=\sum_{i=r}^{N}\frac{1}{i!(N-i)!}\text{per}\begin{bmatrix}
    F_{|h_1|}\left(x\right) & 1-F_{|h_1|}\left(x\right)\\
    \vdots & \vdots\\
    \underbrace{F_{|h_N|}\left(x\right)}_{i\text{ columns}} & \underbrace{1-F_{|h_N|}\left(x\right)}_{(N-i)\text{ columns}}
    \end{bmatrix},
  \label{eq:proof_F_Hk}
\end{align}
}%
where $S_i$ denotes all possible subsets of size $i$ from $\{1,\dots,N\}$, and $\text{per}(\cdot)$ denotes the permanent of a matrix.

%% file: app_pdf_zn1_zn2.tex
\section{Closed-form PDFs of $z_{n,(1)}$ and $z_{n,(2)}$}\label{app:pdf_zn1_zn2}
This appendix derives the closed-form \gls{pdf}s for the signal-plus-interference decision variables $z_{n,(1)}$ and $z_{n,(2)}$ defined in \autoref{sec:probability_an_bnk}%\cite[IV-A]{zhang2025error_noma}
. The $z_{n,(1)}$ and $z_{n,(2)}$ are functions of other random variables, specifically:
{\small
\begin{equation}
\begin{aligned}
    z_{n,(1)} &= \frac{|\xi_{n,(1)}| + 2\mathfrak{R}\{\zeta_{n,(1)}\}}{\sqrt{2N_0}}, \\
    z_{n,(2)} &= \frac{|\xi_{n,(2)}| + 2\mathfrak{R}\{\zeta_{n,(2)}\}}{\sqrt{2N_0}},
\end{aligned}
\end{equation}
}%
where the component random variables are defined as:
{\small
\begin{equation}
\begin{aligned}
    |\xi_{n,(1)}| &= \sqrt{p_{(1)}}|h_{n,(1)}|\Delta_{n,(1)}, \\
    \mathfrak{R}\{\zeta_{n,(1)}\} &= \sqrt{p_{(2)}}\mathfrak{R}\{h_{m,(2)}\}x_{m,(2)}, \\
    |\xi_{n,(2)}| &= \sqrt{p_{(2)}}|h_{n,(2)}|\Delta_{n,(2)}, \\
    \mathfrak{R}\{\zeta_{n,(2)}\} &= \sqrt{p_{(1)}}\mathfrak{R}\{h_{m,(1)}\}\Delta_{m,(1)}.
\end{aligned}
\end{equation}
}%
Here, $p_{(k)}$ denotes the power coefficient, $h_{n,(k)}$ represents the channel coefficient, $\Delta_{n,(k)}$ is the Euclidean distance between transmitted symbols, and $x_{m,(k)}$ is the transmitted symbol from \gls{ue} $m$ with decoding order $k$.

The \gls{pdf}s of these component random variables can be derived using the Jacobian transformation \cite{lovett2019differential}. Given the \gls{pdf}s $f_{|h_{n,(k)}|}(x)$ and $f_{\mathfrak{R}\{h_{m,(k)}\}}(x)$ for $k \in \{1,2\}$, we obtain:
{\small
\begin{align}
    f_{|\xi_{n,(1)}|}(x) &= \frac{f_{|h_{n,(1)}|}\left(\frac{x}{\sqrt{p_{(1)}}|\Delta_{n,(1)}|}\right)}{\sqrt{p_{(1)}}|\Delta_{n,(1)}|}, \label{eq:pdf_f_xi_1}\\
    f_{\mathfrak{R}\{\zeta_{n,(1)}\}}(x) &= \frac{f_{\mathfrak{R}\{h_{m,(2)}\}}\left(\frac{x}{\sqrt{p_{(2)}}\mathfrak{R}\{x_{m,(2)}\}}\right)}{\sqrt{p_{(2)}}\mathfrak{R}\{x_{m,(2)}\}}, \label{eq:pdf_f_Rzeta_1}\\
    f_{|\xi_{n,(2)}|}(x) &= \frac{f_{|h_{n,(2)}|}\left(\frac{x}{\sqrt{p_{(2)}}|\Delta_{n,(2)}|}\right)}{\sqrt{p_{(2)}}|\Delta_{n,(2)}|}, \label{eq:pdf_f_xi_2}\\
    f_{\mathfrak{R}\{\zeta_{n,(2)}\}}(x) &= \frac{f_{\mathfrak{R}\{h_{m,(1)}\}}\left(\frac{x}{\sqrt{p_{(1)}}\mathfrak{R}\{\Delta_{m,(1)}\}}\right)}{\sqrt{p_{(1)}}\mathfrak{R}\{\Delta_{m,(1)}\}}. \label{eq:pdf_f_Rzeta_2}
\end{align}
}%
These expressions follow from the transformation theorem for probability densities, where the scaling factor in the denominator accounts for the Jacobian of the transformation.

To obtain the \gls{pdf}s of $z_{n,(1)}$ and $z_{n,(2)}$, which depend on both $|\xi_{n,(k)}|$ and $\mathfrak{R}\{\zeta_{n,(k)}\}$, we need to derive their joint distributions and then compute the marginal \gls{pdf}s through integration. These \gls{pdf}s are essential for computing the probability $P\left(A_{n}\mid B_{n,(k)}\right)$ in the main text.

As shown in the main paper, $z_{n,(1)}$ comprises $|h_{n,(1)}|$ and $\mathfrak{R}\left\{h_{m,(2)}\right\}$, while $z_{n,(2)}$ consists of $|h_{n,(2)}|$ and $\mathfrak{R}\left\{h_{m,(1)}\right\}$. Since these component random variables are independent, we derive the joint \gls{pdf}s by analyzing each component separately. The following subsections present detailed derivations for both $f_{z_{n,(1)}}(z)$ and $f_{z_{n,(2)}}(z)$.

We consider Rayleigh fading channels for uplink transmissions, where $|h_{n}|\sim\mathcal{R}\left(\sigma_{n}\right)$ and $|h_{m}|\sim\mathcal{R}\left(\sigma_{m}\right)$, with $\sigma_{n}\neq\sigma_{m}$ and $n,m\in\{1,2\}$, $n\neq m$. The optimal decoding order is determined by the instantaneous channel gains of all \gls{ue}s. 

For clarity, we adopt the following notation: when \gls{ue} $n$ has a larger channel gain than \gls{ue} $m$ (i.e., $|h_n|\geq|h_m|$), we denote this ordered channel gain as $|h_{n,(1)}|$. Conversely, when \gls{ue} $n$ has a smaller channel gain than \gls{ue} $m$ (i.e., $|h_n|<|h_m|$), we denote it as $|h_{n,(2)}|$. Consequently, $|h_{n,(k)}|$ for $k\in\{1,2\}$ follows a truncated Rayleigh distribution with bounds determined by the other \gls{ue}'s channel gain. According to the derivations in \cref{app:pdf_hnk}, the \gls{pdf}s are:
{\small
\begin{align}\label{eq:pdf_breve_h_1}
  f_{|h_{n,(1)}|}(x)=\frac{2x}{\sigma_{n}^2-\sigma_{m}^2}\left[\exp\left(-\frac{x^2}{\sigma_{n}^2}\right)-\exp\left(-\frac{x^2}{\sigma_{m}^2}\right)\right],
\end{align}
}%
and
{\small
\begin{align}\label{eq:pdf_breve_h_2}
  f_{|h_{n,(2)}|}(x)=\frac{2x\exp\left(-\frac{x^2}{\sigma_{n}^2}\right)}{\sigma_{n}^2}.
\end{align}
}

To verify the accuracy of the derived $f_{|h_{n,(1)}|}(x)$ and $f_{|h_{n,(2)}|}(x)$, we validate these \gls{pdf}s through Monte Carlo simulations. We consider two \gls{ue}s, indexed by $n=1$ and $m=2$, with channel gains modeled as $|h_1|\sim\mathcal{R}\left(\sigma_1\right)$ and $|h_2|\sim\mathcal{R}\left(\sigma_2\right)$. The channel variances are set to $\sigma_1^2=20\,\mathrm{dB}$ and $\sigma_2^2=7.96\,\mathrm{dB}$.

\begin{figure*}[!ht]
  \captionsetup[subfloat]{labelfont=scriptsize,textfont=scriptsize}
  \subfloat[Channel distribution of \gls{ue} 1 when $|h_1|>|h_2|$]{%
    \includegraphics[width=0.23\textwidth, trim=50 170 50 180, clip]{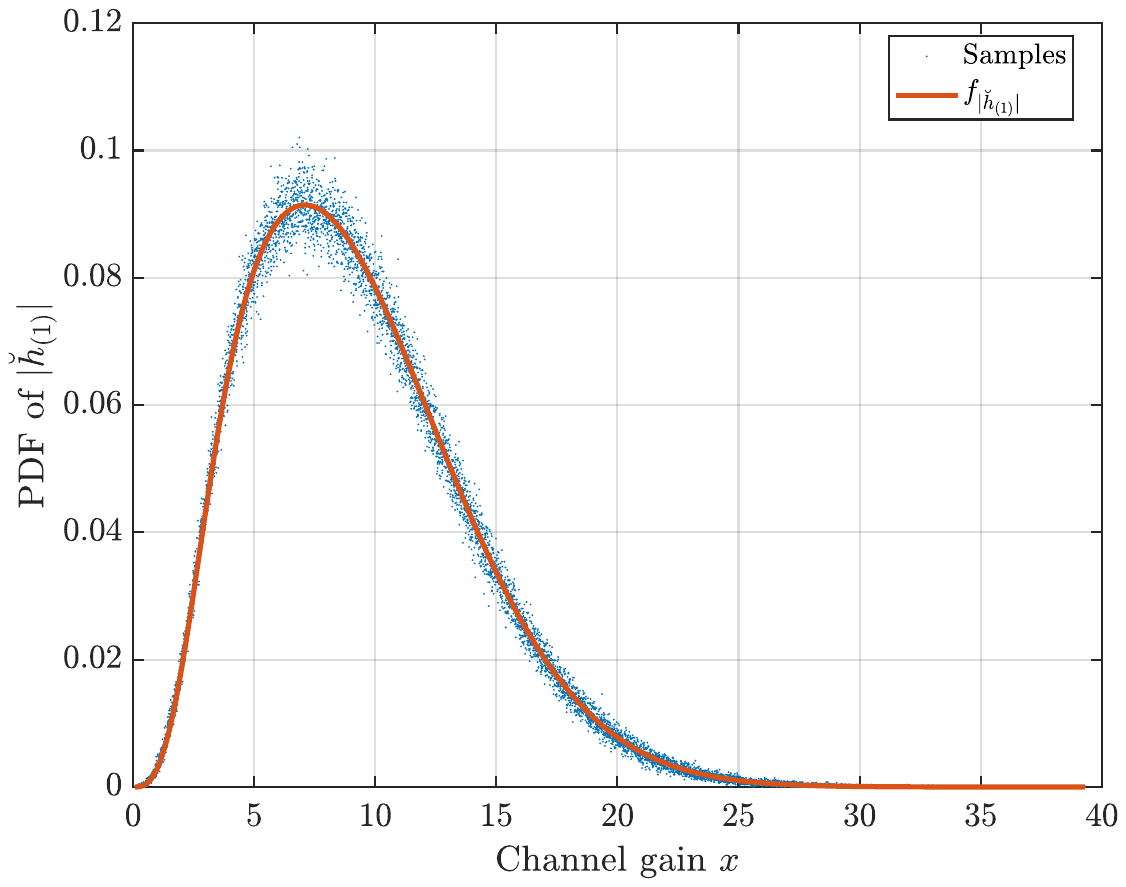}
    \label{fig:sorted_ch_ue1_order1}
  }
  \hfill
  \subfloat[Channel distribution of \gls{ue} 1 when $|h_1|\leq|h_2|$]{%
    \includegraphics[width=0.23\textwidth, trim=50 170 50 180, clip]{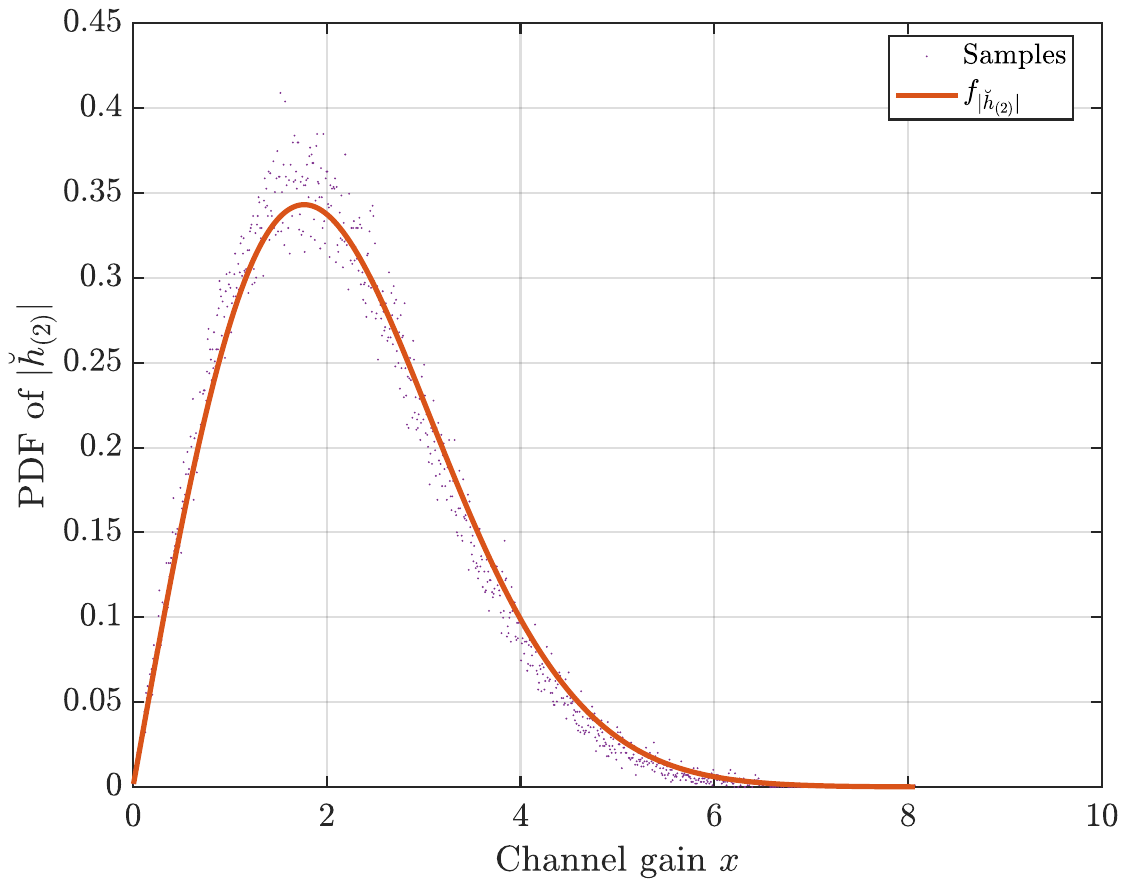}
    \label{fig:sorted_ch_ue1_order2}
  }
  \hfill
  \subfloat[Channel distribution of \gls{ue} 2 when $|h_2|>|h_1|$]{%
    \includegraphics[width=0.23\textwidth, trim=50 170 50 180, clip]{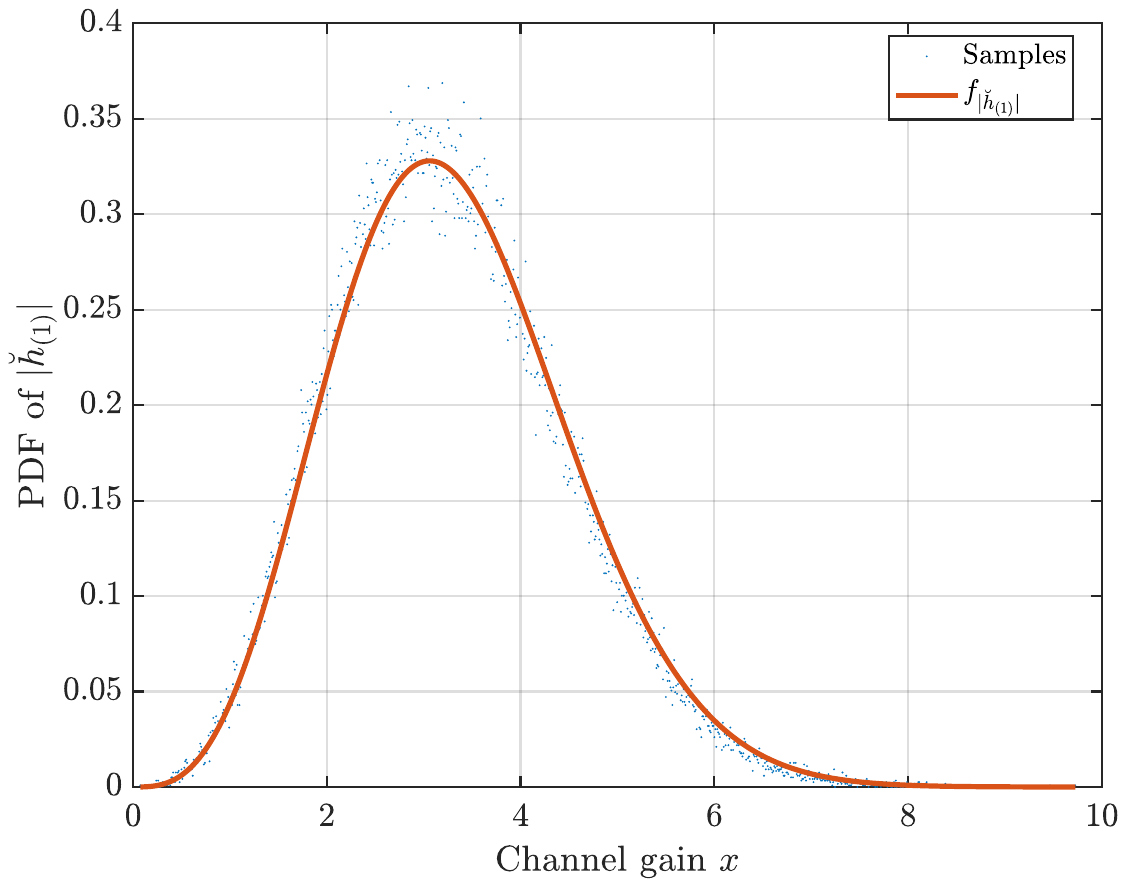}
    \label{fig:sorted_ch_ue2_order1}
  }
  \hfill
  \subfloat[Channel distribution of \gls{ue} 2 when $|h_2|\leq|h_1|$]{%
    \includegraphics[width=0.23\textwidth, trim=50 170 50 180, clip]{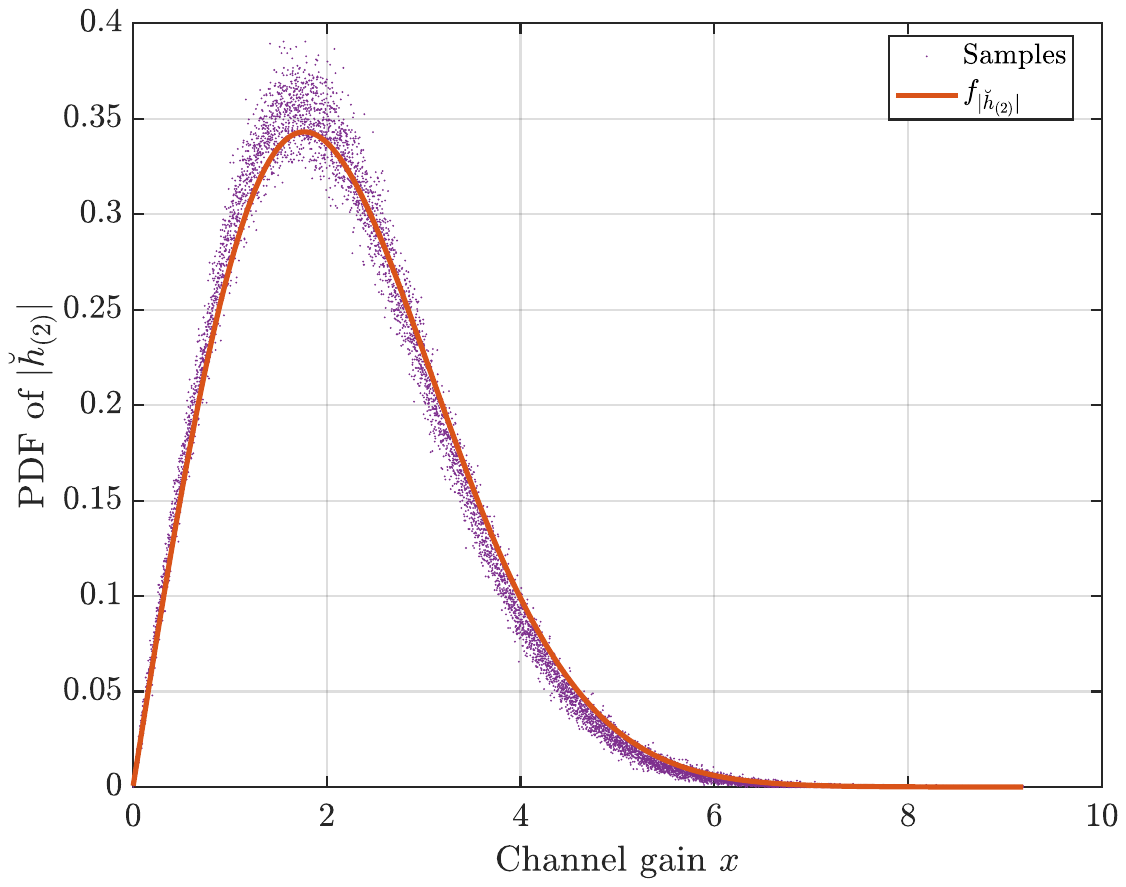}
    \label{fig:sorted_ch_ue2_order2}
  }
  \caption{Sorted channel distributions of 2 \gls{ue}s where $\sigma_1^2=20\,\text{dB}$ and $\sigma_2^2=7.96\,\text{dB}$}
  \label{fig:sorted_channel_distributions}
\end{figure*}

\autoref{fig:sorted_channel_distributions} presents the \gls{pdf}s of the sorted channel gains, where the dots represent Monte Carlo samples and the solid red lines show the analytical expressions derived from $f_{|h_{1,(1)}|}$, $f_{|h_{1,(2)}|}$, $f_{|h_{2,(1)}|}$, and $f_{|h_{2,(2)}|}$. \autoref{fig:sorted_channel_distributions} shows the distributions of sorted channel gains for both \gls{ue}s under different ordering conditions: (a)-(b) depict \gls{ue} 1's channel for $|h_1| \geq |h_2|$ and $|h_1| < |h_2|$, respectively; (c)-(d) show \gls{ue} 2's channel for $|h_2| > |h_1|$ and $|h_2| \leq |h_1|$, respectively.

When $\sigma_1 \geq \sigma_2$, a special consideration is required for computing $f_{|h_{2,(1)}|}(x)$ (the \gls{pdf} of \gls{ue} 2 under the condition $|h_2| \geq |h_1|$). In this scenario, we use a modified variance $\sigma_m$ where $|\sigma_m - \sigma_1| < \epsilon$ with $\epsilon = 10^{-5}$. This adjustment is necessary because when $\sigma_1 \geq \sigma_2$, only a small fraction of samples satisfy the condition $|h_2| \geq |h_1|$, and among these samples, the channel gain $|h_1|$ tends to be close to zero. As shown in \cref{eq:pdf_hn1}, when $|h_m| \to 0$, the truncated distribution approaches a standard Rayleigh distribution as given in \cref{eq:pdf_rayleigh}. Consequently, the distribution of $|h_1|$ conditioned on $|h_1| \leq |h_2|$ can be well approximated by the distribution of $|h_2|$.

The next step is to derive the \gls{pdf} of $\mathfrak{R}\left\{h_{m,(k)}\right\}$ for $k\in\{1,2\}$. Since the real part of $h_m$ follows $\mathfrak{R}\{h_{m}\}\sim\mathcal{N}\left(0,\frac{\sigma_{m}^2}{2}\right)$, and similar to $|h_{n,(k)}|$, the random variable $\mathfrak{R}\left\{h_{m,(k)}\right\}$ has a truncated Gaussian distribution whose support is determined by the channel gains of other \gls{ue}s. However, as discussed in \cref{app:truncated_Rhm}, the complexity of these truncated distributions makes it challenging to derive closed-form expressions for the marginal \gls{pdf}s $f_{\mathfrak{R}\{h_{m,(k)}\}}\left(x\right)$ directly. 

To overcome this challenge, we approximate $f_{\mathfrak{R}\{h_{m,(k)}\}}$ as a weighted sum of Gaussian functions using the curve fitting method proposed in \cite{5999593}. Specifically, the approximation takes the form
{\small
\begin{align}
  f_{\mathfrak{R}\{h_{m,(k)}\}}\left(x\right)\approx\sum_{i=1}^{N_{\mathcal{G}}}a_{i}\exp\left(-\frac{\left(x-b_i\right)^2}{c_i^2}\right),
  \label{eq:curve_fit_gaussian_func}
\end{align}
}%
where $N_{\mathcal{G}}$ denotes the number of Gaussian basis functions, and $\{a_i, b_i, c_i\}_{i=1}^{N_{\mathcal{G}}}$ are the fitting coefficients. These coefficients are determined by fitting the empirical \gls{pdf}s of $\mathfrak{R}\{h_{m,(1)}\}$ and $\mathfrak{R}\{h_{m,(2)}\}$ obtained from Monte Carlo simulations. The fitting procedure is implemented using MATLAB's \texttt{fit} function with Gaussian models\cite{Gaussian8:online}.

To validate the accuracy of our Gaussian approximation method, we conduct numerical experiments using samples from the distributions of $\mathfrak{R}\{h_{1,(1)}\}$, $\mathfrak{R}\{h_{1,(2)}\}$, $\mathfrak{R}\{h_{2,(1)}\}$, and $\mathfrak{R}\{h_{2,(2)}\}$. For illustration, we consider channel variances $\sigma_{1}=10$ and $\sigma_{2}=2.5$. We perform the fitting using MATLAB's \texttt{fit} function with \texttt{gauss1} (single Gaussian) and \texttt{gauss3} (three Gaussians) models, depending on the complexity of each distribution. Using $8.3\times10^6$ samples ensures statistical reliability of the fitted parameters.

\begin{figure*}[htb]
  \centering
  % Set caption formatting for all subfigures
  \captionsetup[subfloat]{labelfont=scriptsize,textfont=scriptsize}
  
  % First row: Real parts of h_1
  \subfloat[PDF of $\mathfrak{R}\{h_{1,(1)}\}$\label{subfig:fRh1_1}]{%
    \includegraphics[width=0.23\textwidth, trim=50 170 50 180, clip]{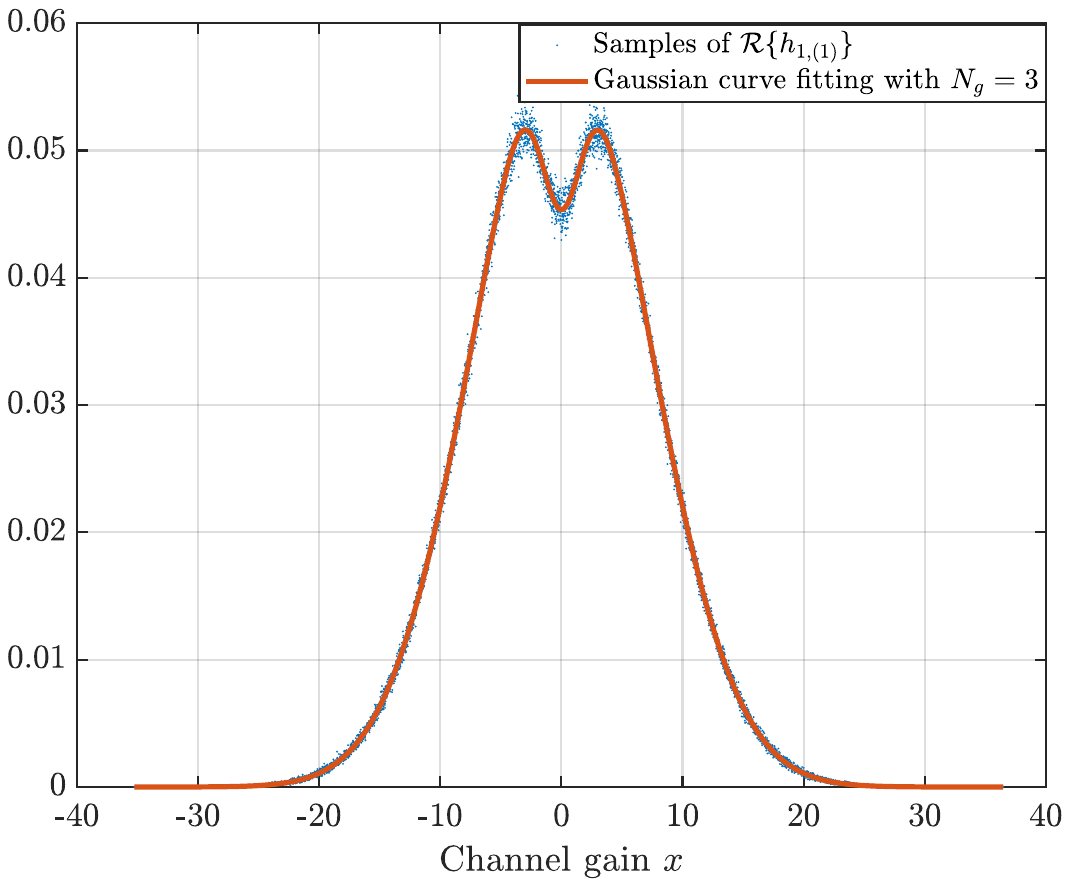}%
  }%
  \hspace{0.02\textwidth}% Add small spacing between subfigures
  \subfloat[PDF of $\mathfrak{R}\{h_{1,(2)}\}$\label{subfig:fRh1_2}]{%
    \includegraphics[width=0.23\textwidth, trim=50 170 50 180, clip]{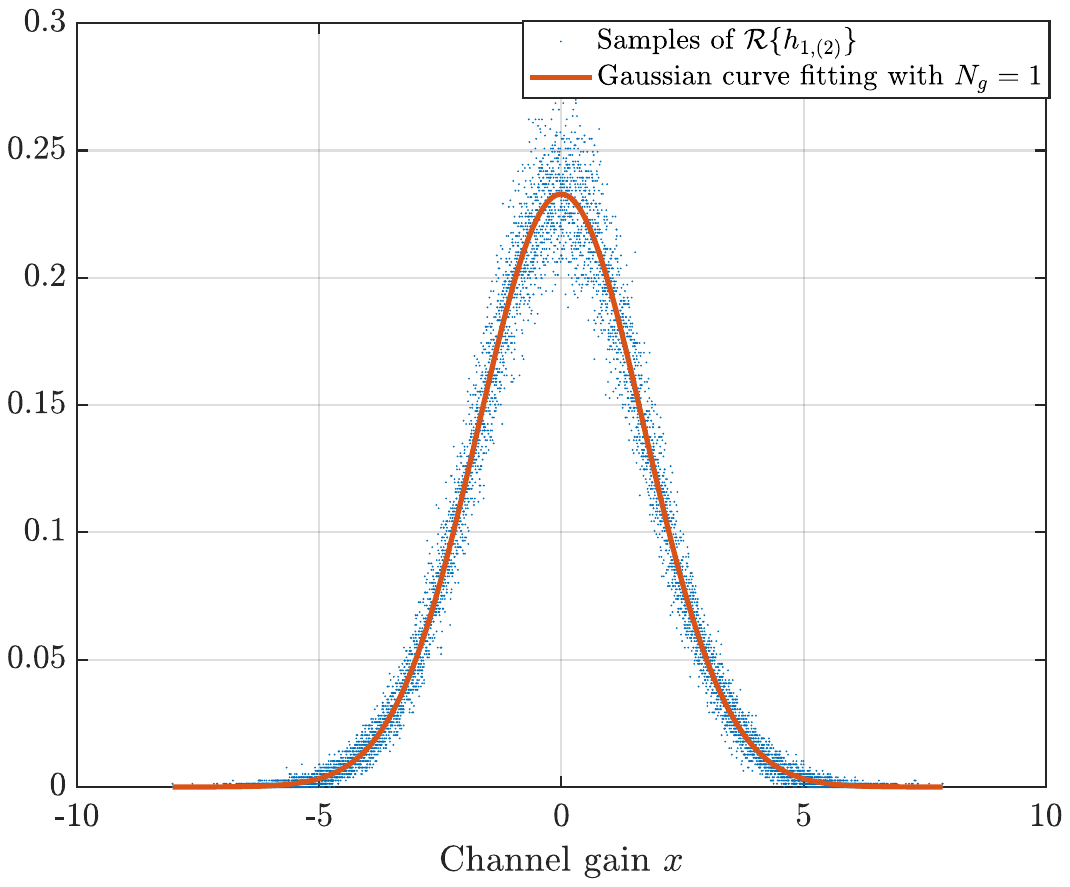}%
  }%
  \hspace{0.02\textwidth}%
  % Second row: Real parts of h_2
  \subfloat[PDF of $\mathfrak{R}\{h_{2,(1)}\}$\label{subfig:fRh2_1}]{%
    \includegraphics[width=0.23\textwidth, trim=50 170 50 180, clip]{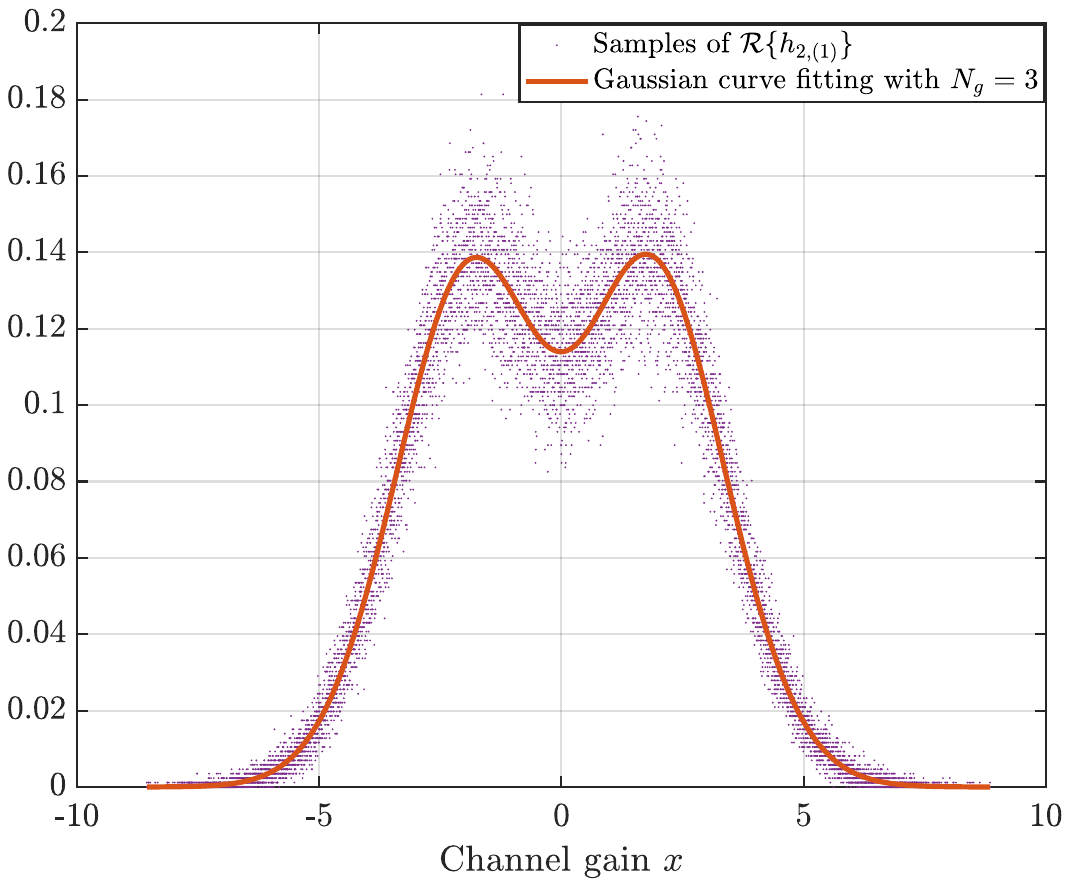}%
  }%
  \hspace{0.02\textwidth}%
  \subfloat[PDF of $\mathfrak{R}\{h_{2,(2)}\}$\label{subfig:fRh2_2}]{%
    \includegraphics[width=0.23\textwidth, trim=50 170 50 180, clip]{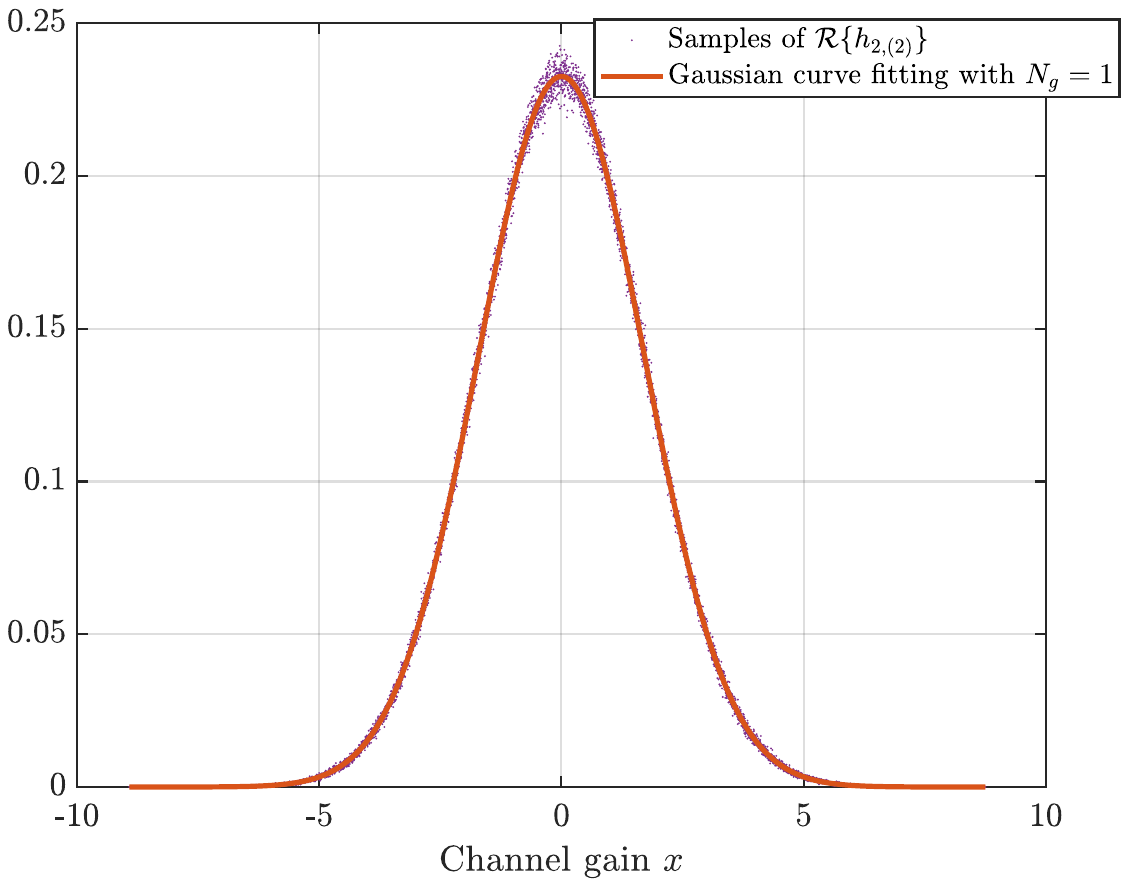}%
  }%
  
  \caption{Probability density functions of real channel components: 
    Monte Carlo samples (dots) versus Gaussian curve fitting approximations (solid lines)}
  \label{fig:pdf_samples_vs_curve_fitted}
\end{figure*}

\autoref{tab:ex_curve_fit_ceoff} presents the fitted coefficients for $f_{\mathfrak{R}\{h_{m,(1)}\}}$ and $f_{\mathfrak{R}\{h_{m,(2)}\}}$ with their 95\% confidence bounds, where $m\in\{1,2\}$. The visual comparison in \autoref{fig:pdf_samples_vs_curve_fitted} confirms that the fitted curves accurately capture the empirical \gls{pdf}s. Note that for different channel variance configurations, the fitting coefficients must be recalculated. The choice of $N_{\mathcal{G}}$ offers a design parameter to balance approximation accuracy against computational complexity.

\begin{table*}[htbp]
  \centering
  \caption{Fitted Gaussian coefficients for approximating $f_{\mathfrak{R}\{h_{m,(1)}\}}$ and $f_{\mathfrak{R}\{h_{m,(2)}\}}$ with 95\% confidence intervals. Parameters: $m\in\{1,2\}$, $\sigma_{1}=10$, $\sigma_{2}=2.5$}
  \label{tab:ex_curve_fit_ceoff}%
  \begin{tabular}{@{}llcccc@{}}
    \toprule
    \textbf{Target PDF} & \boldmath$N_{\mathcal{G}}$\unboldmath & \textbf{Coeff.} & \boldmath$a_i$\unboldmath & \boldmath$b_i$\unboldmath & \boldmath$c_i$\unboldmath \\
    \midrule
    \multirow{6}{*}{$f_{\mathfrak{R}\{h_{1,(1)}\}}$} & \multirow{6}{*}{3} 
      & $i=1$ & $0.02045$ & $-2.471$ & $9.590$ \\
      & & & {\scriptsize $[-0.161, 0.202]$} & {\scriptsize $[-15.31, 10.37]$} & {\scriptsize $[7.959, 11.22]$} \\
    \cmidrule{3-6}
      & & $i=2$ & $0.04136$ & $1.197$ & $9.747$ \\
      & & & {\scriptsize $[-0.134, 0.217]$} & {\scriptsize $[-8.692, 11.09]$} & {\scriptsize $[8.282, 11.21]$} \\
    \cmidrule{3-6}
      & & $i=3$ & $-0.01452$ & $0.003346$ & $2.423$ \\
      & & & {\scriptsize $[-0.0146, -0.0144]$} & {\scriptsize $[-0.0064, 0.0131]$} & {\scriptsize $[2.404, 2.442]$} \\
    \midrule
    \multirow{2}{*}{$f_{\mathfrak{R}\{h_{1,(2)}\}}$} & \multirow{2}{*}{1} 
      & $i=1$ & $0.2329$ & $-0.003983$ & $2.422$ \\
      & & & {\scriptsize $[0.2324, 0.2334]$} & {\scriptsize $[-0.0081, 0.0001]$} & {\scriptsize $[2.416, 2.428]$} \\
    \midrule
    \multirow{6}{*}{$f_{\mathfrak{R}\{h_{2,(1)}\}}$} & \multirow{6}{*}{3} 
      & $i=1$ & $-0.1442$ & $0.2475$ & $2.051$ \\
      & & & {\scriptsize $[-1.438, 1.149]$} & {\scriptsize $[-0.166, 0.661]$} & {\scriptsize $[0.649, 3.453]$} \\
    \cmidrule{3-6}
      & & $i=2$ & $0.2247$ & $1.152$ & $2.416$ \\
      & & & {\scriptsize $[-0.901, 1.351]$} & {\scriptsize $[-4.179, 6.483]$} & {\scriptsize $[1.148, 3.684]$} \\
    \cmidrule{3-6}
      & & $i=3$ & $0.1414$ & $-1.746$ & $2.241$ \\
      & & & {\scriptsize $[0.0464, 0.2365]$} & {\scriptsize $[-2.338, -1.153]$} & {\scriptsize $[2.038, 2.445]$} \\
    \midrule
    \multirow{2}{*}{$f_{\mathfrak{R}\{h_{2,(2)}\}}$} & \multirow{2}{*}{1} 
      & $i=1$ & $0.2326$ & $0.001333$ & $2.426$ \\
      & & & {\scriptsize $[0.2325, 0.2327]$} & {\scriptsize $[0.0003, 0.0023]$} & {\scriptsize $[2.424, 2.427]$} \\
    \bottomrule
  \end{tabular}%
  \begin{tablenotes}
    \small
    \item Numbers in brackets indicate 95\% confidence intervals. Each PDF is approximated as $f(x) \approx \sum_{i=1}^{N_{\mathcal{G}}} a_i \exp\left(-\frac{(x-b_i)^2}{c_i^2}\right)$.
  \end{tablenotes}
\end{table*}%

% MATLAB code is: C:\Users\heqzh\Documents\GitHub\noma-im-sim\matlab\main\noma_ul_sic\pdf_check\curve_fit_truncated_pdf.m
% Results is: \\wsl.localhost\Ubuntu-22.04\home\heqzha\workspace\github\heqzha\ul-noma-sic-pep-paper\data\gaussian_fit_pdf_Rh

Once the component PDFs are obtained, we proceed to derive the PDFs of $z_{n,(k)}$ for $k\in\{1,2\}$. 

\subsection{PDF of $z_{n,(1)}$}
Given $z_{n,(1)}=\frac{|\xi_{n,(1)}|+2\mathfrak{R}\{\zeta_{n,(1)}\}}{\sqrt{2N_0}}$, we define the auxiliary variables:
{\small
\begin{equation}
\begin{aligned}
  X_{n,(1)} &= |\xi_{n,(1)}|+2\mathfrak{R}\{\zeta_{n,(1)}\}, \\
  Y_{n,(1)} &= 2\mathfrak{R}\{\zeta_{n,(1)}\}.
\end{aligned}
\end{equation}
}

Using the transformation of random variables, the PDF of $z_{n,(1)}$ is
{\small
\begin{align}\label{eq:pdf_z_1}
  f_{\mathcal{Z}_{n,(1)}}(z) &= \sqrt{2N_0}f_{X_{n,(1)}}\left(\sqrt{2N_0}z\right)\nonumber\\
  &= \sqrt{2N_0}\int_{-\infty}^{\infty} f_{|\xi_{n,(1)}|}\left(\sqrt{2N_0}z-y\right)f_{Y_{n,(1)}}(y)\,dy,
\end{align}
}%
where $f_{Y_{n,(1)}}(y) = \frac{1}{2}f_{\mathfrak{R}\{\zeta_{n,(1)}\}}(y/2)$ accounts for the scaling factor.

To obtain the closed-form expression:
\begin{enumerate}
  \item Substitute \cref{eq:pdf_breve_h_1} into \cref{eq:pdf_f_xi_1} to get $f_{|\xi_{n,(1)}|}(x)$
  \item Use \cref{eq:curve_fit_gaussian_func} with coefficients from \cref{eq:pdf_f_Rzeta_2} to approximate $f_{\mathfrak{R}\{\zeta_{n,(1)}\}}(x)$
  \item Evaluate the convolution integral in \cref{eq:pdf_z_1} analytically\footnote{The closed-form expression was derived using Mathematica \cite{Mathematica}.}
\end{enumerate}

The resulting closed-form expression is presented in the main paper as \cref{eq:cf_pdf_z_1}%\cite[(21)]{zhang2025error_noma}
.

\subsection{PDF of $z_{n,(2)}$}
Similarly, for $z_{n,(2)}=\frac{|\xi_{n,(2)}|+2\mathfrak{R}\{\zeta_{n,(2)}\}}{\sqrt{2N_0}}$, we define:
{\small
\begin{equation}
\begin{aligned}
  X_{n,(2)} &= |\xi_{n,(2)}|+2\mathfrak{R}\{\zeta_{n,(2)}\}, \\
  Y_{n,(2)} &= 2\mathfrak{R}\{\zeta_{n,(2)}\}.
\end{aligned}
\end{equation}
}

We consider two cases based on whether interference is present:

\textbf{Case 1: With interference ($\zeta_{n,(2)}\neq0$).} The PDF is given by
{\small
\begin{align}\label{eq:pdf_z_2}
  f_{\mathcal{Z}_{n,(2)}}(z) &= \sqrt{2N_0}f_{X_{n,(2)}}\left(\sqrt{2N_0}z\right)\nonumber\\
  &= \sqrt{2N_0}\int_{-\infty}^{\infty} f_{|\xi_{n,(2)}|}\left(\sqrt{2N_0}z-y\right)f_{Y_{n,(2)}}(y)\,dy.
\end{align}
}

The closed-form expression is obtained by:
\begin{enumerate}
  \item Substituting \cref{eq:pdf_breve_h_2} into \cref{eq:pdf_f_xi_2} for $f_{|\xi_{n,(2)}|}(x)$
  \item Using \cref{eq:curve_fit_gaussian_func} with $N_{\mathcal{G}}=1$ to approximate $f_{\mathfrak{R}\{h_{m,(1)}\}}(x)$
  \item Substituting into \cref{eq:pdf_f_Rzeta_1} and evaluating the integral
\end{enumerate}

This yields the expression in the main paper as \cref{eq:cf_pdf_z_2_incor}%\cite[(22)]{zhang2025error_noma}
.

\textbf{Case 2: Without interference ($\zeta_{n,(2)}=0$).} Here $Y_{n,(2)}=0$ and $X_{n,(2)}=|\xi_{n,(2)}|$, so
{\small
\begin{align}
  f_{\mathcal{Z}_{n,(2)}}(z) = \sqrt{2N_0}f_{|\xi_{n,(2)}|}\left(\sqrt{2N_0}z\right).
\end{align}
}

The closed-form expression follows directly from $f_{|\xi_{n,(2)}|}(x)$ and is given in the main paper as \cref{eq:cf_pdf_z_2_cor}%\cite[(23)]{zhang2025error_noma}
.

%% file: app_cf_pep_2_ues.tex
\section{Closed-form of the PEP Expressions of Two-\gls{ue} Case}\label{app:cf_pep_2_ues}
This appendix presents the closed-form expressions for the \gls{pep} of \gls{ue}~$n$ in a two-\gls{ue} scenario, considering different decoding orders.

\subsection{Derivation Approach}

The closed-form expressions are obtained by evaluating the integral of the Q-function with the \gls{pdf} of $\mathcal{Z}_{n,(k)}$, as defined in \cref{eq:upep_integ} %\cite[(9)]{zhang2025error_noma}
. The relevant \gls{pdf}s are given in \cref{eq:cf_pdf_z_1}% \cite[(21)]{zhang2025error_noma}
and \cref{eq:cf_pdf_z_2_incor}%  \cite[(22)]{zhang2025error_noma}
. These expressions were derived using symbolic computation tools~\cite{Mathematica}.

\subsection{Closed-Form \gls{pep} Expressions}

\subsubsection{Case 1: \gls{ue}~$n$ with Decoding Order $(1)$}
When \gls{ue}~$n$ is decoded first, its \gls{pep} is given by:
\input{cf_pep_z1.tex}

\subsubsection{Case 2: \gls{ue}~$n$ with Decoding Order $(2)$}
When \gls{ue}~$n$ is decoded second, the \gls{pep} depends on whether the first-decoded \gls{ue}~$m$ was decoded correctly:

\paragraph{Sub-case 2a: \gls{ue}~$m$ Decoded Incorrectly}
If the first-decoded \gls{ue}~$m$ experiences a decoding error, the \gls{pep} for \gls{ue}~$n$ is:
\input{cf_pep_z2.tex}

\paragraph{Sub-case 2b: \gls{ue}~$m$ Decoded Correctly}
If the first-decoded \gls{ue}~$m$ is decoded correctly, the \gls{pep} for \gls{ue}~$n$ is:
{\small
\begin{align}
  P\left(A_{n} \mid B_{n,(2)}\right) &= \int_{0}^{\infty} Q(z) f_{\mathcal{Z}_{n,(2)}}(z) \, dz \nonumber\\
  &= \frac{1}{12 + 6\Delta_{n,(2)}^2\sigma_{n}^2} + \frac{3}{12 + 8\Delta_{n,(2)}^2\sigma_{n}^2}.
  \label{eq:cf_pep_z_2_cor}
\end{align}
}

%% file: cf_pep_z1.tex
% Find the equation from C:\Users\heqzh\Documents\GitHub\math-note\PEP_u1_rayleigh_bpsk_to_reformat.nb
{\small
\begin{align}\label{eq:cf_pep_z_1}
  P\left(A_{n}\mid B_{n,(1)}\right) &= \int_{0}^{\infty}Q\left(z\right)f_{\mathcal{Z}_{n,(1)}}\left(z\right)dz \\
  &= \frac{\sqrt{N_0\pi}\mathfrak{R}\left\{x_{m,(2)}\right\}a_1\exp\left(-\frac{b_1^2}{c_1^2}\right)}{3\sqrt{2}\Delta_{n,(1)}\left(\sigma_{n}^2-\sigma_{m}^2\right)} \notag \\
  &\quad \times \left[G_1\left(\sigma_{n}\right)+G_2\left(\sigma_{n}\right)-G_1\left(\sigma_{m}\right)-G_2\left(\sigma_{m}\right)\right] \notag,
\end{align}
}%

\noindent where $G_1\left(\sigma\right)$ is given by:
{\small
\begin{multline}\label{eq:G1}
  G_1\left(\sigma\right) = \gamma_1(\sigma) D_1\left(\sigma\right) \\
  \times \left(\frac{\sqrt{3}}{2\sqrt{2}}\sqrt{p_{(1)}}|\Delta_{n,(1)}|\sigma\alpha_1(\sigma) - \beta_1(\sigma)\right) \\
  - \delta_1(\sigma) F_1\left(\sigma\right)
\end{multline}
}%

\noindent and $G_2\left(\sigma\right)$ is:
{\small
\begin{multline}\label{eq:G2}
  G_2\left(\sigma\right) = \gamma_2(\sigma) D_2\left(\sigma\right) \\
  \times \left(p_{(1)}|\Delta_{n,(1)}|^2\sigma\alpha_1(\sigma) - \beta_2(\sigma)\right) \\
  - \delta_2(\sigma) F_1\left(\sigma\right)
\end{multline}
}%

\noindent with $F_1\left(\sigma\right)$ and $F_2\left(\sigma\right)$ defined as:
{\small
\begin{equation}
\begin{aligned}
  F_1\left(\sigma\right) &= \frac{e_2\left(\sigma\right)}{2\sqrt{I\left(\sigma\right)\left(3N_0+I\left(\sigma\right)\right)}} \\
  &\quad \times \left(1+E_1\left(\sigma\right)\left(\lambda_1\sigma^2+\lambda_2\right)\right), \\
  F_2\left(\sigma\right) &= \frac{e_3\left(\sigma\right)}{\sqrt{I\left(\sigma\right)\left(4N_0+I\left(\sigma\right)\right)}} \\
  &\quad \times \left(1+E_2\left(\sigma\right)\left(\lambda_3\sigma^2+\lambda_4\right)\right).
\end{aligned}
\end{equation}
}%

\noindent The functions $D_1\left(\sigma\right)$ and $D_2\left(\sigma\right)$ are given by:
{\small
\begin{equation}
\begin{aligned}
  D_1\left(\sigma\right) &= 6b_1e_2\left(\sigma\right)\sqrt{N_0\pi p_{(2)}}\mathfrak{R}\left\{x_{m,(2)}\right\}\left(1+E_1\left(\sigma\right)\right) \\
  &\quad +\sqrt{3I\left(\sigma\right)\left(3N_0+I\left(\sigma\right)\right)},\\
  D_2\left(\sigma\right) &= 4b_1e_3\left(\sigma\right)\sqrt{N_0\pi p_{(2)}}\mathfrak{R}\left\{x_{m,(2)}\right\}\left(1+E_2\left(\sigma\right)\right) \\
  &\quad +\sqrt{I\left(\sigma\right)\left(4N_0+I\left(\sigma\right)\right)}.
\end{aligned}
\end{equation}
}%

\noindent The intermediate terms used above are defined as:
{\small
\begin{equation}
\begin{aligned}
  \alpha_1(\sigma) &= \sqrt{\frac{4N_0}{p_{(1)}|\Delta_{n,(1)}|^2\sigma^2}+\frac{N_0}{c_1^2p_{(2)}\left(\mathfrak{R}\left\{x_{m,(2)}\right\}\right)^2}}\\
  \beta_1(\sigma) &= \frac{\sqrt{6N_0p_{(2)}}\mathfrak{R}\left\{x_{m,(2)}\right\}|c_1|}{\sqrt{I\left(\sigma\right)}}, \\
  \beta_2(\sigma) &= \frac{4\sqrt{N_0p_{(2)}}\mathfrak{R}\left\{x_{m,(2)}\right\}|c_1|}{\sqrt{I\left(\sigma\right)}}, \\
  \gamma_1(\sigma) &= \frac{3c_1^2e_1\left(\sigma\right)\sigma}{\sqrt{I\left(\sigma\right)\left(3N_0+I\left(\sigma\right)\right)^3}}, \\
  \gamma_2(\sigma) &= \frac{\sqrt{2}c_1^2e_1\left(\sigma\right)\sigma}{\sqrt{I\left(\sigma\right)\left(4N_0+I\left(\sigma\right)\right)^3}}, \\
  \delta_1(\sigma) &= \frac{6b_1p_{(1)}|\Delta_{n,(1)}|^2e_1\left(\sigma\right)\sigma^3|c_1|}{\sqrt{I\left(\sigma\right)^3}}, \\
  \delta_2(\sigma) &= \frac{2b_1p_{(1)}|\Delta_{n,(1)}|^2e_1\left(\sigma\right)\sigma^3|c_1|}{\sqrt{I\left(\sigma\right)}}
\end{aligned}
\end{equation}
}%

\noindent and:
{\small
\begin{equation}
\begin{aligned}
  \lambda_1 &= \sqrt{\frac{3 \pi}{2}}p_{(1)}|\Delta_{n,(1)}|^2, \\
  \lambda_2 &= 2\sqrt{6\pi}c_1^2p_{(2)}\left(\mathfrak{R}\left\{x_{m,(2)}\right\}\right)^2, \\
  \lambda_3 &= \sqrt{\frac{\pi}{2}}p_{(1)}|\Delta_{n,(1)}|^2, \\
  \lambda_4 &= 2\sqrt{2\pi}c_1^2p_{(2)}\left(\mathfrak{R}\left\{x_{m,(2)}\right\}\right)^2. 
\end{aligned}
\end{equation}
}%

\noindent Finally, the auxiliary functions are defined as:
{\small
\begin{equation}
\begin{aligned}
  e_1\left(\sigma\right) &= \exp\left(\frac{b_1^2p_{(1)}|\Delta_{n,(1)}|^2\sigma^2}{c_1^2I\left(\sigma\right)}\right), \\
  e_2\left(\sigma\right) &= \exp\left(\frac{12b_1^2N_0p_{(2)}\left(\mathfrak{R}\left\{x_{m,(2)}\right\}\right)^2}{I\left(\sigma\right)\left(3N_0+I\left(\sigma\right)\right)}\right), \\
  e_3\left(\sigma\right) &= \exp\left(\frac{16b_1^2N_0p_{(2)}\left(\mathfrak{R}\left\{x_{m,(2)}\right\}\right)^2}{I\left(\sigma\right)\left(4N_0+I\left(\sigma\right)\right)}\right), \\
  E_1\left(\sigma\right) &= \text{erf}\left(\frac{2\sqrt{3}b_1\sqrt{N_0p_{(2)}}\mathfrak{R}\left\{x_{m,(2)}\right\}}{\sqrt{I\left(\sigma\right)\left(3N_0+I\left(\sigma\right)\right)}}\right), \\
  E_2\left(\sigma\right) &= \text{erf}\left(\frac{4b_1\sqrt{N_0p_{(2)}}\mathfrak{R}\left\{x_{m,(2)}\right\}}{\sqrt{I\left(\sigma\right)\left(4N_0+I\left(\sigma\right)\right)}}\right), \\
  I\left(\sigma\right) &= p_{(1)}|\Delta_{n,(1)}|^2\sigma^2+4c_1^2p_{(2)}\left(\mathfrak{R}\left\{x_{m,(2)}\right\}\right)^2.
\end{aligned}
\end{equation}
}

% Alternative: For two-column documents, if you want equations to span both columns:
% \usepackage{cuted}
% \begin{strip}
%   % Your long equations here
% \end{strip}

%% file: cf_pep_z2.tex
% Find the equation from C:\Users\heqzh\Documents\GitHub\math-note\PEP_u1_rayleigh_bpsk_to_reformat.nb
{\small
\begin{align}\label{eq:cf_pep_z_2_inc}
  P\left(A_{n}\mid B_{n,(2)}\right) &= \int_{0}^{\infty}Q\left(z\right)f_{\mathcal{Z}_{n,(2)}}\left(z\right)dz \\
  &= \sum_{i=1}^{3}\left[\mathcal{S}_{i,1} + \mathcal{S}_{i,2} + \mathcal{S}_{i,3}\right] \notag,
\end{align}
}%

\noindent where $\mathcal{S}_{i,1}$ is given by:
{\small
\begin{align}\label{eq:S_i1}
    \mathcal{S}_{i,1} &= \frac{a_i D_2}{72\sqrt{2}\Omega_i^{3/2}} \left[T_{i,1}^{(1)} + T_{i,2}^{(1)} - T_{i,3}^{(1)}\right] \notag \\
    &\quad + \frac{a_i D_2}{16\sqrt{3}\Omega_i^{3/2}} \left[T_{i,1}^{(2)} + T_{i,2}^{(2)} - T_{i,3}^{(2)}\right],
\end{align}
}%

\noindent with the first set of terms:
{\small
\begin{equation}
\begin{aligned}
    T_{i,1}^{(1)} &= \frac{b_i e_{i,1}\pi |c_i| D_1 \left(1+E_{i,1}\right)}{\sqrt{1+\frac{2}{c_i^2D_1^2}}}, \\
    T_{i,2}^{(1)} &= \frac{6 c_i^2 e_{i,1} \sqrt{\frac{\pi}{2}} S_{i,1} D_1^2 |c_i|\left(1+E_{i,1}\right)}{1}, \\
    T_{i,3}^{(1)} &= \frac{e^{-\frac{b_i^2}{c_i^2}} \sqrt{\pi} |c_i| D_1 \Phi_{i,1}}{\sqrt{2+\frac{4}{c_i^2D_1^2}} \left(2+c_i^2D_1^2\right)}
\end{aligned}
\end{equation}
}%

\noindent and the second set:
{\small
\begin{equation}
\begin{aligned}
    T_{i,1}^{(2)} &= \frac{b_i e_{i,2} \pi |c_i| D_1 \left(1+E_{i,2}\right)}{\sqrt{2+\frac{3}{c_i^2D_1^2}}}, \\
    T_{i,2}^{(2)} &= \frac{2 c_i^2 e_{i,2} \sqrt{3\pi} S_{i,2} D_1^2 |c_i|\left(1+E_{i,2}\right)}{1}, \\
    T_{i,3}^{(2)} &= \frac{c_i^2 e^{-\frac{b_i^2}{c_i^2}} \sqrt{\pi} D_1^2 \Phi_{i,2}}{\Omega_i\left(3+2 c_i^2D_1^2\right)},
\end{aligned}
\end{equation}
}%

\noindent where:
{\small
\begin{equation}
\begin{aligned}
    \Phi_{i,1} &= 2 b_i e_{i,3} \sqrt{\pi}+ \sqrt{2+\frac{4}{c_i^2D_1^2}} c_i^2 D_1 \\
    &\quad +2 b_i e_{i,3} \sqrt{\pi} E_{i,1}, \\
    \Phi_{i,2} &= b_i e_{i,4} \sqrt{3 \pi}+ \sqrt{2+\frac{3}{D_1^2}} c_i^2 D_1 \\
    &\quad +b_i e_{i,4} \sqrt{3 \pi} E_{i,2}.
\end{aligned}
\end{equation}
}%

\noindent $\mathcal{S}_{i,2}$ is defined as:
{\small
\begin{align}\label{eq:S_i2}
    \mathcal{S}_{i,2} &= \frac{a_i D_2 e_{i,5}}{12} \left[\Theta_{i,1} + \Theta_{i,2}\right] \notag \\
    &\quad + \frac{a_i D_2 e_{i,5}}{8} \left[\Theta_{i,3} + \Theta_{i,4}\right],
\end{align}
}%

\noindent with:
{\small
\begin{equation}
\begin{aligned}
    \Theta_{i,1} &= -\frac{b_i e_{i,6}\sqrt{\pi}|c_i|D_1S_{i,3}\left(1+E_{i,3}\right)}{\Omega_i^{3/2}}, \\
    \Theta_{i,2} &= \frac{b_i\sqrt{\pi}|c_i|\Psi_{i,1}}{\Omega_i\Xi_{i,1}^{3/2}}, \\
    \Theta_{i,3} &= -\frac{\sqrt{3}b_i e_{i,7}\pi |c_i|D_1S_{i,4}\left(1+E_{i,4}\right)}{\Omega_i^{3/2}}, \\
    \Theta_{i,4} &= \frac{3\sqrt{\pi}|c_i|\Psi_{i,2}}{\Omega_i\Xi_{i,2}^{3/2}},
\end{aligned}
\end{equation}
}%

\noindent where:
{\small
\begin{equation}
\begin{aligned}
    \Psi_{i,1} &= b_ie_{i,6}\sqrt{2\pi}D_1+\sqrt{\Omega_i\Xi_{i,1}}, \\
    &\quad +b_ie_{i,6}\sqrt{2\pi}D_1E_{i,3}, \\
    \Psi_{i,2} &= b_ie_{i,7}\sqrt{3\pi}D_1+\sqrt{\Omega_i\Xi_{i,2}}, \\
    &\quad +b_ie_{i,7}\sqrt{3\pi}D_1E_{i,4}, \\
    \Xi_{i,1} &= 2+D_2^2+c_i^2D_1^2, \\
    \Xi_{i,2} &= 3+2D_2^2+2c_i^2D_1^2.
\end{aligned}
\end{equation}
}%

\noindent Finally, $\mathcal{S}_{i,3}$ is:
{\small
\begin{align}\label{eq:S_i3}
    \mathcal{S}_{i,3} &= \frac{a_i D_2 e_{i,8}}{8\Omega_i^{3/2}} \left[\Lambda_{i,1} + \Lambda_{i,2}\right] \notag \\
    &\quad + \frac{a_i D_2 e_{i,8}}{24\Omega_i^{3/2}} \left[\Lambda_{i,3} + \Lambda_{i,4}\right],
\end{align}
}%

\noindent with:
{\small
\begin{equation}
\begin{aligned}
    \Lambda_{i,1} &= -\frac{b_i^2 e_{i,9}\sqrt{\frac{\pi^2}{3}}|c_i|D_1\Upsilon_i\left(1+E_{i,5}\right)}{\sqrt{S_{i,5}}S_{i,7}}, \\
    \Lambda_{i,2} &= -\frac{c_i^2D_1^2\sqrt{3\pi}\Sigma_{i,1}}{2S_{i,6}\sqrt{\Omega_i\Sigma_{i,2}}}, \\
    \Lambda_{i,3} &= -\frac{b_i^2 e_{i,10}\sqrt{\frac{2\pi^2}{3}}|c_i|D_1\Upsilon_i\left(1+E_{i,6}\right)}{\sqrt{S_{i,8}}S_{i,7}}, \\
    \Lambda_{i,4} &= -\frac{c_i^2D_1^2\sqrt{\frac{\pi}{3}}\Sigma_{i,3}}{S_{i,8}\sqrt{2S_{i,9}}},
\end{aligned}
\end{equation}
}%

\noindent where:
{\small
\begin{equation}
\begin{aligned}
    \Upsilon_i &= 4D_2^2+3c_i^2D_1^2, \\
    \Sigma_{i,1} &= \sqrt{3\Sigma_{i,2}}|c_i|+b_ie_{i,9}\sqrt{\pi}\Upsilon_i\left(1+E_{i,5}\right), \\
    \Sigma_{i,2} &= S_{i,6}+D_2^2c_i^2D_1^2\left(7+4c_i^2D_1^2\right), \\
    \Sigma_{i,3} &= \sqrt{6S_{i,8}}|c_i|+2b_1e_{i,10}\sqrt{\pi}\Upsilon_i\left(1+E_{i,6}\right).
\end{aligned}
\end{equation}
}%

\noindent The common term $\Omega_i$ is defined as:
{\small
\begin{align}
    \Omega_i &= D_2^2+c_i^2D_1^2 \label{eq:Omega_i}.
\end{align}
}%

\noindent The parameters $D_1$ and $D_2$ are defined as:
{\small
\begin{equation}
\begin{aligned}
  D_1 &= \sqrt{p_{(1)}}|\Delta_{m,(1)}|, \\
  D_2 &= \sqrt{p_{(2)}}|\Delta_{n,(2)}|\sigma_{n}.
\end{aligned}
\end{equation}
}%

\noindent The error functions $E_{i,j}$ are:
{\small
\begin{equation}
\begin{aligned}
  E_{i,1} &= \text{erf}\left[\frac{b_i}{\sqrt{\frac{1}{2}+\frac{1}{c_i^2D_1^2}}c_i^2D_1}\right], \\
  E_{i,2} &= \text{erf}\left[\frac{b_i}{\sqrt{\frac{2}{3}+\frac{1}{c_i^2D_1^2}}c_i^2D_1}\right], \\
  E_{i,3} &= \text{erf}\left[\frac{\sqrt{2}b_iD_1}{\sqrt{\Omega_i\Xi_{i,1}}}\right], \\
  E_{i,4} &= \text{erf}\left[\frac{\sqrt{3}b_iD_1}{\sqrt{\Omega_i\Xi_{i,2}}}\right], \\
  E_{i,5} &= \text{erf}\left[\frac{\Upsilon_i b_i}{\sqrt{3S_{i,5}}|c_i|}\right], \\
  E_{i,6} &= \text{erf}\left[\frac{\Upsilon_i b_i}{\sqrt{\frac{3S_{i,8}}{2}}|c_i|}\right].
\end{aligned}
\end{equation}
}%

\noindent The exponential terms $e_{i,j}$ are:
{\small
\begin{equation}
\begin{aligned}
  e_{i,1} &= \exp\left(-\frac{b_i^2}{c_i^2}+\frac{2 b_i^2}{c_i^2\left(2 + c_i^2 D_1^2\right)}\right), \\
  e_{i,2} &= \exp\left(-\frac{b_i^2}{c_i^2}+\frac{3 b_i^2}{c_i^2\left(3 +2 c_i^2 D_1^2\right)}\right), \\
  e_{i,3} &= \exp\left(\frac{2 b_i^2}{c_i^2\left(2 +c_i^2 D_1^2\right)}\right), \\
  e_{i,4} &= \exp\left(\frac{3 b_i^2}{c_i^2\left(3 +2 c_i^2 D_1^2\right)}\right), \\
  e_{i,5} &= \exp\left(-\frac{b_i^2}{c_i^2}+\frac{b_1^2D_2^2D_1^2}{\Omega_i c_i^2D_1^2}\right), \\
  e_{i,6} &= \exp\left(\frac{2b_i^2D_1^2}{\Omega_i\Xi_{i,1}}\right), \\
  e_{i,7} &= \exp\left(\frac{3b_i^2D_1^2}{\Omega_i\Xi_{i,2}}\right), \\
  e_{i,8} &= \exp\left(-\frac{b_i^2}{c_i^2}-\frac{b_1^2D_2^2D_1^2}{3\Omega_i c_i^2D_1^2}\right), \\
  e_{i,9} &= \exp\left(\frac{b_i^2\Upsilon_i^2}{3c_i^2S_{i,5}}\right), \\
  e_{i,10} &= \exp\left(\frac{2b_i^2\Upsilon_i^2}{3c_i^2S_{i,8}}\right).
\end{aligned}
\end{equation}
}%

\noindent Finally, the auxiliary terms $S_{i,j}$ are defined as:
{\small
\begin{equation}
\begin{aligned}
  S_{i,1} &= \sqrt{\frac{\frac{1}{D_2^2}+\frac{1}{c_i^2D_1^2}}{1+\frac{2}{c_i^2D_1^2}}}, \\
  S_{i,2} &= \sqrt{\frac{\frac{1}{D_2^2}+\frac{1}{c_i^2D_1^2}}{2+\frac{3}{c_i^2D_1^2}}}, \\
  S_{i,3} &= \sqrt{\frac{\Omega_i\pi}{2\Xi_{i,1}}}, \\
  S_{i,4} &= \sqrt{\frac{\Omega_i\pi}{\Xi_{i,2}}},   \\
  S_{i,5} &= \Omega_i S_{i,6}, \\
  S_{i,6} &= 2D_2^2\left(2+c_i^2D_1^2\right)+c_i^2D_1^2\left(3+2c_i^2D_1^2\right), \\
  S_{i,7} &= \left(-\frac{2b_i}{c_i^2D_1}-\frac{2b_iD_2^2D_1}{3\Omega_i c_i^2D_1^2}\right)|c_i|, \\
  S_{i,8} &= \Omega_i S_{i,9}, \\
  S_{i,9} &= 3c_i^2D_1^2\left(2+c_i^2D_1^2\right)+D_2^2\left(8+3c_i^2D_1^2\right).
\end{aligned}
\end{equation}
}

% One important note: If this is for a two-column document (like an IEEE journal paper), and you want the equations to span both columns, you might need to use the strip environment from the cuted package as an alternative to figure*. For example:
% \usepackage{cuted}
% \begin{strip}
%   % Your equations here
% \end{strip}
% This allows the content to span both columns without creating a floating object.

%% file: app_cf_P_Bnk.tex
\section{Derivation of the Closed-Form Expression for $P\left(B_{n,(k)}\right)$ in the Two-\gls{ue} Case}\label{app:cf_P_Bnk}
According to the definition of $P\left(B_{n,(k)}\right)$ in \autoref{sec:P_Bnk} %\cite[III-B]{zhang2025error_noma}
and \cref{eq:P_Bnk}%\cite[(10)]{zhang2025error_noma}
, we derive the closed-form expressions for $P\left(B_{n,(k)}\right)$ where $n,k \in \{1,2\}$ in a two-\gls{ue} system. Recall that $P\left(B_{n,(k)}\right)$ represents the probability that \gls{ue} $n$ has the $k$-th strongest channel gain among all \gls{ue}s.

\subsection{Derivation for \gls{ue} 1}

\subsubsection{Case 1: $P\left(B_{1,(1)}\right)$ - \gls{ue} 1 has the strongest channel}
The probability $P\left(B_{1,(1)}\right)$ represents the event where \gls{ue} 1 has the strongest channel gain, i.e., $|h_1| \geq |h_2|$:
{\small
\begin{align}
  P\left(B_{1,(1)}\right) &= P\left(|h_1| \geq |h_2|\right) \nonumber\\
  &= \int_{0}^{\infty} P\left(|h_2| \leq x \mid |h_1| = x\right) f_{|h_1|}(x) dx \nonumber\\
  &= \int_{0}^{\infty} F_{|h_2|}(x) f_{|h_1|}(x) dx.
\end{align}
}

Since $|h_i| \sim \text{Rayleigh}(\sigma_i)$, we have:
{\small
\begin{align}
  f_{|h_i|}(x) &= \frac{x}{\sigma_i^2} \exp\left(-\frac{x^2}{2\sigma_i^2}\right), \\
  F_{|h_i|}(x) &= 1 - \exp\left(-\frac{x^2}{2\sigma_i^2}\right).
\end{align}
}

Substituting these expressions:
{\small
\begin{align}
  P\left(B_{1,(1)}\right) &= \int_{0}^{\infty} \left[1 - \exp\left(-\frac{x^2}{2\sigma_2^2}\right)\right] \frac{x}{\sigma_1^2} \exp\left(-\frac{x^2}{2\sigma_1^2}\right) dx \nonumber\\
  &= 1 - \int_{0}^{\infty} \frac{x}{\sigma_1^2} \exp\left(-\frac{x^2}{2}\left[\frac{1}{\sigma_1^2} + \frac{1}{\sigma_2^2}\right]\right) dx.
\end{align}
}

Let $u = \frac{x^2}{2}$, then $du = x dx$:
{\small
\begin{align}
  P\left(B_{1,(1)}\right) &= 1 - \frac{1}{\sigma_1^2} \int_{0}^{\infty} \exp\left(-u\left[\frac{1}{\sigma_1^2} + \frac{1}{\sigma_2^2}\right]\right) du \nonumber\\
  &= 1 - \frac{1}{\sigma_1^2} \cdot \frac{1}{\frac{1}{\sigma_1^2} + \frac{1}{\sigma_2^2}} \nonumber\\
  &= 1 - \frac{\sigma_2^2}{\sigma_1^2 + \sigma_2^2} \nonumber\\
  &= \frac{\sigma_1^2}{\sigma_1^2 + \sigma_2^2}.
\end{align}
}

\subsubsection{Case 2: $P\left(B_{1,(2)}\right)$ - \gls{ue} 1 has the weakest channel}
The probability $P\left(B_{1,(2)}\right)$ represents the event where \gls{ue} 1 has the weakest channel gain, i.e., $|h_1| \leq |h_2|$:
{\small
\begin{align}
  P\left(B_{1,(2)}\right) &= P\left(|h_1| \leq |h_2|\right) \nonumber\\
  &= \int_{0}^{\infty} P\left(|h_2| \geq x \mid |h_1| = x\right) f_{|h_1|}(x) dx \nonumber\\
  &= \int_{0}^{\infty} \left[1 - F_{|h_2|}(x)\right] f_{|h_1|}(x) dx \nonumber\\
  &= \int_{0}^{\infty} \exp\left(-\frac{x^2}{2\sigma_2^2}\right) \frac{x}{\sigma_1^2} \exp\left(-\frac{x^2}{2\sigma_1^2}\right) dx.
\end{align}
}

Following similar steps as above:
{\small
\begin{align}
  P\left(B_{1,(2)}\right) &= \frac{1}{\sigma_1^2} \int_{0}^{\infty} x \exp\left(-\frac{x^2}{2}\left[\frac{1}{\sigma_1^2} + \frac{1}{\sigma_2^2}\right]\right) dx \nonumber\\
  &= \frac{\sigma_2^2}{\sigma_1^2 + \sigma_2^2}.
\end{align}
}

\subsection{Derivation for \gls{ue} 2}

By symmetry and following the same approach:

\subsubsection{Case 1: $P\left(B_{2,(1)}\right)$ - \gls{ue} 2 has the strongest channel}
{\small
\begin{align}
  P\left(B_{2,(1)}\right) = P\left(|h_2| \geq |h_1|\right) = \frac{\sigma_2^2}{\sigma_1^2 + \sigma_2^2}.
\end{align}
}

\subsubsection{Case 2: $P\left(B_{2,(2)}\right)$ - \gls{ue} 2 has the weakest channel}
{\small
\begin{align}
  P\left(B_{2,(2)}\right) = P\left(|h_2| \leq |h_1|\right) = \frac{\sigma_1^2}{\sigma_1^2 + \sigma_2^2}.
\end{align}
}

\subsection{Summary}
For the two-\gls{ue} case, we have:
{\small
\begin{equation}
\begin{aligned}
  P\left(B_{1,(1)}\right) &= P\left(B_{2,(2)}\right) = \frac{\sigma_1^2}{\sigma_1^2 + \sigma_2^2}, \\
  P\left(B_{1,(2)}\right) &= P\left(B_{2,(1)}\right) = \frac{\sigma_2^2}{\sigma_1^2 + \sigma_2^2}.
\end{aligned}
\end{equation}
}

Note that $P\left(B_{1,(1)}\right) + P\left(B_{1,(2)}\right) = 1$ and $P\left(B_{2,(1)}\right) + P\left(B_{2,(2)}\right) = 1$, which confirms the validity of our derivations.

%% file: app_ber_of_m_qam.tex
\section{BER Analysis for $\mathcal{M}$-QAM Modulation Schemes}\label{app:ber_of_m_qam}

This appendix provides detailed derivations of the \gls{ber} expressions for Gray-coded 4\gls{qam} and 64\gls{qam} constellations, examining the error distances and decision boundaries for each bit position. The 16\gls{qam} analysis is presented in~\autoref{sec:gray_16qam}.

\subsection{4QAM Analysis}
The 4\gls{qam} (or \gls{qpsk}) constellation represents the simplest form of quadrature amplitude modulation, where each symbol encodes two bits $b_1b_2$. The constellation points are positioned at $\{\pm d, \pm jd\}$ in the complex plane, with Gray coding ensuring adjacent symbols differ by only one bit.

In the 4\gls{qam} constellation:
\begin{itemize}
	\item Bit $b_1$ determines the sign of the in-phase (I) component
	\item Bit $b_2$ determines the sign of the quadrature (Q) component
\end{itemize}

Due to the perfect symmetry of the 4\gls{qam} constellation, the error analysis for both bits is identical. The decision boundaries are the I-axis ($\mathfrak{R}\{x\}=0$) for bit $b_1$ and the Q-axis ($\mathfrak{I}\{x\}=0$) for bit $b_2$.

\begin{figure}[!t]
	\centering
	\includegraphics[width=0.5\columnwidth]{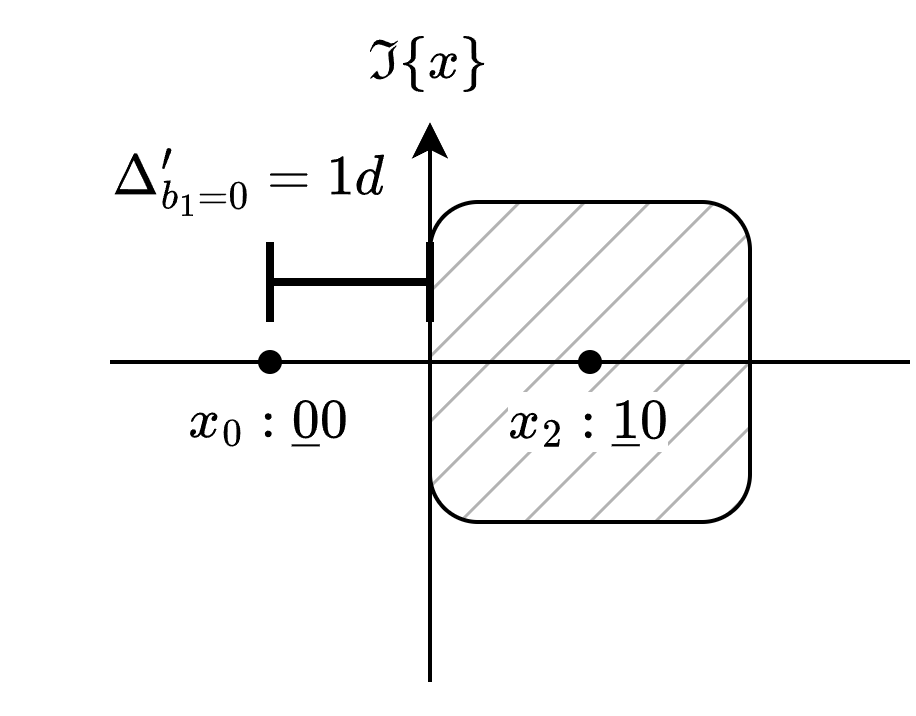}
	\caption{Error distances of 4\gls{qam} symbols with $b_1=0$ to their decision boundary. The minimum distance $2d$ determines the bit error probability.}
	\label{fig:4QAM_b1-0}
\end{figure}

For bit $b_1=0$, the transmitted symbols belong to the set $\{00, 01\}$ with $\mathfrak{R}\{x\}=d$. An error occurs when the received signal crosses the decision boundary at $\mathfrak{R}\{x\}=0$. The minimum error distance is $\Delta_{b_1}=2d$, as illustrated in \autoref{fig:4QAM_b1-0}.

The conditional error probabilities for both bits are:
{\small
	\begin{align}
		P_{(k),b_1}=P_{(k),b_2}=Q\left(\frac{2d\rho_{(k)}+\mathcal{I}_{(k)}}{\sqrt{2N_0}}\right).
	\end{align}
}

Since all symbols have the same error distance to their respective decision boundaries, the average \gls{ber} for $\mathcal{M}_{(k)}=4$ simplifies to:
{\small
	\begin{align}
		P_{(k)}&=\frac{1}{|\mathbb{S}|}\sum_{\mathbf{S}_{i}\in\mathbb{S}}P_{(k),b_1}\nonumber\\
		&=\frac{1}{|\mathbb{S}|}\sum_{\mathbf{S}_{i}\in\mathbb{S}}Q\left(\frac{2d\rho_{(k)}+\mathcal{I}_{(k)}}{\sqrt{2N_0}}\right).
	\end{align}
}

\subsection{64QAM Analysis}
The 64\gls{qam} constellation employs 6-bit symbols $b_1b_2b_3b_4b_5b_6$, where the bits are mapped according to Gray coding to minimize bit errors. The constellation structure divides the bits into two groups:
\begin{itemize}
	\item Bits $b_1b_2b_3$ determine the in-phase (I) component position
	\item Bits $b_4b_5b_6$ determine the quadrature (Q) component position
\end{itemize}

Due to the quadrature symmetry of the constellation, the error analysis for the I-component bits applies equally to the Q-component bits. Therefore, we focus our analysis on bits $b_1b_2b_3$, with the understanding that $P_{(k),b_i} = P_{(k),b_{i+3}}$ for $i \in \{1,2,3\}$.

\subsubsection{Analysis of bit $b_1$}
Bit $b_1$ determines whether the constellation point lies in the left or right half of the constellation. The decision boundary is at $\mathfrak{R}\{x\} = 0$.

For symbols with $b_1 = 0$, four distinct error distances exist, as illustrated in \autoref{subfig:64QAM_b1-0}:
\begin{itemize}
	\item $\Delta_{b_1}' = d$: for symbols closest to the decision boundary
	\item $\Delta_{b_1}'' = 3d$: for symbols in the second column from the boundary
	\item $\Delta_{b_1}''' = 5d$: for symbols in the third column
	\item $\Delta_{b_1}'''' = 7d$: for symbols farthest from the boundary
\end{itemize}

The conditional error probability for bit $b_1$ is obtained by averaging over all possible transmitted symbols:
{\small
	\begin{align}
		P_{(k),b_1}&=P_{(k),b_4}=\nonumber\\
		&\frac{1}{2}\left[Q\left(\frac{2d\rho_{(k)}+\mathcal{I}_{(k)}}{\sqrt{2N_0}}\right)+Q\left(\frac{6d\rho_{(k)}+\mathcal{I}_{(k)}}{\sqrt{2N_0}}\right)\right. \nonumber\\
		&\left.+Q\left(\frac{10d\rho_{(k)}+\mathcal{I}_{(k)}}{\sqrt{2N_0}}\right)+Q\left(\frac{14d\rho_{(k)}+\mathcal{I}_{(k)}}{\sqrt{2N_0}}\right)\right].
	\end{align}
}

\subsubsection{Analysis of bit $b_2$}
\begin{figure}[!ht]
	\captionsetup[subfloat]{labelfont=scriptsize,textfont=scriptsize}
	\subfloat[Error distances of symbols where $b_1=0$ to their decision boundary \label{subfig:64QAM_b1-0}]{%
		\includegraphics[width=0.45\textwidth]{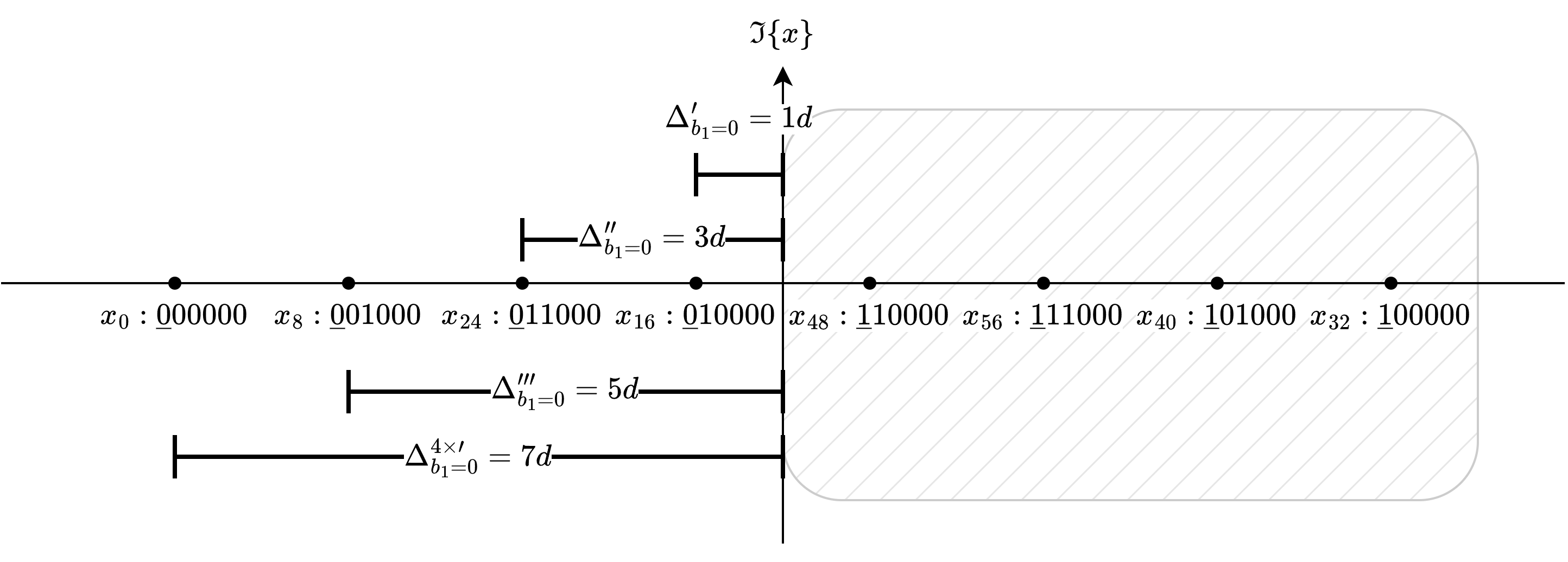}
	}
	\hfill
	\subfloat[Error distances of symbols where $b_{2}=1$ to the decision boundaries\label{subfig:64QAM_b2-1}]{%
		\includegraphics[width=0.45\textwidth]{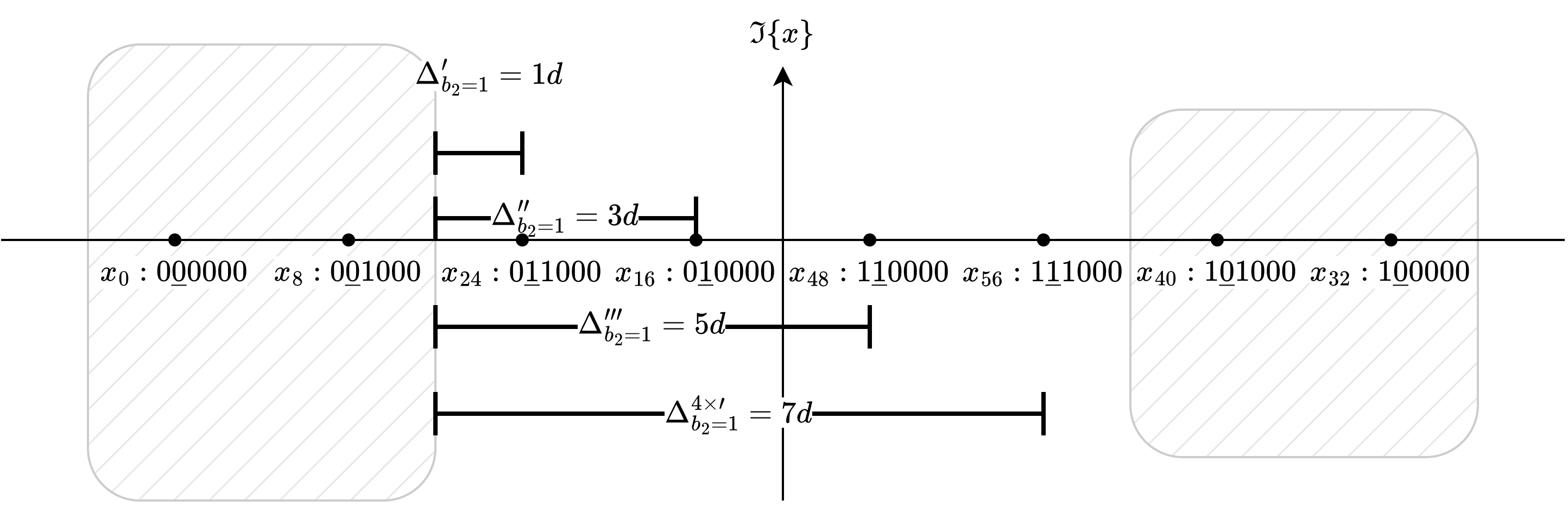}
	}
	\hfill
	\subfloat[Error distances of symbols where $b_{2}=0$ to the decision boundaries\label{subfig:64QAM_b2-0}]{%
		\includegraphics[width=0.45\textwidth]{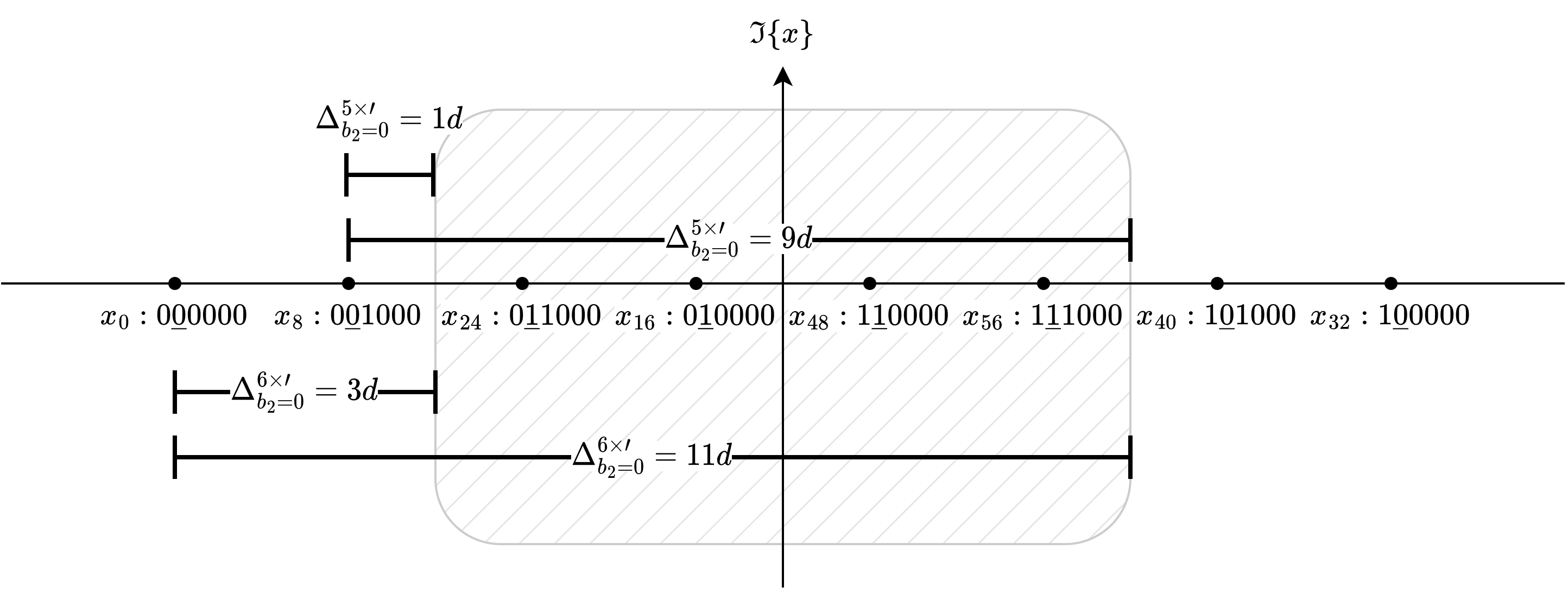}
	}
	\hfill
	\subfloat[Error distances of symbols where $b_{3}=1$ to the decision boundaries\label{subfig:64QAM_b3-1}]{%
		\includegraphics[width=0.45\textwidth]{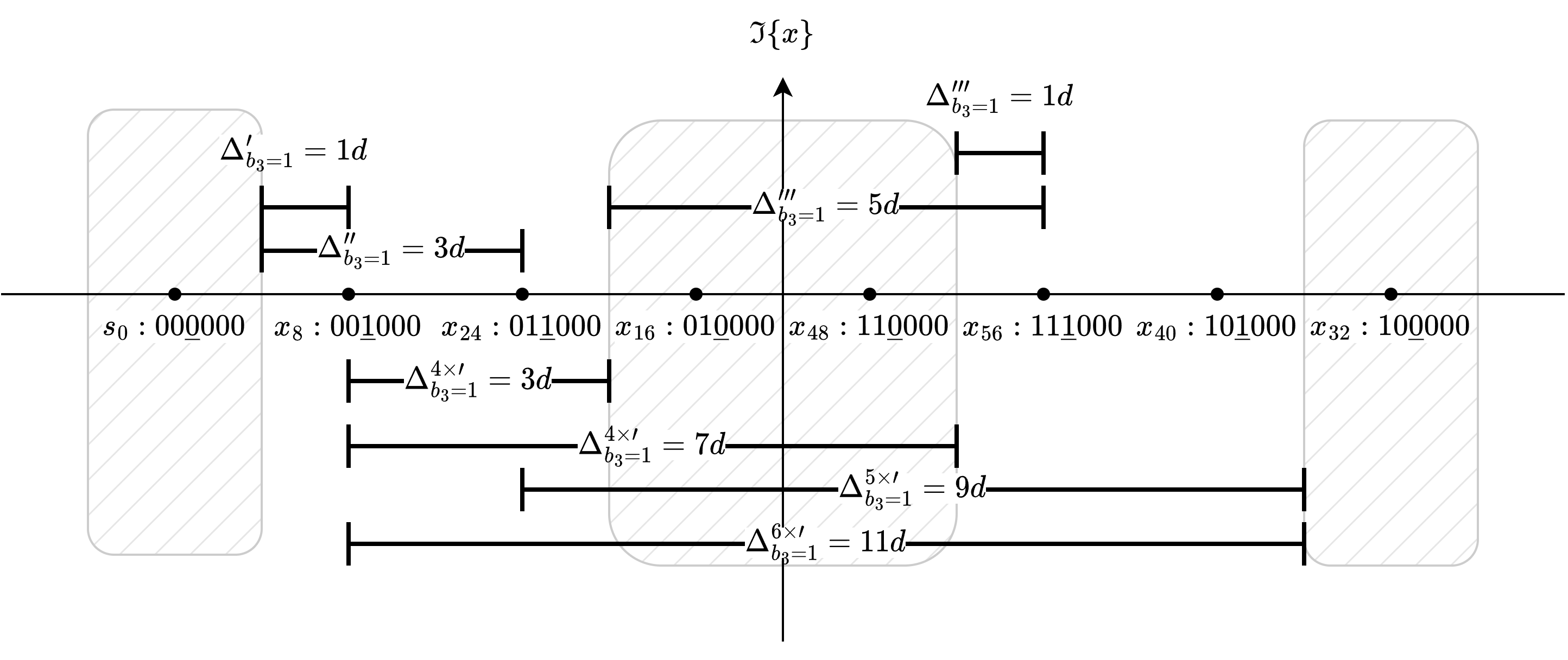}
	}
	\hfill
	\subfloat[Error distances of symbols where $b_{3}=0$ to the decision boundaries\label{subfig:64QAM_b3-0}]{%
		\includegraphics[width=0.45\textwidth]{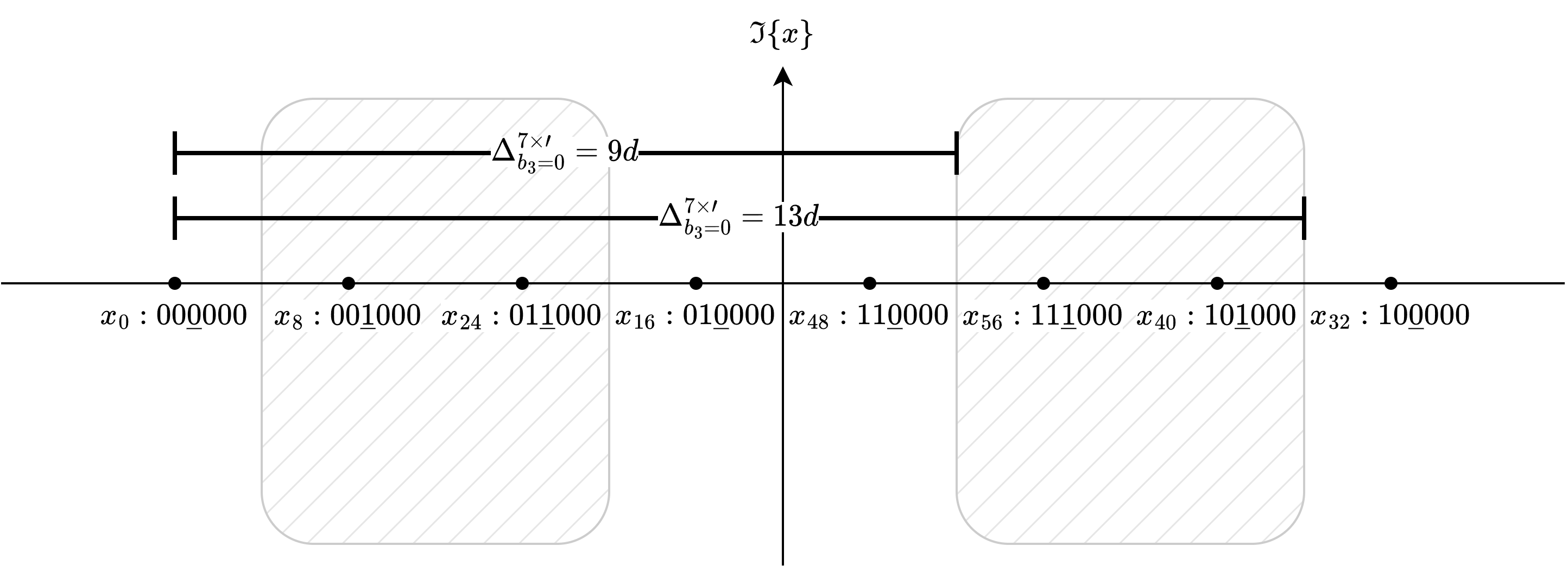}
	}
	\caption{Error distances between transmitted 64\gls{qam} symbols and their decision boundaries for different bit positions in Gray-coded constellation mapping}
	\label{fig:64QAM_err_dist}
\end{figure}

Bit $b_2$ represents the middle bit for the I-component and exhibits more complex error behavior due to multiple decision boundaries. The analysis considers two error scenarios:

\textbf{Case 1: $b_2 = 1$ decoded as 0}\\
As shown in \autoref{subfig:64QAM_b2-1}, symbols with $b_2 = 1$ can be erroneously decoded as $b_2 = 0$ when the received signal crosses the decision boundaries at $\mathfrak{R}\{x\} = \pm 4d$. This occurs when $\Delta_{b_2} \geq 4d$ or $\Delta_{b_2} \leq -4d$.

\textbf{Case 2: $b_2 = 0$ decoded as 1}\\
As illustrated in \autoref{subfig:64QAM_b2-0}, symbols with $b_2 = 0$ are erroneously decoded as $b_2 = 1$ when the received signal falls within the region $-4d < \Delta_{b_2} < 4d$, where $b_2 = 1$ symbols are located.

The conditional error probability for bit $b_2$ accounts for all possible error transitions:
{\small
	\begin{align}
		P_{(k),b_2}&=P_{(k),b_5}=\nonumber\\
		&\frac{1}{2}\bigg[2Q\left(\frac{2d\rho_{(k)}+\mathcal{I}_{(k)}}{\sqrt{2N_0}}\right)+2Q\left(\frac{6d\rho_{(k)}+\mathcal{I}_{(k)}}{\sqrt{2N_0}}\right)\nonumber\\
		&+Q\left(\frac{10d\rho_{(k)}+\mathcal{I}_{(k)}}{\sqrt{2N_0}}\right)+Q\left(\frac{14d\rho_{(k)}+\mathcal{I}_{(k)}}{\sqrt{2N_0}}\right)\nonumber\\
		&-Q\left(\frac{18d\rho_{(k)}+\mathcal{I}_{(k)}}{\sqrt{2N_0}}\right)-Q\left(\frac{22d\rho_{(k)}+\mathcal{I}_{(k)}}{\sqrt{2N_0}}\right)\bigg].
	\end{align}
}

\subsubsection{Analysis of bit $b_3$}
Bit $b_3$ exhibits the most intricate error pattern due to alternating decision regions. The error analysis reveals multiple decision boundaries at $\mathfrak{R}\{x\} = \{-6d, -2d, 2d, 6d\}$.

The conditional error probability for bit $b_3$ is:
{\small
	\begin{align}
		P_{(k),b_3}&=P_{(k),b_6}=\nonumber\\
		&\frac{1}{2}\bigg[2Q\left(\frac{2d\rho_{(k)}+\mathcal{I}_{(k)}}{\sqrt{2N_0}}\right)+2Q\left(\frac{6d\rho_{(k)}+\mathcal{I}_{(k)}}{\sqrt{2N_0}}\right)\nonumber\\
		&-Q\left(\frac{10d\rho_{(k)}+\mathcal{I}_{(k)}}{\sqrt{2N_0}}\right)-Q\left(\frac{14d\rho_{(k)}+\mathcal{I}_{(k)}}{\sqrt{2N_0}}\right)\nonumber\\
		&+2Q\left(\frac{18d\rho_{(k)}+\mathcal{I}_{(k)}}{\sqrt{2N_0}}\right)+Q\left(\frac{22d\rho_{(k)}+\mathcal{I}_{(k)}}{\sqrt{2N_0}}\right)\nonumber\\
		&-Q\left(\frac{26d\rho_{(k)}+\mathcal{I}_{(k)}}{\sqrt{2N_0}}\right)\bigg].
	\end{align}
}

\subsubsection{Average BER Calculation}
As demonstrated in \autoref{tab:mqam_err_dist} %\cite[TABLE I]{zhang2025error_noma}
, the error distances associated with bit $b_1$ form a proper subset of those for bits $b_2$ and $b_3$. This observation is crucial for computational efficiency: the error events captured by $b_1$ are already accounted for in the analysis of $b_2$ and $b_3$.

Therefore, the average \gls{ber} calculation requires only the contributions from bits $b_2$ and $b_3$:
{\small
	\begin{align}
		P_{(k)}&=\frac{1}{|\mathbb{S}|}\sum_{\mathbf{S}_{i}\in\mathbb{S}}\left[P_{(k),b_2}+P_{(k),b_3}\right]\nonumber\\
		&=\frac{1}{|\mathbb{S}|}\sum_{\mathbf{S}_{i}\in\mathbb{S}}\bigg[4Q\left(\frac{2d\rho_{(k)}+\mathcal{I}_{(k)}}{\sqrt{2N_0}}\right)+4Q\left(\frac{6d\rho_{(k)}+\mathcal{I}_{(k)}}{\sqrt{2N_0}}\right)\nonumber\\
		&+Q\left(\frac{18d\rho_{(k)}+\mathcal{I}_{(k)}}{\sqrt{2N_0}}\right)-Q\left(\frac{26d\rho_{(k)}+\mathcal{I}_{(k)}}{\sqrt{2N_0}}\right)\bigg].
	\end{align}
}

%% file: app_pdf_hnk.tex
\section{Closed-form PDFs of $|h_{n,(k)}|$, $k=1,2$}\label{app:pdf_hnk}
Consider a scenario where $|h_n|$ is a random variable representing the channel gain of \gls{ue} $n$, while $|h_m|$ is treated as a constant representing the channel gain of \gls{ue} $m$. Based on the channel ordering, we define:
\begin{itemize}
  \item $h_{n,(1)}$: the channel gain $|h_n|$ conditioned on $|h_n| \geq |h_m|$, with support on $[|h_m|, +\infty)$
  \item $h_{n,(2)}$: the channel gain $|h_n|$ conditioned on $|h_n| \leq |h_m|$, with support on $[0, |h_m|]$
\end{itemize}
These conditional distributions are examples of truncated distributions. The \gls{pdf} of a truncated random variable can be expressed as \cite{doi:10.1080/03610919108812975}:
{\small
\begin{align}
  f_{a\leq|h|\leq b}\left(x\right)=\frac{f_{|h|}\left(x\right)}{F_{|h|}\left(a\right)-F_{|h|}\left(b\right)},
  \label{eq:truncated_pdf}
\end{align}
}%
where $a$ and $b$ denote the lower and upper truncation boundaries, respectively, while $f_{|h|}\left(x\right)$ and $F_{|h|}\left(x\right)$ represent the original (untruncated) \gls{pdf} and \gls{cdf} of $|h|$. 

Special cases arise at the extremes of the ordering: for the strongest channel gain $|h_{(1)}|$ among all \gls{ue}s, the upper boundary extends to $b \to +\infty$, while for the weakest channel gain $|h_{(N)}|$, the lower boundary begins at $a = 0$.

To obtain the truncated Rayleigh distribution, we assume the channel gain of \gls{ue} $n$ satisfies $|h_{n}|\sim \mathcal{R}\left(\sigma_{n}\right)$, where $\sigma_{n}$ is the scale parameter for \gls{ue} $n$. The \gls{pdf} and \gls{cdf} of $|h_{n}|$ are respectively given by \cite{cavers2000mobile}:
{\small
\begin{align}
  f_{|h_n|}\left(x\right)&=\frac{2x}{\sigma_n^2}\exp\left(-\frac{x^2}{\sigma_n^2}\right), \quad x \geq 0,
  \label{eq:pdf_rayleigh}\\
  F_{|h_n|}\left(x\right)&=1-\exp\left(-\frac{x^2}{\sigma_n^2}\right), \quad x \geq 0.
  \label{eq:cdf_rayleigh}
\end{align}
}

For the truncated distribution conditioned on $|h_{n}|\geq|h_{m}|$, we have the support $x \in [|h_{m}|, +\infty)$. Substituting \Autoref{eq:pdf_rayleigh,eq:cdf_rayleigh} into \cref{eq:truncated_pdf} with $a=|h_{m}|$ and $b\to+\infty$:
{\small
\begin{align}
  {f}_{|h_{n}|\geq|h_{m}|}\left(x\right)&=\frac{f_{|h_{n}|}\left(x\right)}{F_{|h_{n}|}\left(|h_{m}|\right)-F_{|h_{n}|}\left(+\infty\right)}\nonumber\\
  &=\frac{2x}{\sigma_{n}^2}\exp\left(-\frac{x^2-|h_{m}|^2}{\sigma_{n}^2}\right).
  \label{eq:pdf_hn1}
\end{align}
}

Similarly, for the truncated distribution conditioned on $|h_{n}|\leq|h_{m}|$, we have the support $x \in [0, |h_{m}|]$. With $a=0$ and $b=|h_{m}|$:
{\small
\begin{align}
  f_{|h_{n}|\leq|h_{m}|}\left(x\right)&=\frac{f_{|h_{n}|}\left(x\right)}{F_{|h_{n}|}\left(0\right)-F_{|h_{n}|}\left(|h_{m}|\right)}\nonumber\\
  &=\frac{2x\exp\left(-\frac{x^2}{\sigma_{n}^2}\right)}{\sigma_{n}^2\left(1-\exp\left(-\frac{|h_{m}|^2}{\sigma_{n}^2}\right)\right)}.
  \label{eq:pdf_hn2}
\end{align}
}

The corresponding truncated \gls{pdf}s for $|h_{m}|$ conditioned on $|h_{m}|\geq|h_{n}|$ and $|h_{m}|\leq|h_{n}|$ can be derived analogously by interchanging the indices $n$ and $m$. For brevity and without loss of generality, we focus on \gls{ue} $n$ for the subsequent derivations.

So far, we have obtained the truncated \glspl{pdf} of the channel gains for orders $k=1,2$. However, treating the boundaries of these truncated \glspl{pdf} as constants is overly restrictive, since these boundaries are random variables that depend on the channel gains of other \glspl{ue}. To address this limitation, we derive the marginal \glspl{pdf} of $|h_{n,(1)}|$ and $|h_{n,(2)}|$ by integrating the joint \glspl{pdf} over their respective random boundaries.

Consider a random variable $y\sim\mathcal{R}(\sigma_m)$ representing $|h_m|$. For the case where $|h_n| \geq y$, the marginal \gls{pdf} is obtained by integrating over the range $0 \leq y \leq x$:
{\small
\begin{align}\label{eq:cf_pdf_hn1}
  f_{|h_{n,(1)}|}(x) &= \int_0^{x} f_{|h_{n}|\geq y}(x) f_{|h_{m}|}(y) \, dy\nonumber\\
  &= \int_0^{x}\frac{2x}{\sigma_{n}^2}\exp\left(-\frac{x^2-y^2}{\sigma_{n}^2}\right)\frac{2y}{\sigma_{m}^2}\exp\left(-\frac{y^2}{\sigma_{m}^2}\right) \, dy\nonumber\\
  &= \frac{2x}{\sigma_{n}^2-\sigma_{m}^2}\left[\exp\left(-\frac{x^2}{\sigma_{n}^2}\right)-\exp\left(-\frac{x^2}{\sigma_{m}^2}\right)\right]\footnotemark.
\end{align}
}
\footnotetext{The closed-form solution in \cref{eq:cf_pdf_hn1} was obtained using symbolic computation tools \cite{Mathematica}.}

Similarly, for the case where $|h_n| \leq y$, the marginal \gls{pdf} is obtained by integrating over the range $x \leq y < \infty$\footnote{The integration limits are effectively $(0, \infty)$ since the truncated \gls{pdf} $f_{|h_{n}|\leq y}(x)$ inherently enforces the condition $x < y$.}:
{\small
\begin{align}\label{eq:cf_pdf_hn2}
  f_{|h_{n,(2)}|}(x) &= \int_x^{\infty} f_{|h_{n}|\leq y}(x) f_{|h_{m}|}(y) \, dy\nonumber\\
  &= \int_0^{\infty}\frac{2x\exp\left(-\frac{x^2}{\sigma_{n}^2}\right)}{\sigma_{n}^2\left(1-\exp\left(-\frac{y^2}{\sigma_{n}^2}\right)\right)}\frac{2y}{\sigma_{m}^2}\exp\left(-\frac{y^2}{\sigma_{m}^2}\right) \, dy\nonumber\\
  &= \frac{2x\exp\left(-\frac{x^2}{\sigma_{n}^2}\right)}{\sigma_{m}^2\sigma_{n}^2}\underbrace{\int_0^{\infty}\frac{y \exp\left(-\frac{y^2}{\sigma_{m}^2}\right)}{1-\exp\left(-\frac{y^2}{\sigma_{n}^2}\right)} \, dy}_{\text{cf. \cite{zwillinger2007table}, 3.311.4 with $a=1$}}\nonumber\\
  &= \frac{2x\exp\left(-\frac{x^2}{\sigma_{n}^2}\right)}{\sigma_{m}^2\sigma_{n}^2}\lim_{a\to 0}\sum_{k=0}^{\infty}a^k\frac{\sigma_{m}^2\sigma_{n}^2}{\sigma_{n}^2+k\sigma_{m}^2}\nonumber\\
  &\approx \frac{2x\exp\left(-\frac{x^2}{\sigma_{n}^2}\right)}{\sigma_{n}^2}.
\end{align}
}

%% file: app_truncated_Rhm.tex
\section{Truncated Distribution of the Real Part of \( h_m \)}\label{app:truncated_Rhm}
This appendix analyzes the distribution of $\mathfrak{R}\{h_{m,(k)}\}$ and discusses the challenges in deriving its closed-form \gls{pdf}.

\subsection{Background}
Given that $h_m \sim \mathcal{CN}(0, \sigma_{m}^2)$, the real part follows $\mathfrak{R}\{h_{m}\} \sim \mathcal{N}(0, \sigma_{m}^2/2)$. The ordered statistic $\mathfrak{R}\{h_{m,(k)}\}$ represents the real part of $h_m$ under specific ordering conditions on the channel magnitudes.

\subsection{Case 1: Strong Channel Condition ($|h_{m}| \geq |h_{n}|$)}
Under this condition, the constraint $|h_{m}|^2 \geq |h_{n}|^2$ imposes bounds on $\mathfrak{R}\{h_{m}\}$:
{\small
\begin{equation}
\mathfrak{R}\{h_{m}\}^2 + \mathfrak{I}\{h_{m}\}^2 \geq |h_{n}|^2.
\end{equation}
}

This yields the following truncation bounds:
\begin{itemize}
    \item If $\mathfrak{R}\{h_{m}\} \geq 0$: $\mathfrak{R}\{h_{m}\} \geq \mathcal{L}_{m} = \sqrt{|h_{n}|^2 - \mathfrak{I}\{h_{m}\}^2}$,
    \item If $\mathfrak{R}\{h_{m}\} < 0$: $\mathfrak{R}\{h_{m}\} \leq \mathcal{U}_{m} = -\sqrt{|h_{n}|^2 - \mathfrak{I}\{h_{m}\}^2}$.
\end{itemize}

Therefore, $\mathfrak{R}\{h_{m,(1)}\}$ follows:
{\small
\begin{equation}
\mathfrak{R}\{h_{m,(1)}\} \sim \mathcal{TN}\left(0, \frac{\sigma_{m}^2}{2}, a_{m}, b_{m}\right),
\end{equation}
}%
where $(a_{m}, b_{m}) = (\mathcal{L}_{m}, +\infty)$ if $\mathfrak{R}\{h_{m}\} \geq 0$ and $(a_{m}, b_{m}) = (-\infty, \mathcal{U}_{m})$ if $\mathfrak{R}\{h_{m}\} < 0$ \cite{robert1995simulation}.

\subsection{Case 2: Weak Channel Condition ($|h_{m}| \leq |h_{n}|$)}
For this condition, $\mathfrak{R}\{h_{m}\}$ is bounded within the interval $[-\mathcal{B}_m, \mathcal{B}_m]$ where:
{\small
\begin{equation}
\mathcal{B}_m = \sqrt{|h_{n}|^2 - \mathfrak{I}\{h_{m}\}^2}.
\end{equation}
}

Thus:
{\small
\begin{equation}
\mathfrak{R}\{h_{m,(2)}\} \sim \mathcal{TN}\left(0, \frac{\sigma_{m}^2}{2}, -\mathcal{B}_m, \mathcal{B}_m\right).
\end{equation}
}

\subsection{Challenges in Deriving the Closed-Form PDF}
The \gls{pdf} of $\mathfrak{R}\{h_{m,(k)}\}$ requires integrating the joint \gls{pdf} over random boundaries. The primary challenges are:
\begin{enumerate}
    \item The truncation bounds depend on two random variables: $|h_{n}|$ and $\mathfrak{I}\{h_{m}\}$
    \item The bounds themselves are nonlinear functions of these random variables
    \item The resulting integral involves the product of multiple probability densities with interdependent limits
\end{enumerate}

These factors make analytical evaluation intractable, necessitating numerical methods or approximations for practical applications.